\documentclass[11pt]{article}
\usepackage{amsfonts}
\usepackage{multicol}
\usepackage{amssymb,mathrsfs,latexsym}

\begin{document}
\newcommand{\qed}{\hphantom{.}\hfill $\Box$\medbreak}
\newcommand{\Proof}{\noindent{\bf Proof \ }}

\newtheorem{theorem}{Theorem}[section]
\newtheorem{lemma}[theorem]{Lemma}
\newtheorem{corollary}[theorem]{Corollary}
\newtheorem{remark}[theorem]{Remark}
\newtheorem{example}[theorem]{Example}
\newtheorem{definition}[theorem]{Definition}
\newtheorem{construction}[theorem]{Construction}

\title{\large{\bf Determination of
sizes of optimal three-dimensional optical orthogonal codes of
weight three with the AM-OPP restriction\footnote{Supported by the
NSFC under Grant 11401582 and the NSFHB under Grant A2015507019 (L.
Wang), and the NSFC under Grants 11271042 and 11431003 (Y. Chang)}}}

\author
 {
       Lidong Wang$^{1,2}$, Yanxun Chang$^{1,}$\thanks{corresponding author. E-mail address: yxchang@bjtu.edu.cn}\\
     \small $^1$Institute of Mathematics, Beijing Jiaotong University, \\ \small Beijing 100044, P. R.
     China\\
     \small $^2$Department of Basic Courses, Chinese People's Armed Police
      Force Academy,\\ \small Langfang 065000, P. R. China
  }
\date{}

\maketitle

\noindent {\bf Abstract:} In this paper, we further investigate the
constructions on three-dimensional $(u\times v\times w,k,1)$ optical
orthogonal codes with the at most one optical pulse per
wavelength/time plane restriction (briefly AM-OPP $3$-D $(u\times
v\times w,k,1)$-OOCs) by way of the corresponding designs. Several
new auxiliary designs such as incomplete holey group divisible
designs and incomplete group divisible packings are introduced and
therefore new constructions are presented. As a consequence, the
exact number of codewords of an optimal AM-OPP $3$-D $(u\times
v\times w,3,1)$-OOC is finally determined for any positive integers
$v,w$ and $u\geq3$.


\section{Introduction}

Optical code-division multiple access (OCDMA) is one of the
attractive multiple-access schemes for optical networks. Good
features of OCDMA include accommodation of burst traffic,
asynchronous transmission, etc. Optical orthogonal codes (OOCs) have
been designed for OCDMA. A one-dimensional ($1$-D) optical
orthogonal code ($1$-D OOC) is a set of binary sequences having good
auto and cross-correlations. $1$-D OOCs were first suggested by
\cite{csw} in $1989$. Since then there are many researches on $1$-D
OOCs (see, e.g., \cite{ab,be,cfm,fm,gy,mc1,mc2,y}). $1$-D OOCs
spread optical pulses only in time domain. One limitation of $1$-D
OOCs is that the length of the sequence increases rapidly when the
number of users or the weight of the code is increased, which means
large bandwidth expansion is required if a big number of codewords
is needed. To lessen this problem, two-dimensional ($2$-D) optical
orthogonal codes ($2$-D OOCs) are proposed \cite{yk}. Optical pulses
in $2$-D OOCs are spread in both time and wavelength domain. The
performance of the OCDMA system is much improved. There is a
considerable literature on $2$-D OOC constructions. We refer the
readers to the survey \cite{okeb} for more details. To further
improve the performance of $2$-D OOCs, three-dimensional ($3$-D)
optical orthogonal codes ($3$-D OOCs) are introduced \cite{kyp}. In
$3$-D OOCs, optical pulses are spread in space, time and wavelength
domain. We define a $3$-D OOC formally as follows.

Let $u,v,w,k$ and $\lambda$ be positive integers, where $u,v$ and
$w$ are the number of spatial channels, wavelengths, and time slots,
respectively.  A  {\em three-dimensional $(u\times v\times
w,k,\lambda)$ optical orthogonal code} (briefly $3$-D $(u\times
v\times w,k,\lambda)$-OOC), $\cal{C}$, is a family of $u\times
v\times w$ $(0, 1)$ arrays (called {\em codewords}) of Hamming
weight $k$ satisfying: for any two arrays $A=[a(i,j,l)]$,
$B=[b(i,j,l)]\in\cal{C}$ and any integer $\tau$:
$$\sum_{i=0}^{u-1}\sum_{j=0}^{v-1}\sum_{l=0}^{w-1}a(i,j,l)b(i,j,l+\tau)\leq\lambda,$$
where either $A\neq B$ or $\tau\not\equiv0\hspace{-0.3cm}\pmod{w}$,
and the arithmetic $l+\tau$ is reduced modulo $w$.

A wavelength/time plane is also called a \emph{spatial plane}. There
are several classes of $3$-D OOCs of interests. The following two
additional restrictions on the placement of pulses are often placed
within the codewords of a $3$-D OOC:
\begin{itemize}
\item the one-pulse per plane (OPP) restriction: each
codeword contains exactly one optical pulse per spatial plane.
\item the at-most one-pulse per plane (AM-OPP) restriction: each
codeword contains at most one optical pulse per spatial plane.
\end{itemize}

It is clear that $3$-D OOCs which satisfy the OPP restriction must
satisfy the AM-OPP restriction. In what follows, a $3$-D $(u\times
v\times w,k,\lambda)$-OOC with the AM-OPP restriction is denoted by
an AM-OPP $3$-D $(u\times v\times w,k,\lambda)$-OOC.

The number of codewords of a $3$-D OOC is called its \emph{size}.
Let $\Phi(u\times v\times w,k,\lambda)$ denote the largest possible
size of an AM-OPP $3$-D $(u\times v\times w,k,\lambda)$-OOC. Based
on the Johnson bound \cite{j} for constant weight codes, an upper
bound of $\Phi(u\times v\times w,k,\lambda)$ is given by Shum
\cite{s}.

\begin{lemma}\cite{s}\label{johnson}
$\Phi(u\times v\times w,k,\lambda)\leq J(u\times v\times
w,k,\lambda)$ holds for any positive integers $u\geq k>\lambda$,
where
$$J(u\times v\times
w,k,\lambda)=\lfloor\frac{uv}{k}\lfloor\frac{vw(u-1)}{k-1}
\cdots\lfloor\frac{vw(u-\lambda)}{k-\lambda}\rfloor\cdots\rfloor\rfloor.$$
\end{lemma}

An AM-OPP $3$-D $(u\times v\times w,k,\lambda)$-OOC with
$\Phi(u\times v\times w,k,\lambda)$ codewords is said to be
\emph{optimal}. Furthermore, we say that an optimal AM-OPP $3$-D
$(u\times v\times w,k,\lambda)$-OOC is \emph{perfect} if
$$\Phi(u\times v\times w,k,\lambda)=v^{\lambda+1}w^{\lambda}\frac{u(u-1)\cdots(u-\lambda)}{k(k-1)\cdots(k-\lambda)}.$$

Some work has been done for optimal AM-OPP $3$-D $(u\times v\times
w,3,1)$-OOCs. Shum \cite{s} has solved the existence problem of
perfect AM-OPP $3$-D $(u\times v\times w,3,1)$-OOCs.

\begin{theorem}\cite{s}\label{perfect}
A perfect AM-OPP $3$-D $(u\times v\times w,3,1)$-OOC exists if and
only if

$(1)$ $u=3:$ $w$ is odd, or $w$ is even and $v$ is even;

$(2)$ $u\geq4:$ $(u-1)vw\equiv0\hspace{-0.3cm}\pmod{2}$,
$u(u-1)vw\equiv0\hspace{-0.3cm}\pmod{3}$, and
$v\equiv0\hspace{-0.3cm}\pmod{2}$ when
$u\equiv2,3\hspace{-0.3cm}\pmod{4}$ and
$w\equiv2\hspace{-0.3cm}\pmod{4}$.
\end{theorem}

The authors \cite{wc} studied the constructions for general optimal
AM-OPP $3$-D $(u\times v\times w,k,1)$-OOCs. We improved the upper
bound on $\Phi(u\times v\times w,3,1)$ and determined the size of
optimal AM-OPP $3$-D $(u\times v\times w,3,1)$-OOCs with some
possible exceptions.

\begin{lemma}\label{bound}\cite{wc}
$\Phi(u\times v\times w,3,1)\leq J^{*}(u\times v\times w,3,1)$,
where
$$
    J^\ast(u\times v\times w,3,1)=\left\{
          \begin{array}{ll}
               J(u\times v\times w,3,1)-1, & {\rm{if}}\
               v\equiv1\hspace{-0.3cm}\pmod2,
               w\equiv0\hspace{-0.3cm}\pmod4,\\ & ~~~{\rm{and}}~ u=3,\\
                &~\hspace{-0.12cm}{\rm{or}}~v\equiv1\hspace{-0.3cm}\pmod2,w\equiv2\hspace{-0.3cm}\pmod4,
                \\ & ~~~{\rm{and}}~ u\equiv3,6,7,10\hspace{-0.3cm}\pmod{12},\\
                &~\hspace{-0.12cm}{\rm{or}}~vw\equiv6\hspace{-0.3cm}\pmod{12},w\equiv2\hspace{-0.3cm}\pmod{4},
                \\ & ~~~{\rm{and}}~ u\equiv2,11\hspace{-0.3cm}\pmod{12},\\
                &~\hspace{-0.12cm}{\rm{or}}~(u-1)v^2w\equiv4\hspace{-0.3cm}\pmod{6},
                \\ & ~~~{\rm{and}}~ u\equiv2\hspace{-0.3cm}\pmod{3},\\
                &~\hspace{-0.12cm}{\rm{or}}~w=2,u(u-1)v^2\equiv8\hspace{-0.3cm}\pmod{12},\\

               J(u\times v\times w,3,1), & \rm{otherwise}.
          \end{array}
       \right.
$$.
\end{lemma}

\begin{theorem}\cite{bl,wc,y1}\label{ieee ooc}
An optimal AM-OPP $3$-D $(u\times v\times w,3,1)$-OOC with
$J^\ast(u\times v\times w,3,1)$ codewords exists if either $w=1$,
$v\geq1$ and $u\geq3$, or $w>1$, $v\geq1$, $u\geq3$ and
$u\not\equiv2\hspace{-0.3cm}\pmod{6}$, except possibly for
$u\equiv11\hspace{-0.3cm}\pmod{12}$,
$v\equiv1,5\hspace{-0.3cm}\pmod{6}$ and $w\equiv10\pmod{12}$.
\end{theorem}

In this paper, we further study the combinatorial constructions for
general optimal AM-OPP $3$-D $(u\times v\times w,k,1)$-OOCs. We
introduce some new auxiliary designs and new constructions. As a
consequence, we determine the sizes of optimal AM-OPP $3$-D
$(u\times v\times w,3,1)$-OOCs for any positive integers $v,w$ and
$u\geq3$.

The rest of this paper is organized as follows. In Section II, we
give a combinatorial description of AM-OPP $3$-D $(u\times v\times
w,k,1)$-OOCs. In Sections III and IV, we introduce three types of
auxiliary designs and their existence results respectively, which
will be used in the constructions of AM-OPP $3$-D $(u\times v\times
w,k,1)$-OOCs. In Section V, we obtain the main result of this paper.
Section VI gives a concluding remark. Note that in order to save
space, we do not list most of the date used in this paper. The
interested reader may access them in the Supporting
Information\footnote{Supporting Information, arXiv:}.

\section{Combinatorial description}

Optimal AM-OPP $3$-D $(u\times v\times w,k,1)$-OOCs are closely
related to combinatorial designs. In this section, we define some
terminologies and notations in design theory and state the link
between AM-OPP $3$-D $(u\times v\times w,k,1)$-OOCs and the designs.
We always assume that $I_n=\{0,1,\ldots,n-1\}$ and denote by $Z_n$
the additive group of integers modulo $n$ throughout this paper.

Let $K$ be a set of positive integers. A \emph{group divisible
packing} ($K$-GDP) is a triple $(X,\mathcal{G},\mathcal{B})$ which
satisfying the following properties:\vspace{0.1cm}

(1) $X$ is a finite set of \emph{points};\vspace{0.1cm}

(2) $\mathcal{G}$ is a partition of $X$ into subsets (called
\emph{groups});\vspace{0.1cm}

(3) $\mathcal{B}$ is a collection of subsets of $X$ (called
\emph{blocks}), each of size from $K$, such that any pair of $X$
appears in a group or in at most one block, but not
both.\vspace{0.1cm}

If $\mathcal{G}$ contains $u_i$ groups of size $g_i$, $1\leq i\leq
r$, then we call $g_{1}^{u_1} g_{2}^{u_2}\cdots g_{r}^{u_r}$ the
\emph{group type} (or \emph{type}) of the GDP. If $K=\{k\}$, we
further omit the braces to simply write $k$ for $K$. If every
admissible pair appears in exactly one block, this GDP is also
referred to as a \emph{group divisible design} and denoted by a
$K$-GDD. A $K$-GDD of type $1^u$ is also called a \emph{pairwise
balance design}, or PBD$(u,K)$ for short.

We record some results concerning GDDs and PBDs, which will be used
later.

\begin{lemma}\label{gtw1}\cite{chr}
Let $g$, $t$ and $w$ be positive integers. Then there exists a
$3$-GDD of type $g^tw^1$ if and only if the following conditions are
all satisfied: $(i)$ $t\geq3$, or $t=2$ and $w=g$; $(ii)$ $w\leq
g(t-1)$; $(iii)$ $g(t-1)+w\equiv0\hspace{-0.3cm}\pmod 2$; $(iv)$
$gt\equiv0\hspace{-0.3cm}\pmod 2$; and $(v)$
$g^2t(t-1)+2gtw\equiv0\hspace{-0.3cm}\pmod6$.
\end{lemma}

\begin{lemma}\label{b(4,5)}\cite{abg}
A PBD$(u,\{4,5,6\})$ exists for any positive integer $u\geq4$ and
$u\not\in\{7,8,9$, $10,11,12,14,15,18,19,23\}$.
\end{lemma}

Suppose $(X,\mathcal{G},\mathcal{B})$ is a GDD. If there exists a
partition
$\mathcal{P}=\{\mathcal{P}_1,\mathcal{P}_2,\ldots,\mathcal{P}_s\}$
of $\mathcal{B}$ such that each $\mathcal{P}_i$ (called a
\emph{parallel class}) is a partition of $X$ for $1\leq i\leq s$,
then this GDD is said to be \emph{resolvable} and denoted by an
RGDD. If the set of block size of a GDD (PBD) is denoted by
$K\cup\{k^*\}$, we means that this GDD (PBD) has exactly one block
of size $k$.\vspace{0.2cm}

Now we use an RGDD to construct a PBD which has exactly one block of
size $5$.

\begin{lemma}\label{5^*}
There exists a PBD$(u,\{3,4,5^*\})$ for any positive integer
$u\equiv2\hspace{-0.3cm}\pmod{6}$ and $u\geq14$.
\end{lemma}

\Proof A PBD$(14,\{3,4,5^*\})$ exists by Lemma $5.3$ of \cite{h}.
When $u\geq20$, a $3$-RGDD of type $3^{(u-5)/3}$ can be found in
\cite{gm}, which contains $(u-8)/2$ parallel classes. Adjoining $5$
infinite points to complete five parallel classes, then we obtain a
$\{3,4\}$-GDD of type $3^{(u-5)/3}5^1$, which is also a
PBD$(u,\{3,4,5^*\})$. \qed

Next we introduce the concept of $w$-cyclic GDPs, which are very
useful in the constructions of AM-OPP $3$-D OOCs.\vspace{0.2cm}

Let $w$ be a common divisor of $g_i$, say $g_i=v_iw$, $1\leq i\leq
r$. Suppose $(X,\mathcal{G},\mathcal{B})$ is a $K$-GDP of type
$g_{1}^{u_1} g_{2}^{u_2}\cdots g_{r}^{u_r}$. If there is a
permutation $\pi$ on $X$ which is the product of
$\sum_{i=1}^{r}v_iu_i$ disjoint $w$-cycles, fixes every group, and
leaves $\mathcal{B}$ invariant, then this design is said to be
\emph{$w$-cyclic}. A $w$-cyclic $k$-GDP (GDD) of type $w^u$ is
called a \emph{semi-cyclic} $k$-GDP (GDD) of type $w^u$ and denoted
by a $k$-SCGDP (SCGDD) of type $w^u$.

Without loss of generality,  for a $w$-cyclic $K$-GDP of type
$(vw)^u$, we always identify $X=I_u\times I_v\times Z_w$,
$\mathcal{G}=\{\{i\}\times I_v\times Z_w:i\in I_u\}$ and the
permutation $\pi:$ $(i,j,x)\mapsto(i,j,x+1)$ mod $(-,-,w)$.

Any permutation $\pi$ partitions $\mathcal{B}$ into equivalence
classes called the \emph{block orbits} under $\pi$. A set of base
blocks is an arbitrary set of representatives for these block orbits
of $\mathcal{B}$. A $w$-cyclic $k$-GDP of type $(vw)^u$ is called
\emph{optimal} if it contains the largest possible number of base
blocks.

Obviously, the following lemma holds. We omit its proof.

\begin{lemma}\label{scgdp to w-cyclic gdp}
If there exists a $k$-SCGDP of type $(vw)^u$ with $b$ base blocks,
then there exists a $w$-cyclic $k$-GDP of type $(vw)^u$ with $vb$
base blocks.
\end{lemma}

The authors \cite{wc} established the equivalence between an optimal
AM-OPP 3-D $(u\times v\times w,k,1)$-OOC and an optimal $w$-cyclic
$k$-GDP of type $(vw)^u$.

\begin{theorem}\label{equi}\cite{wc}
An optimal AM-OPP $3$-D $(u\times v\times w, k, 1)$-OOC is
equivalent to an optimal $w$-cyclic $k$-GDP of type $(vw)^u$.
\end{theorem}

Combining with Theorems \ref{perfect} and \ref{ieee ooc}, we have
the following result.

\begin{corollary}\label{w-cyclic 3gdp}
$(1)$ A $w$-cyclic $3$-GDD of type $(vw)^u$ exists if and only if
the conditions given in Theorem \ref{perfect} are satisfied.

$(2)$ An optimal $w$-cyclic $3$-GDP of type $(vw)^u$ with
$J^*(u\times v\times w,3,1)$ base blocks exists for the conditions
shown in Theorem \ref{ieee ooc}.
\end{corollary}

\section{Holey group divisible designs}
In this section, we introduce our first auxiliary design called a
\emph{holey group divisible design} (HGDD), which will play an
important role in constructions of optimal $w$-cyclic $3$-GDPs.

A $k$-HGDD is defined to be a quadruple
($X,\mathcal{G},\mathcal{H},\mathcal{B}$) which satisfies the
following properties:\vspace{0.15cm}

(1) $X$ is an $(u\sum_{i=1}^tg_i)$-set (of
\emph{points});\vspace{0.15cm}

(2) $\mathcal{G}=\{G_1,\cdots,G_u\}$ is a partition of $X$ into $u$
subsets (called \emph{groups}) of points $\sum_{i=1}^tg_i$
each;\vspace{0.15cm}

(3) $\mathcal{H}=\{H_1,\cdots,H_t\}$ is another partition of $X$
into $t$ subsets (called \emph{holes}) of points $ug_i$ each such
that $|H_i\cap G_j|=g_i$ for any $1\leq i\leq t$ and $1\leq j\leq
u$;\vspace{0.15cm}

(4) $\mathcal{B}$ is a collection of $k$-subsets of $X$ (called
\emph{blocks}) such that any pair of $X$ from two distinct groups
appears in a hole or exactly in one block but not both, and no other
pairs of $X$ occur in any block.\vspace{0.1cm}

If $\mathcal{H}$ contains $t_i$ holes of size $ug_i$, $1\leq i\leq
r$, we then use an ``exponential" notation $g_1^{t_1}\cdots
g_r^{t_r}$ to denote the multiset $T=\{g_j:j=1,2,\ldots,t\}$ and
call $(u,T)$ the type of the deisgn.

The necessary and sufficient conditions on the existence of
$3$-HGDDs of type $(u,g^tw^1)$ have been determined by Wang and Yin
\cite{wy1}.

\begin{theorem}\cite{wy1}\label{3hgdd}
Let $u,t,g$ and $w$ be nonnegative integers. The necessary and
sufficient conditions for the existence of a $3$-HGDD of type
$(u,g^tw^1)$ are that $u\geq3$, $t=2$ and $g=w$; or $t\geq3$, $0\leq
w\leq g(t-1)$, $gt(u-1)\equiv0\hspace{-0.3cm}\pmod2$,
$(u-1)(w-g)\equiv0\hspace{-0.3cm}\pmod2$ and
$gtu(u-1)(g(t-1)-w)\equiv0\hspace{-0.3cm}\pmod3$.
\end{theorem}

For the purpose of constructing optimal $w$-cyclic $k$-GDP of type
$(vw)^u$, the HGDDs must admit some special
permutations.\vspace{0.1cm}

Suppose $(X,\mathcal{G},\mathcal{H},\mathcal{B})$ is a $k$-HGDD of
type $(u,(gh)^{mt})$. If there is a permutation $\pi$ on $X$ which
is the product of $uhm$ disjoint $gt$-cycles, fixes every group,
leaves $\mathcal{B}$ invariant, and partitions the $mt$ elements of
$\mathcal{H}$ into $m$ equivalence classes, then this design is said
to be \emph{$(h,gt,m)$-cyclic}.

Without loss of generality, we always identify $X=I_u\times
I_{h}\times I_{m}\times Z_{gt}$, ${\mathcal{G}}=\{\{i\}\times
I_{h}\times I_{m}\times Z_{gt}: i\in I_u\}$ and
${\mathcal{H}}=\{I_u\times
I_{h}\times\{i\}\times\{0+j,t+j,\ldots,(g-1)t+j\}:(i,j)\in
I_{m}\times Z_t\}$. In this case, the permutation $\pi$ can be taken
as $(x,y,z,w)\mapsto(x,y,z,w+1)$ mod $(-,-,-,gt)$.

A $(1,gt,1)$-cyclic $k$-HGDD of type $(u,g^t)$ is also referred to
as a \emph{semi-cyclic} $k$-HGDD of type $(u,g^t)$ and denoted by
$k$-SCHGDD of type $(u,g^t)$. Clearly, under the action of $\pi$,
$\mathcal{B}$ can be partitioned into equivalence classes called the
\emph{block orbits}. A set of base blocks is a set of
representatives for these block orbits of $\mathcal{B}$.

\begin{example}\label{(5,3^3)}
There exists a $(3,3,1)$-cyclic $3$-HGDD of type $(5,3^3)$.
\end{example}

\Proof Let $X=Z_5\times Z_3\times Z_3$, $\mathcal{G}=\{\{i\}\times
Z_3\times Z_3:i\in Z_5\}$, and $\mathcal{H}=\{Z_5\times Z_3\times
\{j\}:j\in Z_3\}$. Developing the following $4$ initial base blocks
by $(+i,+j,-)$ mod $(5,3,-)$ yields all $60$ base blocks of the
required design, where $(i,j)\in Z_5\times Z_3$.\vspace{0.2cm}

{\small\hspace{2cm}$\{(0,0,0),(1,0,1),(2,1,2)\}$,
$\{(0,0,0),(1,0,2),(2,2,1)\}$,

\hspace{2cm}$\{(0,0,0),(1,2,1),(3,0,2)\}$,
$\{(0,0,0),(1,1,2),(3,1,1)\}$}. \qed

We give our first ``Filling Construction" via $(h,gt,1)$-cyclic
$k$-HGDDs.

\begin{construction}\cite{wc}{\rm{(Filling Construction-I)}}\label{fill1}
Suppose that the following designs exist:\vspace{0.2cm}

\hspace{-0.57cm}$(1)$ an $(h,gt,1)$-cyclic $k$-HGDD of type
$(u,(gh)^t)$ with $b$ base blocks;\vspace{0.1cm}

\hspace{-0.57cm}$(2)$ a $g$-cyclic $k$-GDP of type $(gh)^u$ with $f$
base blocks.\vspace{0.2cm}

\hspace{-0.58cm}Then, there exists a $gt$-cyclic $k$-GDP of type
$(ght)^u$ with $b+f$ base blocks.
\end{construction}

For applying Construction \ref{fill1}, we need some
$(h,gt,m)$-cyclic $k$-HGDDs. Below, we investigate its constructive
methods and apply these constructions to obtain some existence
results. Constructions \ref{hgdd to hgdd} and \ref{inf1} can be
found in \cite{wc}.

\begin{construction}\label{hgdd to hgdd}
If there exists a $k$-SCHGDD of type $(u,(gh)^t)$, then there exists
an $(h,g,t)$-cyclic $k$-HGDD of type $(u,(gh)^t)$.
\end{construction}

Construction \ref{hgdd to hgdd} indicates that $k$-SCHGDDs are
useful in the construction of $(h,g,t)$-cyclic $k$-HGDDs. We quote
the result on $3$-SCHGDDs for later use.

\begin{theorem}\label{schgdd}\cite{f,fww}
There exists a $3$-SCHGDD of type $(u,g^t)$ if and only if $u,t\geq
3$, $(t-1)(u-1)g\equiv 0\hspace{-0.3cm}\pmod {2}$, and
$(t-1)u(u-1)g\equiv 0\hspace{-0.3cm}\pmod {6}$ except when

$(1)$ $u\equiv 3,7\hspace{-0.3cm}\pmod {12}$, $g\equiv
1\hspace{-0.3cm}\pmod {2}$ and $t\equiv 2\hspace{-0.3cm}\pmod {4}$;
$(2)$ $u=3$, $g\equiv 1\hspace{-0.3cm}\pmod {2}$ and $t\equiv
0\hspace{-0.3cm}\pmod {2}$; $(3)$ $u=t=3$, $g\equiv
0\hspace{-0.3cm}\pmod {2}$; $(4)$ $(u,g,t)\in\{(5,1,4),(6,1,3)\}$;

\hspace{-0.57cm}and possibly when

$(1)$ $t=8$, either $g\equiv 1\hspace{-0.3cm}\pmod {2}$ and $u\equiv
1,3\hspace{-0.3cm}\pmod {6}$ and $u\geq7$, or $g\equiv
3\hspace{-0.3cm}\pmod {6}$ and $u\equiv5\hspace{-0.3cm}\pmod {6}$;
$(2)$ $u\equiv1,9\hspace{-0.3cm}\pmod {12}$, $g\equiv
1\hspace{-0.3cm}\pmod {2}$ and $t\equiv 2\hspace{-0.3cm}\pmod {4}$;
$(3)$ $u\equiv5\pmod {6}$ and $u\geq11$, either $g\equiv
3\hspace{-0.3cm}\pmod {6}$ and $t\equiv 2\hspace{-0.3cm}\pmod {4}$,
or $g\equiv 1,5\hspace{-0.3cm}\pmod {6}$ and $t\equiv
10\hspace{-0.3cm}\pmod {12}$.
\end{theorem}

\Proof Reference \cite{f} deals with the case when $u=8$, $g\equiv
2,10\hspace{-0.3cm}\pmod {12}$ and $t\equiv 7,10\pmod {12}$. Other
cases can be found in \cite{fww}.\qed

\begin{construction}\label{inf1}{\rm{(Inflation-I)}}
Suppose that there exists an $(h,gt,m)$-cyclic $k$-HGDD of type
$(u,(gh)^{mt})$. If there exists a $w$-cyclic $l$-GDD of type
$(vw)^k$, then there exists an $(hv,gtw,m)$-cyclic $l$-HGDD of type
$(u,(ghvw)^{mt})$.
\end{construction}

Constructions \ref{pbd1} and \ref{fillhgdd} are simple
generalizations of Constructions $3.2$ and $3.1$ of \cite{fwc},
respectively. It is a routine matter of checking their correctness.

\begin{construction}\label{pbd1}
Suppose there exists a PBD$(u,K)$. If there exists an
$(h,gt,m)$-cyclic $l$-HGDD of type $(k,(gh)^{mt})$ for any $k\in K$,
then there exists an $(h,gt,m)$-cyclic $l$-HGDD of type
$(u,(gh)^{mt})$.
\end{construction}

\begin{construction}\label{fillhgdd}
If there exist an $(h,gt,1)$-cyclic $k$-HGDD of type $(u,(gh)^{t})$
and an $(h,g,1)$-cyclic $k$-HGDD $(u,h^{g})$, then there exists an
$(h,gt,1)$-cyclic $k$-HGDD of type $(u,h^{gt})$.
\end{construction}

Now we are ready to apply these constructions to obtain some
infinite classes of $(h,gt,m)$-cyclic $3$-HGDDs, which are essential
to our work.

\begin{lemma}\label{(v,w,1)-cyclic hgdd1}
Suppose $u\equiv0,1\hspace{-0.3cm}\pmod{3}$ and $u\geq3$. Let $g$ be
a positive integer. Then there exists a $(g,t,1)$-cyclic $3$-HGDD of
type $(u,g^{t})$ for any $t\equiv1\hspace{-0.3cm}\pmod{2}$ and
$t\geq3$, except for $(u,g,t)=(6,1,3)$.
\end{lemma}

\Proof By Theorem \ref{schgdd}, there is a $3$-SCHGDD of type
$(u,1^{t})$, which is also a $(1,t,1)$-cyclic $3$-HGDD of type
$(u,1^{t})$. Inflate it by a $1$-cyclic $3$-GDD of type $g^3$ from
Corollary \ref{w-cyclic 3gdp}. Hence, by Construction \ref{inf1}, we
obtain a $(g,t,1)$-cyclic $3$-HGDD of type $(u,g^{t})$. \qed

\begin{lemma}\label{(v,w,1)-cyclic hgdd}
Suppose $u\equiv2\hspace{-0.3cm}\pmod{6}$ and $u\geq8$. Let $h$ be a
positive integer. Then there exists an $(h,gt,1)$-cyclic $3$-HGDD of
type $(u,(gh)^{t})$ if $g,t$ satisfy one of following
conditions.\vspace{0.1cm}

$(1)$ $g=1$, $t\equiv1\hspace{-0.3cm}\pmod{6}$ and
$t\geq7$;\hspace{0.3cm} $(2)$ $g=2$,
$t\equiv1\hspace{-0.3cm}\pmod{3}$ and $t\geq4$;

$(3)$ $g=3$, $t\equiv1\hspace{-0.3cm}\pmod{2}$ and
$t\geq3$;\hspace{0.3cm} $(4)$ $g=4$,
$t\equiv1\hspace{-0.3cm}\pmod{3}$ and $t\geq4$;

$(5)$ $g=6$, $t\equiv1\hspace{-0.3cm}\pmod{2}$ and $t\geq3$.
\end{lemma}

\Proof The proof is similar to that of Lemma \ref{(v,w,1)-cyclic
hgdd1}. The required $3$-SCHGDDs of type $(u,g^{t})$ and  $1$-cyclic
$3$-GDDs of type $h^3$ can be found in Theorem \ref{schgdd} and
Corollary \ref{w-cyclic 3gdp} respectively. \qed

\begin{lemma}\label{(v,w,1)-cyclic hgdd v=6t+3}
Suppose $u\equiv2\hspace{-0.3cm}\pmod{6}$ and $u\geq8$. Let
$g\in\{1,2\}$. Then there exists an $(h,gt,1)$-cyclic $3$-HGDD of
type $(u,(gh)^{t})$ for any $h\equiv3\hspace{-0.3cm}\pmod{6}$,
$t\equiv1,5\hspace{-0.3cm}\pmod{6}$ and $t\geq5$.
\end{lemma}

\Proof When $h=3$, by Theorem \ref{schgdd}, there exists a
$3$-SCHGDD of type $(u,(3g)^{t})$. Since $(3,gt)=1$, $Z_{3gt}$ is
isomorphic to $Z_3\times Z_{gt}$. Hence, we have a $(3,gt,1)$-cyclic
$3$-HGDD of type $(u,(3g)^{t})$.

When $h\geq9$, apply Construction \ref{inf1} with a $1$-cyclic
$3$-GDD of type $(h/3)^3$ from Corollary \ref{w-cyclic 3gdp} to
obtain an $(h,gt,1)$-cyclic $3$-HGDD of type $(u,(gh)^{t})$. \qed

\begin{lemma}\label{(3,3,1)-cyclic}
Let $u\equiv2\hspace{-0.3cm}\pmod{6}$ and $u\geq8$. Then there
exists a $(3,3,1)$-cyclic $3$-HGDD of type $(u,3^3)$.
\end{lemma}

\Proof When $u=8$, let $X=I\times Z_3\times Z_3$,
$\mathcal{G}=\{\{i\}\times Z_3\times Z_3:i\in I\}$,
$\mathcal{H}=\{I\times Z_3\times \{j\}:j\in Z_3\}$, where
$I=Z_7\cup\{\infty\}$. Developing the following $8$ initial base
blocks by $(+i,+j,-)$ mod $(7,3,-)$ yields all $168$ base blocks of
the required design, where $(i,j)\in Z_7\times Z_3$ and
$\infty+1=\infty$.\vspace{0.2cm}

{\small$\{(0,0,0),(1,1,2),(3,0,1)\}$,
$\{(0,0,0),(1,0,2),(\infty,0,1)\}$, $\{(0,0,0),(1,2,2),(3,2,1)\}$,

$\{(0,0,0),(1,0,1),(3,0,2)\}$, $\{(0,0,0),(2,1,2),(\infty,2,1)\}$,
$\{(0,0,0),(1,1,1),(3,2,2)\}$,

$\{(0,0,0),(1,2,1),(3,1,2)\}$,
$\{(0,0,0),(3,1,1),(\infty,2,2)\}$.}\vspace{0.2cm}

When $u\geq14$, start with a PBD$(u,\{3,4,5^*\})$ from Lemma
\ref{5^*}. Applying Construction \ref{pbd1} with a $(3,3,1)$-cyclic
$3$-HGDD of type $(k,3^3)$ for each $k\in\{3,4,5\}$, which exists by
Example \ref{(5,3^3)} and Lemma \ref{(v,w,1)-cyclic hgdd1}, we then
obtain the required designs. \qed

\begin{lemma}\label{(v,w,1)-cyclic hgdd v=6t+3 II}
Let $u\equiv2\hspace{-0.3cm}\pmod{6}$ and $u\geq8$. Then there
exists a $(g,t,1)$-cyclic $3$-HGDD of type $(u,g^{t})$ for any
$g\equiv t\equiv3\hspace{-0.3cm}\pmod{6}$.
\end{lemma}

\Proof When $t=3$, by Lemma \ref{(3,3,1)-cyclic}, there exists a
$(3,3,1)$-cyclic $3$-HGDD of type $(u,3^3)$. Inflating it by a
$1$-cyclic $3$-GDD of type $(g/3)^3$ from Corollary \ref{w-cyclic
3gdp}, we then obtain a $(g,3,1)$-cyclic $3$-HGDD of type $(u,g^3)$
by Construction \ref{inf1}.

When $t\geq9$, by Lemma \ref{(v,w,1)-cyclic hgdd}, there exists a
$(g,t,1)$-cyclic $3$-HGDD of type $(u,(3g)^{(t/3)})$. Filling in the
holes with above $(g,3,1)$-cyclic $3$-HGDD of type $(u,g^3)$, we
then obtain the required designs by Construction \ref{fillhgdd}.
\qed

Dai, \emph{et al}. \cite{dc} discussed the existence of a
$(1,g,t)$-cyclic $3$-HGDDs of type $(u,g^t)$, which is called a
$g$-cyclic $3$-HGDDs of type $(u,g^t)$. We quote the result as
follows.

\begin{theorem}\label{w-cyclic hgdd (u,w^t)}\cite{dc}
A $(1,g,t)$-cyclic $3$-HGDD of type $(u,g^t)$ exists if and only if
$u,t\geq 3$, $(t-1)(u-1)g\equiv 0\hspace{-0.3cm}\pmod {2}$ and
$t(t-1)u(u-1)g\equiv 0\hspace{-0.3cm}\pmod{6}$, except for $u=t=3$
and $g\equiv 0\hspace{-0.3cm}\pmod{2}$.
\end{theorem}

In fact, when $g\neq w$, we also need the $k$-HGDDs of type
$(u,g^{t}w^1)$ admitting a permutation. But the general form of the
permutation will be very complex. We here only permit the $k$-HGDDs
having a simple permutation.\vspace{0.2cm}

Let $h$ be a common divisor of $g$ and $w$, say $g=g^\prime h$ and
$w=w^\prime h$. Suppose $(X,\mathcal{G},\mathcal{H},\mathcal{B})$ is
a $k$-HGDD of type $(u,g^{t}w^1)$. If there is a permutation $\pi$
on $X$ which is the product of $u(g^\prime t+w^\prime)$ disjoint
$h$-cycles, fixes every group and every hole, leaves $\mathcal{B}$
invariant, then this design is said to be \emph{$h$-cyclic}.
Similarly, $\mathcal{B}$ can be partitioned into equivalence classes
under the action of $\pi$. A set of base blocks is a set of
representatives for these equivalence classes.

The following result will be used in the next section to construct
another auxiliary design.

\begin{lemma}\label{h-cyclic hgdd}
Let $u\equiv2\hspace{-0.3cm}\pmod{6}$, $u\geq8$ and
$h\equiv1\hspace{-0.3cm}\pmod{2}$. Then there exists an $h$-cyclic
$3$-HGDD of type $(u,(3h)^{2t}(sh)^1)$ for any $t\geq2$ and
$s\in\{1,5\}$.
\end{lemma}

\Proof Start with a $3$-HGDD of type $(u,3^{2t}s^1)$, which exists
by Theorem \ref{3hgdd}. Inflate it by a $3$-SCGDD of type $h^3$ from
Corollary \ref{w-cyclic 3gdp} to obtain an $h$-cyclic $3$-HGDD of
type $(u,(3h)^{2t}(sh)^1)$.\qed

\vspace{0.2cm}We next introduce a new auxiliary design, which can be
considered as a generalization of a $k$-HGDD.\vspace{0.25cm}

An \emph{incomplete} HGDD ($k$-IHGDD) of type $(u,e,g^t)$ is defined
to be a quintuple ($X,Y,\mathcal{G},\mathcal{H},\mathcal{B}$) which
satisfies the following properties:\vspace{0.1cm}

(1) $X$ is an ($ugt$)-set (of \emph{points});\vspace{0.1cm}

(2) $\mathcal{G}$ is a partition of $X$ into $u$ subsets (called
\emph{groups}) of points $gt$ each;\vspace{0.1cm}

(3) $\mathcal{H}$ is another partition of $X$ into $t$ subsets
(called \emph{holes}) of points $ug$ each such that $|G\cap H|=g$
for any $G\in\mathcal{G}$, $H\in\mathcal{H}$;\vspace{0.1cm}

(4) $Y$ is an union of $e$ groups of $\mathcal{G}$;\vspace{0.1cm}

(5) $\mathcal{B}$ is a collection of $k$-subsets of $X$ (called
\emph{blocks}) such that any pair of $X$ from two distinct groups
with the exception of those in which both lie in $Y$ appears in a
hole or exactly in one block but not both, and no other pairs of $X$
occur in any block.\vspace{0.1cm}

We also need the $k$-IHGDDs admitting a special
permutation.\vspace{0.2cm}

A $k$-IHGDD of type $(u,e,(gh)^{mt})$,
$(X,Y,\mathcal{G},\mathcal{H},\mathcal{B})$, is said to be
\emph{$(h,gt,m)$-cyclic}, if there is a permutation $\pi$ on $X$
which is the product of $uhm$ disjoint $gt$-cycles, fixes every
group, leaves $\mathcal{B}$ invariant, and partitions the $mt$
elements of $\mathcal{H}$ into $m$ equivalence classes. A
$(1,gt,1)$-cyclic $k$-IHGDD of type $(u,e,g^t)$ is also referred to
as a \emph{semi-cyclic} $k$-IHGDD of type $(u,e,g^t)$ and denoted by
$k$-SCIHGDD of type $(u,e,g^t)$.\vspace{0.1cm}

Without loss of generality, we always identify $X=I_u\times
I_{h}\times I_{m}\times Z_{gt}$, $Y=I_e\times I_{h}\times
I_{m}\times Z_{gt}$, ${\mathcal{G}}=\{\{i\}\times I_{h}\times
I_{m}\times Z_{gt}: i\in I_u\}$ and ${\mathcal{H}}=\{I_u\times
I_{h}\times\{i\}\times\{0+j,t+j,\ldots,(g-1)t+j\}:(i,j)\in
I_{m}\times Z_t\}$. In this case, the permutation can be taken as
$(x,y,z,w)\mapsto(x,y,z,w+1)$ mod $(-,-,-,gt)$. Clearly, under the
action of $\pi$, $\mathcal{B}$ can be partitioned into equivalence
classes called the \emph{block orbits}. A set of base blocks is a
set of representatives for these block orbits of $\mathcal{B}$.

\begin{example}\label{3-cyclic ihgdd}
There exists a $3$-SCIHGDD of type $(8,2,1^{3})$.
\end{example}

\Proof Let $X=I_8\times Z_3$, $\mathcal{G}=\{\{i\}\times Z_3:i\in
I_8\}$, $Y=\{0,4\}\times Z_3$, and
$\mathcal{H}=\{I_8\times\{j\}:j\in Z_3\}$. Only base blocks are
listed below.\vspace{0.1cm}

{\small\hspace{1.5cm}$\{(0,0),(1,1),(2,2)\}$,
$\{(0,0),(5,2),(7,1)\}$, $\{(0,0),(2,1),(3,2)\}$,

\hspace{1.5cm}$\{(0,0),(5,1),(6,2)\}$, $\{(0,0),(1,2),(3,1)\}$,
$\{(0,0),(6,1),(7,2)\}$,

\hspace{1.5cm}$\{(1,1),(5,2),(4,0)\}$, $\{(1,1),(6,0),(3,2)\}$,
$\{(1,1),(6,2),(5,0)\}$,

\hspace{1.5cm}$\{(1,1),(7,0),(4,2)\}$, $\{(1,1),(2,0),(7,2)\}$,
$\{(2,2),(4,1),(6,0)\}$,

\hspace{1.5cm}$\{(2,2),(4,0),(5,1)\}$, $\{(2,2),(7,0),(6,1)\}$,
$\{(2,2),(3,1),(5,0)\}$,

\hspace{1.5cm}$\{(3,0),(4,1),(6,2)\}$, $\{(3,0),(7,1),(4,2)\}$,
$\{(3,0),(5,1),(7,2)\}$.}\qed

\begin{lemma}\label{weil}\cite{fwc}
Let $p\geq5$ be a prime. Then there exists an element $x\in
Z_p\setminus\{0,\pm1\}$ such that $x$ and $x+1$ are both nonsquares
and $x-1$ is a square.
\end{lemma}

\begin{lemma}\label{(8,2,1^p)}
There exists a $3$-SCIHGDD of type $(8,2,1^{p})$ for any odd prime
$p$.
\end{lemma}

\Proof The case of $p=3$ can be found in Example \ref{3-cyclic
ihgdd}. We construct a $3$-SCIHGDD of type $(8,2,1^{p})$ for each
prime $p\geq5$ below. Let $X=I\times Z_p$,
$\mathcal{G}=\{\{i\}\times Z_p:i\in I\}$, $Y=\{a,b\}\times Z_p$, and
$\mathcal{H}=\{I\times\{j\}:j\in Z_p\}$, where $I=Z_6\cup\{a,b\}$.
By Lemma \ref{weil}, we can take $x\in Z_p\setminus\{0,\pm1\}$ such
that $x$ and $x+1$ are both nonsquares, while $x-1$ is a square.
Only initial base blocks are listed below.\vspace{0.1cm}

$p\equiv1\hspace{-0.3cm}\pmod4$:\vspace{0.1cm}

{\small$\{(1+2i,0),(3+2i,x),(a,x+1)\}$,
$\{(2i,0),(1+2i,1),(3+2i,x)\}$,

$\{(2i,0),(3+2i,1),(b,x+1)\}$, $\{(2i,0),(2+2i,1),(a,x)\}$,

$\{(1+2i,0),(2+2i,x),(b,1)\}$,
$\{(2i,0),(1+2i,x),(2+2i,x+1)\}$.\vspace{0.1cm}

$p\equiv3\hspace{-0.3cm}\pmod4$:\vspace{0.1cm}

$\{(2i,0),(1+2i,x+1),(3+2i,1)\}$, $\{(2i,0),(2+2i,1),(a,x)\}$,

$\{(2i,0),(1+2i,1),(2+2i,x)\}$, $\{(2i,0),(3+2i,x),(b,x-1)\}$,

$\{(1+2i,0),(3+2i,x),(a,x+1)\}$,
$\{(1+2i,0),(2+2i,x),(b,1)\}$.}\vspace{0.1cm}

Here, $0\leq i\leq2$. All $9(p-1)$ base blocks are generated by
multiplying above $18$ initial base blocks by $\omega^{2r}$, where
$\omega$ is a primitive element of $Z_p$, $0\leq r\leq (p-3)/2$.\qed

We give two constructions on $k$-SCIHGDDs, which can be considered
as the generalizations of Constructions $3.1$ and $3.4$ of
\cite{fwc} respectively.

\begin{construction}\label{inf ih}
If there exist a $k$-SCIHGDD of type $(u,e,g^{t})$ and a $k$-SCGDD
of type $h^k$, then there exists a $k$-SCIHGDD of type
$(u,e,(gh)^{t})$.
\end{construction}

\begin{construction}\label{fill ih}
If there exist a $k$-SCIHGDD of type $(u,e,(gh)^{t})$ and a
$k$-SCIHGDD of type $(u,e,g^{h})$, then there exists a $k$-SCIHGDD
of type $(u,e,g^{ht})$.
\end{construction}

Applying these constructions, we give an infinite class of
$3$-SCIHGDDs.

\begin{lemma}\label{(8,2,1^w)}
There exists a $3$-SCIHGDD of type $(8,2,1^{t})$ for any odd
integers $t\geq3$.
\end{lemma}

\Proof  Let $t=p_1^{a_1}p_2^{a_2}\cdots p_s^{a_s}$, where $p_i\geq3$
is a prime for each $1\leq i\leq s$. Start from a $3$-SCIHGDD of
type $(8,2,1^{p_1})$, which exists by Lemma \ref{(8,2,1^p)}. Inflate
it by a $3$-SCGDD of type $q^3$ from Corollary \ref{w-cyclic 3gdp},
$q\in\{p_1,p_2,\ldots,p_s\}$. By Construction \ref{inf ih}, we
obtain a $3$-SCIHGDD of type $(8,2,q^{p_1})$. Fill in the holes with
a $3$-SCIHGDD of type $(8,2,1^{q})$, which exists by Lemma
\ref{(8,2,1^p)}. Apply Construction \ref{fill ih} to obtain a
$3$-SCIHGDD of type $(8,2,1^{p_1q})$. Repeating this process will
produce the required designs for any odd integers $t\geq3$.\qed

\begin{lemma}\label{(8,2,1^nh)}
There exists a $(1,t,h)$-cyclic $3$-IHGDD of type $(8,2,1^{th})$ for
any odd positive integers $h$ and $t\geq3$.
\end{lemma}

\Proof  By Lemma \ref{(8,2,1^w)}, there exists a $3$-SCIHGDD of type
$(8,2,1^{th})$, which is also a $(1,t,h)$-cyclic $3$-IHGDD of type
$(8,2,1^{th})$.\qed

\section{Incomplete group divisible packings}

Incomplete group divisible designs (IGDDs) are important auxiliary
designs in the constructions of GDDs. In this section, we generalize
this concept to incomplete group divisible packings (IGDPs), which
will be used to construct $w$-cyclic GDPs.

An \emph{incomplete group divisible packing} ($K$-IGDP) is a
quadruple $(X,Y,\mathcal{G,B})$ where $X$ is a set (of
\emph{points}), $Y$ is a subset (called a \emph{hole}) of $X$,
$\mathcal{G}$ is a partition of $X$ into subsets (called
\emph{groups}), and $\mathcal{B}$ is a collection of subsets (called
\emph{blocks}) of $X$ each of size from $K$ such that\vspace{0.1cm}

(1) each block intersects each group in at most one point;
\vspace{0.1cm}

(2) no pair of distinct points of $Y$ occurs in any
block;\vspace{0.1cm}

(3) every pair of points from distinct groups with the exception of
those in which both lie in $Y$ occurs in at most one block of
$\mathcal{B}$.\vspace{0.1cm}

If $K=\{k\}$, we omit the braces to simply write $k$ for $K$. If
every admissible pair appears in exactly one block, then this IGDP
is called an \emph{incomplete group divisible design}, or a $K$-IGDD
for short.\vspace{0.1cm}

We introduce two types of IGDPs here.

\begin{itemize}
\item A $K$-IGDP of type $(g,t)^{u}$
is an IGDP in which each block has size from $K$ and there are $u$
groups of size $g$, each of which intersects the hole in $t$ points.

\item A $K$-IGDP of type $g^{(u,t)}$
is an IGDP in which each block has size from $K$ and there are $u$
groups of size $g$, wherein $t$ groups constitute the hole.
\end{itemize}

In next subsections, we study the constructions of these two types
of $K$-IGDPs with special permutations and give some existence
results for later use.

\subsection{ $h$-cyclic $K$-IGDPs of type $(gh,th)^u$}

We need the $K$-IGDPs of type $(gh,th)^u$ admitting a special
permutation.\vspace{0.1cm}

A $K$-IGDP of type $(gh,th)^{u}$ $(X,Y,\mathcal{G,B})$ is said to be
\emph{$h$-cyclic} if there is a permutation on $X$ which is the
product of $ug$ disjoint $h$-cycles, fixes every group, and leaves
$Y,\mathcal{B}$ invariant. Without loss of generality, we can always
identify $X=I_u\times I_g\times Z_h$, $\mathcal{G}=\{\{i\}\times
I_g\times Z_h:i\in I_u\}$ and $Y=I_u\times I_t\times Z_h$. In this
case, the permutation can be taken as $(i,j,x)\mapsto(i,j,x+1)$ mod
$(-,-,h)$. $\mathcal{B}$ can be partitioned into some block orbits
under the permutation. A set of base blocks is a set of
representatives for these block orbits of
$\mathcal{B}$.\vspace{0.1cm}

We exhibit our second ``Filling Construction" via $h$-cyclic
$k$-IGDPs of type $(gh,th)^{u}$.

\begin{construction}\cite{wc}{\rm{(Filling Construction-II)}}\label{fill2}
Suppose that the following designs exist:\vspace{0.1cm}

\hspace{-0.57cm}$(1)$ an $h$-cyclic $k$-IGDP of type $(gh,th)^{u}$
with $f$ base blocks;\vspace{0.1cm}

\hspace{-0.57cm}$(2)$ an $h$-cyclic $k$-GDP of type $(th)^{u}$ with
$b$ base blocks.\vspace{0.2cm}

\hspace{-0.58cm}Then, there exists an $h$-cyclic $k$-GDP of type
$(gh)^{u}$ with $b+f$ base blocks.
\end{construction}

For applying Construction \ref{fill2}, we need to construct more
$h$-cyclic $K$-IGDPs of type $(gh,th)^{u}$. At the beginning, we
give some constructive methods. Construction \ref{pbd} can be found
in \cite{wc}. Constructions \ref{inf2} and \ref{igdd2} are simple
generalizations of the corresponding constructions in \cite{wc}.

\begin{construction}{\rm{(Weighting Construction-I)}}\label{pbd}
Suppose there exist a $K$-GDD of type $m^u$ and an $h$-cyclic
$l$-IGDD of type $(gh,th)^{k}$ for each $k\in K$. Then there exists
an $h$-cyclic $l$-IGDD of type $(mgh,mth)^u$.
\end{construction}

\begin{construction}{\rm{(Inflation-II)}}\label{inf2}
Suppose that there exists an $h$-cyclic $K$-IGDD of type
$(gh,th)^u$. Let $\mathcal{F}$ be the base block set of this design.
Suppose $\mathcal{F}$ contains $r_i$ base blocks of size $k_i$,
$1\leq i\leq s$. If there exists a $w$-cyclic $k$-GDP of type
$(vw)^{k_i}$ with $b_i$ base blocks for each $1\leq i\leq s$, then
there exists an $hw$-cyclic $k$-IGDP of type $(ghvw,thvw)^u$ with
$\sum_{i=1}^s r_ib_i$ base blocks.
\end{construction}

\emph{Remark $1$}: For each base block of size $k_i$ in
$\mathcal{F}$, if the input design is a $w$-cyclic $k$-GDD of type
$(vw)^{k_i}$ in Construction \ref{inf2}, then the output design is
an $hw$-cyclic $k$-IGDD of type $(ghvw,thvw)^u$.

\begin{construction}\label{igdd2}
Suppose there exists a $(g,h,t)$-cyclic $k$-HGDD of type $(u,(gh)^t)$
with $b$ base blocks and an $h$-cyclic $k$-IGDP of type
$(gh+eh,eh)^u$ with $f$ base blocks. Then there exists an $h$-cyclic
$k$-IGDP of type $(gth+eh,eh)^u$ with $b+tf$ base blocks.
\end{construction}

\emph{Remark $2$}: If the input design in Construction \ref{igdd2}
is an $h$-cyclic $k$-IGDD of type $(gh+eh,eh)^u$, then the output
design is an $h$-cyclic $k$-IGDD of type
$(gth+eh,eh)^u$.\vspace{0.2cm}

When $w$ is even, Construction \ref{inf2} some times cannot work.
Therefore, we introduce a special kind of $k$-IGDD called an
$h$-perfect $k$-IGDD as follows.\vspace{0.2cm}

Suppose that $\mathcal{F}=\{B_i:1\leq i\leq r\}$ is a set of base
blocks of an $hw$-cyclic $k$-IGDD of type $(ghw,thw)^u$ on
$I_u\times I_g\times Z_{hw}$. Without loss of generality, let
$$B_i=\{(a_{i1},b_{i1},0),(a_{i2},b_{i2},c_{i2}),\ldots,(a_{ik},b_{ik},c_{ik})\}$$
for each $1\leq i\leq r$. Define
$$ele(\mathcal{F})=\cup_{i=1}^{r}\{c_{i2},c_{i3},\ldots,c_{ik}\}.$$ The $hw$-cyclic $k$-IGDD of type $(ghw,thw)^u$
is said to be \emph{$h$-perfect}, denoted by $h$-perfect $hw$-cyclic
$k$-IGDD of type $(ghw,thw)^u$, if
$$ele(\mathcal{F})\subset\{x+yw:0\leq x\leq \lfloor w/2\rfloor,0\leq y\leq
h-1\}.$$ When $h=1$, a $1$-perfect $k$-IGDD is simply called a
perfect $k$-IGDD. Note that any $hw$-cyclic $k$-IGDD of type
$(ghw,thw)^u$ has the $h$-perfect property when $w$ is $2$.

\begin{construction}\label{perfect igdd}
Suppose that the following designs exist:\vspace{0.1cm}

\hspace{-0.57cm}$(1)$ a perfect $w$-cyclic $k$-IGDD of type
$(gw,tw)^u$;\vspace{0.1cm}

\hspace{-0.57cm}$(2)$ an $h$-perfect $hw$-cyclic $k$-IGDD of type
$(ghw,thw)^u$;\vspace{0.1cm}

\hspace{-0.57cm}$(3)$ a $k$-SCHGDD of type $(k,h^m)$.\vspace{0.2cm}

\hspace{-0.58cm}Then, there exists an $hm$-perfect $hmw$-cyclic
$k$-IGDD of type $(ghmw,thmw)^u$.
\end{construction}

\Proof Suppose that
$\mathcal{F}=\{\{(a_{i1},b_{i1},0),(a_{i2},b_{i2},c_{i2}),\ldots,(a_{ik},b_{ik},c_{ik})\}:1\leq
i\leq r\}$ be a set of base blocks of a perfect $w$-cyclic $k$-IGDD
of type $(gw,tw)^u$, where $0\leq c_{ij}\leq\lfloor w/2\rfloor$ for
each $1\leq i\leq r$ and $2\leq j\leq k$.

Let
$\mathcal{A}=\{\{(q_{i1},f_{i1},0),(q_{i2},f_{i2},x_{i2}+y_{i2}w),\ldots,(q_{ik},f_{ik},x_{ik}+y_{ik}w)\}:1\leq
i\leq s\}$ be a set of base blocks of an $h$-perfect $hw$-cyclic
$k$-IGDD of type $(ghw,thw)^u$, where $0\leq x_{ij}\leq \lfloor
w/2\rfloor$ and $0\leq y_{ij}\leq h-1$ for each $1\leq i\leq s$ and
$2\leq j\leq k$.

Let $\mathcal{E}=\{\{(1,d_{1j}),(2,d_{2j}),\ldots,(k,d_{kj})\}:1\leq
j\leq h(m-1)\}$ be the base block set of a $k$-SCHGDD of type
$(k,h^m)$. Then $\{d_{\alpha j}-d_{\beta j}:1\leq j\leq
h(m-1)\}=Z_{hm}\setminus\{0,m,\ldots,(h-1)m\}$ for any
$1\leq\alpha\neq\beta\leq k$.

Now we construct the desired $hm$-perfect $hmw$-cyclic $k$-IGDD of
type $(ghmw,thmw)^u$ on $I_u\times I_g\times Z_{hmw}$ whose base
blocks consists of the following two parts:

$(1)$ For each
$B_i=\{(a_{i1},b_{i1},0),(a_{i2},b_{i2},c_{i2}),\ldots,(a_{ik},b_{ik},c_{ik})\}\in\mathcal{F}$,
construct a family
$$\mathcal{C}_{B_i}^j=\{(a_{i1},b_{i1},0),(a_{i2},b_{i2},c_{i2}+(d_{2j}-d_{1j})w),\ldots,(a_{ik},b_{ik},c_{ik}+(d_{kj}-d_{1j})w)\},$$
where $1\leq j\leq h(m-1)$ and the third coordinates are reduced
modulo $hmw$.

$(2)$ For each
$A_i=\{(q_{i1},f_{i1},0),(q_{i2},f_{i2},x_{i2}+y_{i2}w),\ldots,(q_{ik},f_{ik},x_{ik}+y_{ik}w)\}\in\mathcal{A}$,
let
$$A_{i}^\prime=\{(q_{i1},f_{i1},0),(q_{i2},f_{i2},x_{i2}+y_{i2}mw),\ldots,(q_{ik},f_{ik},x_{ik}+y_{ik}mw)\}.$$

Let
$\mathcal{C}_1=\cup_{i=1}^{r}\cup_{j=1}^{h(m-1)}\mathcal{C}_{B_i}^j$,
$\mathcal{C}_2=\cup_{i=1}^{s}A_{i}^\prime$ and
$\mathcal{C}=\mathcal{C}_1\cup\mathcal{C}_2$. It is readily checked
that $\mathcal{C}$ forms a family of base blocks of the required
$hm$-perfect $hmw$-cyclic $k$-IGDD of type $(ghmw,thmw)^u$. \qed

Applying these constructions, we give some results on $h$-cyclic
$3$-IGDDs of type $(gh,th)^u$ for later use. The results in Lemma
\ref{(2v,2)^u} are taken from Lemmas $4.12-4.15$ of \cite{wc}.

\begin{lemma}\label{(2v,2)^u}\cite{wc}
There exists a $w$-cyclic $3$-IGDD of type $(vw,w)^u$ if $u,v,w$
satisfy one of the following conditions.\vspace{0.1cm}

$(1)$ $(u,v,w)=(5,2,4)$;

$(2)$ $u\in\{3,4\}$, $w=2$ and $v\equiv1\hspace{-0.3cm}\pmod{2}$;

$(3)$ $u=5$, either $w=2$ and $v\equiv1,2\hspace{-0.3cm}\pmod{3}$,
or $w=6$ and $v\equiv1\hspace{-0.3cm}\pmod{2}$.
\end{lemma}

\begin{lemma}\label{(vw,w)^11}
There exists a $w$-cyclic $3$-IGDD of type $(vw,w)^{11}$ for any
$v\equiv1,5\hspace{-0.3cm}\pmod{6}$ and
$w\equiv10\hspace{-0.3cm}\pmod{12}$.
\end{lemma}

\Proof Start with a PBD$(11,\{3,5^*\})$, which is also a $3$-GDD of
type $1^65^1$ from Lemma \ref{gtw1}. Apply Construction \ref{pbd}
with $2$-cyclic $3$-IGDDs of type $(2v,2)^{k}$ for $k\in\{3,5\}$
from Lemma \ref{(2v,2)^u} to obtain a $2$-cyclic $3$-IGDD of type
$(2v,2)^{11}$. Inflate it by a $3$-SCGDD of type $(w/2)^3$ from
Corollary \ref{w-cyclic 3gdp}. By Construction \ref{inf2}, we obtain
a $w$-cyclic $3$-IGDD of type $(vw,w)^{11}$.\qed

\begin{lemma}\label{perfect (2w,w)^8}
Let $w\equiv0\hspace{-0.3cm}\pmod{2}$. Then there exists a
$w$-cyclic $3$-IGDD of type $(2w,w)^8$.
\end{lemma}

\Proof Let $w=2^nw^\prime$, where $n\geq1$ and
$w^\prime\equiv1\hspace{-0.3cm}\pmod{2}$. By Lemma A.1 in Supporting
Information, there exist a perfect $2$-cyclic $3$-IGDD of type
$(4,2)^8$ and a $2$-perfect $4$-cyclic $3$-IGDD of type $(8,4)^8$.
Applying Construction \ref{perfect igdd} with a $3$-SCHGDD of type
$(3,2^{2^{n-2}})$ from Theorem \ref{schgdd} for any $n\geq4$, we
obtain a $2^{n-1}$-perfect $2^{n}$-cyclic $3$-IGDD of type
$(2^{n+1},2^{n})^8$.

Combining the $8$-cyclic $3$-IGDD of type $(16,8)^8$ from Lemma A.1
in Supporting Information, we have a $2^{n}$-cyclic $3$-IGDD of type
$(2^{n+1},2^{n})^8$ for all $n\geq1$. Inflate it by a $3$-SCGDD of
type ${w^{\prime}}^3$ from Corollary \ref{w-cyclic 3gdp}, we obtain
the required $w$-cyclic $3$-IGDD of type $(2w,w)^8$.\qed

\begin{lemma}\label{(vw,w)^8}
Let $v\equiv1,2\hspace{-0.3cm}\pmod{3}$ and $v\geq2$. Then there
exists a $w$-cyclic $3$-IGDD of type $(vw,w)^8$ for any
$w\equiv0\hspace{-0.3cm}\pmod{2}$.
\end{lemma}

\Proof When $v=2$, a $w$-cyclic $3$-IGDD of type $(2w,w)^8$ exists
by Lemma \ref{perfect (2w,w)^8}. When $v\geq4$, start with a
$(1,w,v-1)$-cyclic $3$-HGDD of type $(8,w^{v-1})$ from Theorem
\ref{w-cyclic hgdd (u,w^t)}. Apply Construction \ref{igdd2} with a
$w$-cyclic $3$-IGDD of type $(2w,w)^8$ to obtain the required
$w$-cyclic $3$-IGDD of type $(vw,w)^8$. \qed

\begin{lemma}\label{(8,4)^u u=6t+2}
Let $u\equiv2\hspace{-0.3cm}\pmod{6}$ and $u\geq8$. Then there
exists a $4$-cyclic $3$-IGDD of type $(4v,4)^u$ for any
$v\equiv1,2\hspace{-0.3cm}\pmod{3}$.
\end{lemma}

\Proof $(1)$ $v=2$. For $u=5$, Lemma \ref{(2v,2)^u} provides the
required design. For $u\in\{4,6,8,14\}$, Lemmas A.1 and A.2 in
Supporting Information provide the required designs. For $u\geq20$,
start with a PBD$(u,\{4,5,6\})$, which exists by Lemma \ref{b(4,5)}.
Apply Construction \ref{pbd} with a $4$-cyclic $3$-IGDD of type
$(8,4)^k$ for $k\in\{4,5,6\}$ to obtain the required design.

$(2)$ $v\geq4$. By Theorem \ref{w-cyclic hgdd (u,w^t)}, there exists
a $(1,4,v-1)$-cyclic $3$-HGDD of type $(u,4^{v-1})$. Apply
Construction \ref{igdd2} with a $4$-cyclic $3$-IGDD of type
$(8,4)^u$ to obtain the required design. \qed

\begin{lemma}\label{(2v,2)^8}
Suppose $u\equiv2\hspace{-0.3cm}\pmod{12}$ and $u\geq14$. Then there
exists a $2$-cyclic $3$-IGDD of type $(2v,2)^u$ for any
$v\equiv1,5\hspace{-0.3cm}\pmod{6}$.
\end{lemma}

\Proof Start with a PBD$(u,\{3,4,5^*\})$ from Lemma \ref{5^*}. Apply
Construction \ref{pbd} with $2$-cyclic $3$-IGDDs of type $(2v,2)^k$
for $k\in\{3,4,5\}$, which exists by Lemma \ref{(2v,2)^u}, to obtain
the required design. \qed

\begin{lemma}\label{(vw,sw)^u}
Let $u\equiv2\hspace{-0.3cm}\pmod{12}$ and $u\geq14$. Then there
exists a $w$-cyclic $3$-IGDD of type $(vw,w)^{u}$ for any
$w\equiv10\hspace{-0.3cm}\pmod{12}$ and
$v\equiv1,5\hspace{-0.3cm}\pmod{6}$.
\end{lemma}

\Proof By Lemma \ref{(2v,2)^8}, a $2$-cyclic $3$-IGDD of type
$(2v,2)^{u}$ exists. Inflate it by a $3$-SCGDD of type $(w/2)^3$
from Corollary \ref{w-cyclic 3gdp}. Then we obtain a $w$-cyclic
$3$-IGDD of type $(vw,w)^{u}$ by Construction \ref{inf2}.\qed

\begin{lemma}\label{(6v,6)^u,u=12t+2}
Let $u\equiv2\hspace{-0.3cm}\pmod{12}$ and $u\geq14$. Then there
exists a $6$-cyclic $3$-IGDD of type $(6v,6)^u$ for any
$v\equiv1\hspace{-0.3cm}\pmod{2}$.
\end{lemma}

\Proof For each $k\in\{3,4\}$, take a $2$-cyclic $3$-IGDD of type
$(2v,2)^k$ from Lemma \ref{(2v,2)^u}. Inflate it by a $3$-SCGDD of
type $3^3$ from Corollary \ref{w-cyclic 3gdp} to obtain a $6$-cyclic
$3$-IGDD of type $(6v,6)^k$. When $k=5$, a $6$-cyclic $3$-IGDD of
type $(6v,6)^5$ can be found in Lemma \ref{(2v,2)^u}. Apply
Construction \ref{pbd} with a PBD$(u,\{3,4,5^*\})$ to obtain the
required designs.\qed

\begin{corollary}\label{2-cyclic (6v,6)^u}
Let $u\equiv2\hspace{-0.3cm}\pmod{12}$ and $u\geq14$. Then there
exists a $2$-cyclic $3$-IGDD of type $(6v,6)^u$ for any
$v\equiv1\hspace{-0.3cm}\pmod{2}$.
\end{corollary}

\Proof By Lemma \ref{(6v,6)^u,u=12t+2}, there exists a $6$-cyclic
$3$-IGDD of type $(6v,6)^u$, which is also a $2$-cyclic $3$-IGDD of
type $(6v,6)^u$.\qed

At the end of this subsection, we present two existence results on
$h$-cyclic $3$-IGDPs of type $(gh,th)^u$.

\begin{lemma}\label{(3v,3)^u}
Suppose that $u\in\{8,14\}$ and $s\in\{1,5\}$. Let $v=6t+s$ and
$t\geq1$. Then there exists a $3$-cyclic $3$-IGDP of type
$(3v,3s)^u$ with $u(v-s)(3(u-1)(v+s)-1)/6$ base blocks.
\end{lemma}

\Proof When $t=1$, by Lemma A.3 in Supporting Information, there
exists a $\{3,u-2\}$-IGDD of type $(v,s)^u$, in which the blocks of
size $u-2$ form a partition of the points outside the hole. Give
weight $3$ to each point and input $3$-SCGDDs of type $3^{3}$ or
$3$-SCGDPs of type $3^{u-2}$ with $J^*((u-2)\times1\times3)$ base
blocks, which can be found in Corollary \ref{w-cyclic 3gdp}. By
Construction \ref{inf2}, we obtain a $3$-cyclic $3$-IGDP of type
$(3v,3s)^u$, which has $u(3(u-1)(v+s)-1)$ base blocks.

When $t\geq2$, start from a $3$-cyclic $3$-HGDD of type
$(u,9^{2t}(3s)^1)$, which has $3u(u-1)t(6t+2s-3)$ base blocks and
exists by Lemma \ref{h-cyclic hgdd}. Fill in the $2t$ holes of size
$9u$ using $3$-cyclic $3$-GDPs of type $9^u$ from Lemma
\ref{w=6t+3,v=6t+3}, each of which has $J^*(u\times3\times3,3,1)$
base blocks, to obtain a $3$-cyclic $3$-IGDP of type $(3v,3s)^u$
with $3u(u-1)t(6t+2s-3)+ut(9u-10)= u(v-s)(3(u-1)(v+s)-1)/6$ base
blocks. Note that $J^*(u\times3\times3,3,1)=u(9u-10)/2$. \qed

\begin{lemma}\label{w-cyclic igdp w=6t+5}
Let $u\in\{8,14\}$. Then there exists a $w$-cyclic $3$-IGDP of type
$(vw,w)^u$ with $u(v-1)((u-1)(v+1)w-1)/6$ base blocks for any
$v\equiv1\hspace{-0.3cm}\pmod6$, $v\geq7$ and $w\equiv5\pmod6$.
\end{lemma}

\Proof The construction is similar to that of Lemma \ref{(3v,3)^u}.
When $v=7$, start also with the $\{3,u-2\}$-IGDD of type $(7,1)^u$
from Lemma A.3 in Supporting Information. Give weight $w$ to each
point and input $3$-SCGDDs of type $w^3$ or $3$-SCGDPs of type
$w^{u-2}$ with $J^*((u-2)\times1\times w)$ base blocks, which exist
by Corollary \ref{w-cyclic 3gdp}.

When $v\geq13$, start with a $w$-cyclic $3$-HGDD of type
$(u,(3w)^{(v-1)/3}w^1)$ from Lemma \ref{h-cyclic hgdd}. For each
hole of size $3wu$, construct a $w$-cyclic $3$-GDP of type $(3w)^u$
with $J^*(u\times3\times w,3,1)$ base blocks, which exists by Lemma
\ref{w=6t+5,v=6t+3}. It is readily checked that we obtain a
$w$-cyclic $3$-IGDP of type $(vw,w)^u$ with
$u(v-1)((u-1)(v+1)w-1)/6$ base blocks.\qed

\emph{Remark $3$}: The results of Lemmas \ref{w=6t+3,v=6t+3} and
\ref{w=6t+5,v=6t+3} are used in Lemmas \ref{(3v,3)^u} and
\ref{w-cyclic igdp w=6t+5}, respectively. Note that the
constructions of Lemmas \ref{w=6t+3,v=6t+3} and \ref{w=6t+5,v=6t+3}
only need the conclusions of Section III.

\subsection{$h$-cyclic $K$-IGDPs of type
$(gh)^{(u,t)}$}

We also need the $K$-IGDPs of type $(gh)^{(u,t)}$ admitting a
special permutation.\vspace{0.2cm}

A $K$-IGDP of type $(gh)^{(u,t)}$ $(X,Y,\mathcal{G,B})$ is said to
be \emph{$h$-cyclic} if there is a permutation on $X$ which is also
a product of $ug$ disjoint $h$-cycles, fixes every group, and leaves
$Y$, $\mathcal{B}$ invariant. Without loss of generality, we can
always identify $X=I_u\times I_g\times Z_h$,
${\mathcal{G}}=\{\{i\}\times I_{g}\times Z_{h}: i\in I_u\}$ and
$Y=I_t\times I_{g}\times Z_h$. In this case, the permutation can be
taken as $(i,j,x)\mapsto(i,j,x+1)$ mod $(-,-,h)$. Also,
$\mathcal{B}$ can be partitioned into some block orbits under the
permutation. A set of base blocks is a set of representatives for
these block orbits of $\mathcal{B}$.\vspace{0.2cm}

We exhibit our third ``Filling Construction" via $h$-cyclic
$k$-IGDPs of type $(gh)^{(u,t)}$.

\begin{construction}\cite{wc}{\rm{(Filling Construction-III)}}\label{fill3}
Suppose that the following designs exist:\vspace{0.2cm}

\hspace{-0.57cm}$(1)$ an $h$-cyclic $k$-IGDP of type $(gh)^{(u,t)}$
with $b$ base blocks;\vspace{0.1cm}

\hspace{-0.57cm}$(2)$ an $h$-cyclic $k$-GDP of type $(gh)^{t}$ with
$f$ base blocks.\vspace{0.2cm}

\hspace{-0.58cm}Then, there exists an $h$-cyclic $k$-GDP of type
$(gh)^{u}$ with $b+f$ base blocks.
\end{construction}

We introduce some constructive methods for $h$-cyclic $k$-IGDPs of
type $(gh)^{(u,t)}$. The first three constructions are taken from
\cite{wc}. The last construction is a simple generalization of
Construction $4.19$ of \cite{wc}.

\begin{construction}{\rm{(Inflation-III)}}\label{inf3}
Suppose there exists an $h$-cyclic $K$-IGDD of type $(gh)^{(u,t)}$.
If there exists a $w$-cyclic $l$-GDD of type $(vw)^k$ for each $k\in
K$, then there exists an $hw$-cyclic $l$-IGDD of type
$(ghvw)^{(u,t)}$.
\end{construction}

\begin{construction}{\rm{(Weighting Construction-II)}}\label{weight2}
Suppose there exists a $K$-GDD of type ${g_{1}}^{u_1}\cdots
{g_{r}}^{u_r}$. If there exists an $h$-cyclic $l$-GDD of type
$(ht)^{k}$ for each $k\in K$, then there exists an $h$-cyclic
$l$-GDD of type ${(g_{1}ht)}^{u_1}\cdots {(g_{r}ht)}^{u_r}$.
\end{construction}

\begin{construction}\label{ihgdd}
Suppose that there exists a PBD$(u,K\cup\{k^\ast\})$. If there exist
a $g$-cyclic $l$-IGDD of type $g^{(u,k)}$ and an $l$-SCHGDD of type
$(n,g^t)$ for each $n\in K$, then there exists a $gt$-cyclic
$l$-IGDD of type $(gt)^{(u,k)}$.
\end{construction}

\begin{construction}{\rm{(Filling Construction-IV)}}\label{fill4}
Suppose that the following designs exist:\vspace{0.2cm}

\hspace{-0.57cm}$(1)$ an $h$-cyclic $k$-GDD of type
$\{ght_i:i=1,2,\ldots,r\}$ with $f$ base blocks;\vspace{0.1cm}

\hspace{-0.57cm}$(2)$ an $h$-cyclic $k$-IGDP of type
$(gh)^{(t_i+t,t)}$ with $b_i$ base blocks for each $1\leq i\leq
r-1$.\vspace{0.2cm}

\hspace{-0.58cm}Then there exists an $h$-cyclic $k$-IGDP of type
$(gh)^{(u+t,t_{r}+t)}$ with $f+\sum_{i=1}^{r-1}b_i$ base blocks,
where $u=\sum_{i=1}^{r}t_i$. Furthermore, if an $h$-cyclic $k$-IGDP
of type $(gh)^{(t_{r}+t,t)}$ with $b_r$ base blocks exists, then an
$h$-cyclic $k$-IGDP of type $(gh)^{(u+t,t)}$ with
$f+\sum_{i=1}^{r}b_i$ base blocks exists.
\end{construction}

\emph{Remark $4$}: If the input design in Construction \ref{fill4}
is an $h$-cyclic $k$-IGDD of type $(gh)^{(t_i+t,t)}$ for each $1\leq
i\leq r-1$, then the output design is an $h$-cyclic $k$-IGDD of type
$(gh)^{(u+t,t_r+t)}$. If further input an $h$-cyclic $k$-IGDD of
type $(gh)^{(t_{r}+t,t)}$, then the output is an $h$-cyclic $k$-IGDD
of type $(gh)^{(u+t,t)}$.

\vspace{0.2cm}

We summarize our results on $h$-cyclic $3$-IGDDs of type
$(gh)^{(u,t)}$ as follows.

\begin{lemma}\label{2^{7,3}}\cite{wc}
There exists a $2$-cyclic $3$-IGDD of type $2^{(7,3)}$.
\end{lemma}

\begin{lemma}\cite{zc}\label{sequence}
If $(m,d)\equiv(2,0),(2,1),(1,0),(3,1)\hspace{-0.3cm}\pmod{(4,2)}$
and $m(m-2d+1)+2\geq0$, then $[d,d+3m]\setminus\{d+3m-1\}$ can be
partitioned into triples $\{a_i,b_i,c_i\}$, $1\leq i\leq m$ such
that $a_i+b_i=c_i$.
\end{lemma}

\begin{lemma}\label{2^{12t+8,5}}
There exists a $2$-cyclic $3$-IGDD of type $2^{(u,5)}$ for any
$u\equiv8\hspace{-0.3cm}\pmod{12}$ and $u\geq20$.
\end{lemma}

\Proof For $u\in\{20,32,44,56\}$, the required $2$-cyclic $3$-IGDD
of type $2^{(u,5)}$ can be found in Lemma B.1 in Supporting
Information.

When $u\geq68$, let $d=8$ and $m=(u-14)/3$. Then $m\geq18$,
$(m,d)\equiv(2,0)\pmod{(4,2)}$ and $m(m-2d+1)+2\geq0$. By Lemma
\ref{sequence}, we partition $[8,8+3m]\setminus\{3m+7\}$ into
triples $\{a_i,b_i,c_i\}$, $1\leq i\leq m$ such that $a_i+b_i=c_i$.

Let $X=Z_{2(u-5)}\cup (I_5\times Z_2)$,
$\mathcal{G}=\{\{i,u-5+i\}:0\leq i\leq u-6\}\cup\{\{i\times
Z_2\}:i\in I_5\}$, and $Y=I_5\times Z_2$. Let $\mathcal{F}$ contains
following $m+6$ base blocks:\vspace{0.2cm}

\hspace{3.5cm}$\{0,a_j,c_j\}$, $1\leq j\leq m$;\vspace{0.15cm}

\hspace{3.5cm}$\{0,u-7,(4,0)\}$, $\{0,2,6\}$;\vspace{0.15cm}

\hspace{3.5cm}$\{0,(2i+1),(i,0)\}$, $0\leq i\leq3$.\vspace{0.2cm}

Developing these base blocks by $+1$ mod $2(u-5)$ yields all blocks,
where $(i,x)+1\equiv(i,x+1)$ mod $(-,2)$. It is readily checked that
the designs are isomorphism to $2$-cyclic $3$-IGDDs of type
$2^{(u,5)}$.\qed

\begin{lemma}\label{2^{u,t}}
There exists a $2$-cyclic $3$-IGDD of type $2^{(u,t)}$ if $u,t$
satisfy one of the following conditions.\vspace{0.1cm}

$(1)$ $u\equiv2\hspace{-0.3cm}\pmod{12}$, $u\geq38$ and $t=14$;

$(2)$ $u\equiv11\hspace{-0.3cm}\pmod{12}$, $u\geq23$ and $t=11$.

\end{lemma}

\Proof $(1)$  Start with a $2$-cyclic $3$-GDD of type
$24^{(u-2)/12}$ from Corollary \ref{w-cyclic 3gdp}. Fill in the
groups with a $2$-cyclic $3$-IGDD of type $2^{(14,2)}$ from Lemma
B.2 in Supporting Information. By Construction \ref{fill4}, we
obtain a $2$-cyclic $3$-IGDD of type $2^{(u,14)}$.

$(2)$ By Lemma \ref{gtw1}, there exists a $3$-GDD of type
$2^{(u-11)/4}4^1$. Apply Construction \ref{weight2} with a
$2$-cyclic $3$-GDD of type $4^3$ to obtain a $2$-cyclic $3$-GDD of
type $8^{(u-11)/4}16^1$. Filling in the groups with a $2$-cyclic
$3$-IGDD of type $2^{(7,3)}$ from Lemma \ref{2^{7,3}}, by
Construction \ref{fill4}, then we obtain a $2$-cyclic $3$-IGDD of
type $2^{(u,11)}$. \qed

\begin{lemma}\label{vw^{u,11}}
Let $u\equiv11\hspace{-0.3cm}\pmod{12}$ and $u\geq23$. Then there
exists a $w$-cyclic $3$-IGDD of type $(vw)^{(u,11)}$ for any
$v\equiv1,5\hspace{-0.3cm}\pmod{6}$ and
$w\equiv10\hspace{-0.3cm}\pmod{12}$.
\end{lemma}

\Proof Start with a PBD$(u,\{3,11^*\})$, which is also a $3$-GDD of
type $1^{u-11}11^1$ from Lemma \ref{gtw1}. Apply Construction
\ref{ihgdd} with a $3$-SCHGDD of type $(3,2^{w/2})$ from Theorem
\ref{schgdd} and a $2$-cyclic $3$-IGDD of type $2^{(u,11)}$ from
Lemma \ref{2^{u,t}} to obtain a $w$-cyclic $3$-IGDD of type
$w^{(u,11)}$. Further apply Construction \ref{inf3} with a
$1$-cyclic $3$-GDD of type $v^3$ from Corollary \ref{w-cyclic 3gdp}
to obtain a $w$-cyclic $3$-IGDD of type $(vw)^{(u,11)}$.\qed

\begin{lemma}\label{4^{u,2}}
There exists a $4$-cyclic $3$-IGDD of type $4^{(u,2)}$ for any
$u\equiv2\hspace{-0.3cm}\pmod{6}$ and $u\geq8$.
\end{lemma}

\Proof When $u\in\{8,14\}$, by Lemma B.3 in Supporting Information,
there exists a $4$-cyclic $3$-IGDD of type $4^{(u,2)}$. When
$u\geq20$, start with a $4$-cyclic $3$-GDD of type $24^{(u-2)/6}$
from Corollary \ref{w-cyclic 3gdp}. Apply Construction \ref{fill4}
with a $4$-cyclic $3$-IGDD of type $4^{(8,2)}$ to obtain a
$4$-cyclic $3$-IGDD of type $4^{(u,2)}$. \qed

\begin{lemma}\label{6^{u,2}}
There exists a $6$-cyclic $3$-IGDD of type $6^{(u,6)}$ for any
$u\equiv2\hspace{-0.3cm}\pmod{12}$ and $u\geq14$.
\end{lemma}

\Proof Start with a  $6$-cyclic $3$-GDD of type $24^{(u-2)/4}$ from
Corollary \ref{w-cyclic 3gdp}. Take a $6$-cyclic $3$-IGDD of type
$6^{(6,2)}$, which can be found in Lemma B.4 in Supporting
Information, to fill in the groups. We then obtain the required
design by Construction \ref{fill4}.\qed

\begin{corollary}\label{2-cyclic 6^{u,6}}
Let $u\equiv2\hspace{-0.3cm}\pmod{12}$ and $u\geq14$. Then there
exists a $2$-cyclic $3$-IGDD of type $6^{(u,6)}$.
\end{corollary}

\Proof By Lemma \ref{6^{u,2}}, there exists a $6$-cyclic $3$-IGDD of
type $6^{(u,6)}$, which is also a $2$-cyclic $3$-IGDD of type
$6^{(u,6)}$.\qed

\begin{lemma}\label{vw^{u,5} u=6t+2}
A $w$-cyclic $3$-IGDD of type $(vw)^{(u,5)}$ exists if $u,v,w$
satisfy one of following conditions.

$(1)$ $u\equiv2\hspace{-0.3cm}\pmod{12}$, $u\geq14$,
$v\equiv0\hspace{-0.3cm}\pmod{2}$ and
$w\equiv0\hspace{-0.3cm}\pmod{2}$;

$(2)$ $u\equiv8\hspace{-0.3cm}\pmod{12}$, $u\geq20$, $v\geq1$,
$w\equiv0\hspace{-0.3cm}\pmod{2}$ and $w\not\in\{4,6\}$.

\end{lemma}

\Proof $(1)$ Start with a PBD$(u,\{3,4,5^*\})$, which is also a
$1$-cyclic $\{3,4\}$-IGDD of type $1^{(u,5)}$. Give weight $vw$ to
each point. Input a $w$-cyclic $3$-GDD of type $(vw)^k$ for each
$k\in\{3,4\}$ from Corollary \ref{w-cyclic 3gdp}. By Construction
\ref{inf3}, we then obtain a $w$-cyclic $3$-IGDD of type
$(vw)^{(u,5)}$.

$(2)$ When $w=2$, by Lemma \ref{2^{12t+8,5}}, there exists a
$2$-cyclic $3$-IGDD of type $2^{(u,5)}$. Inflate it by a $1$-cyclic
$3$-GDD of type $v^3$ from Corollary \ref{w-cyclic 3gdp} to obtain
the required design.

When $w\geq8$, similar to the proof of Lemma \ref{vw^{u,11}}, we
start with a PBD$(u,\{3,4,5^*\})$. The required $2$-cyclic $3$-IGDD
of type $2^{(u,5)}$ and $3$-SCHGDDs of type $(k,2^{w/2})$ for
$k\in\{3,4\}$ can be found in Lemma \ref{2^{12t+8,5}} and Theorem
\ref{schgdd}, respectively. \qed

We give an existence result on $h$-cyclic $3$-IGDPs of type
$(gh)^{(u,t)}$ at the end of this subsection.

\begin{lemma}\label{w-cyclic 3-igdp}
Let $u\equiv2\hspace{-0.3cm}\pmod{6}$ and $u\geq20$. Then there
exists a $w$-cyclic $3$-IGDP of type $(vw)^{(u,8)}$ with
$v(u-8)((u+7)vw-1)/6$ base blocks for any odd integers $v$ and
$w\geq3$.
\end{lemma}

\Proof  By Lemma \ref{(8,2,1^nh)}, there exists a $(1,w,v)$-cyclic
$3$-IHGDD of type $(8,2,1^{vw})$, which has $9v(vw-1)$ base blocks.
Take a $3$-GDD of type $2^4$ from Lemma \ref{gtw1}, which has $8$
blocks and can be considered as a $1$-cyclic $3$-IGDP of type
$1^{(8,2)}$. Fill in the holes of the IHGDD to obtain a $w$-cyclic
$3$-IGDP of type $(vw)^{(8,2)}$. Clearly, it has
$9v(vw-1)+8v=v(9vw-1)$ base blocks.

When $u\geq20$, by Corollary \ref{w-cyclic 3gdp}, there exists a
$w$-cyclic $3$-GDD of type $(6vw)^{(u-2)/6}$, which has
$(u-2)(u-8)v^2w/6$ base blocks. Filling in $(u-8)/6$ groups of this
GDD by above $w$-cyclic $3$-IGDP of type $(vw)^{(8,2)}$ with
$v(9vw-1)$ base blocks, then we obtain a $w$-cyclic $3$-IGDP of type
$(vw)^{(u,8)}$ with $v(u-8)((u+7)vw-1)/6$ base blocks by
Construction \ref{fill4}. \qed

\section{Main result}

In this section, applying the filling constructions established in
above sections, we obtain the main result of this paper.

\subsection{The possible exceptions in Corollary \ref{w-cyclic 3gdp}}

We first introduce a new auxiliary design, which plays an important
role in the constructions of some optimal $w$-cyclic $3$-GDPs.
\vspace{0.15cm}

A $3$-SCGDP$^*$ of type $2^{(u,t)}$ is a $3$-SCGDP of type $2^{u}$
on $I_u\times Z_2$ with group set $\{\{i\}\times Z_2:i\in I_u\}$ in
which, for any $a,b\in I_t$ and $x\in Z_2$, all pairs
$\{(a,x),(b,x+1)\}$ are not covered by any block.

\begin{example}\label{small *gdp}
There exists a $3$-SCGDP$^*$ of type $2^{(11,5)}$ with $32$ base
blocks.
\end{example}

\Proof Let $X=I_{11}\times Z_2$, $\mathcal{G}=\{\{i\}\times Z_2:i\in
I_{11}\}$. Only base blocks are listed below. \vspace{0.15cm}

{\small $\{(0,0),(8,1),(10,1)\}$, $\{(0,0),(3,0),(4,0)\}$,
$\{(0,0),(5,0),(6,0)\}$, $\{(0,0),(7,0),(8,0)\}$,

$\{(0,0),(9,0),(10,0)\}$, $\{(0,0),(5,1),(7,1)\}$,
$\{(0,0),(6,1),(9,1)\}$, $\{(0,0),(1,0),(2,0)\}$,

$\{(1,0),(7,1),(10,1)\}$, $\{(1,0),(4,0),(6,0)\}$,
$\{(1,0),(7,0),(9,0)\}$, $\{(1,0),(8,0),(5,1)\}$,

$\{(1,0),(10,0),(6,1)\}$, $\{(1,0),(3,0),(5,0)\}$,
$\{(1,0),(8,1),(9,1)\}$, $\{(2,0),(3,0),(7,0)\}$,

$\{(2,0),(10,0),(9,1)\}$, $\{(2,0),(5,0),(9,0)\}$,
$\{(2,0),(6,0),(7,1)\}$, $\{(2,0),(4,0),(8,0)\}$,

$\{(2,0),(6,1),(10,1)\}$, $\{(2,0),(5,1),(8,1)\}$,
$\{(3,0),(6,0),(8,1)\}$, $\{(3,0),(8,0),(9,1)\}$,

$\{(4,0),(10,0),(8,1)\}$, $\{(4,0),(9,0),(5,1)\}$,
$\{(4,0),(7,0),(9,1)\}$, $\{(4,0),(6,1),(7,1)\}$,

$\{(4,0),(5,0),(10,1)\}$, $\{(3,0),(10,0),(7,1)\}$,
$\{(3,0),(5,1),(10,1)\}$, $\{(3,0),(9,0),(6,1)\}$.} \qed

\begin{lemma}\label{w^11}
Let $w\equiv10\hspace{-0.3cm}\pmod{12}$. Then there exists a
$3$-SCGDP of type $w^{11}$ with $J^{*}(11\times 1\times w,3,1)$ base
blocks.
\end{lemma}

\Proof Start with a PBD$(11,\{3,5^*\})$ $(I_{11},\mathcal{B})$,
which is also a $3$-GDD of type $1^65^1$ from Lemma \ref{gtw1}.
Without loss of generality, let $I_5$ be the block of size $5$. For
each $B\in\mathcal{B}$ of size $3$, construct a $3$-SCHGDD of type
$(3,2^{w/2})$ on $B\times Z_{w}$ with group set $\{\{x\}\times
Z_{w}:x\in B\}$ and hole set $\{B\times\{i,w/2+i\}:0\leq i\leq
w/2-1\}$. For the $B\in\mathcal{B}$ of size $5$, construct a
$3$-SCHGDD of type $(5,1^{w})$ on $B\times Z_{w}$ with group set
$\{\{x\}\times Z_{w}:x\in B\}$ and hole set $\{B\times\{i\}:0\leq
i\leq w-1\}$. Note that the required $3$-SCHGDDs here can be found
in Theorem \ref{schgdd}. Let $\mathcal{A}_B$ be a set of base blocks
for each $B\in\mathcal{B}$. Let
$\mathcal{A}_1=\cup_{B\in\mathcal{B}}\mathcal{A}_B$. Clearly,
$|\mathcal{A}_1|=(55w-100)/3$.

By Example \ref{small *gdp}, there exists a $3$-SCGDP$^*$ of type
$2^{(11,5)}$ on $I_{11}\times Z_2$ with group set $\{\{i\}\times
Z_2:i\in I_{11}\}$, which has $32$ base blocks. Suppose
$\mathcal{F}$ be a family of base blocks of this design. For each
$B\in\mathcal{F}$, let $B^\prime=\{(a,wx/2):(a,x)\in B\}$. Let
$\mathcal{A}_2=\cup_{B\in\mathcal{F}}B^\prime$. It is readily
checked that $\mathcal{A}_1\cup\mathcal{A}_2$ forms a family of base
blocks of the required $3$-SCGDP of type $w^{11}$ with
$J^{*}(11\times 1\times w,3,1)=(55w-4)/3$ base blocks.\qed

\begin{lemma}\label{vw^{12t+11}}
Let $u\equiv11\hspace{-0.3cm}\pmod{12}$. Then there exists a
$w$-cyclic $3$-GDP of type $(vw)^{u}$ with $J^{*}(u\times v\times
w,3,1)$ base blocks for any $v\equiv1,5\hspace{-0.3cm}\pmod{6}$ and
$w\equiv10\hspace{-0.3cm}\pmod{12}$.
\end{lemma}

\Proof When $u=11$, by Lemma \ref{(vw,w)^11}, there exists a
$w$-cyclic $3$-IGDD of type $(vw,w)^{11}$. Filling in the hole with
a $3$-SCGDP of type $w^{11}$ with $J^{*}(11\times 1\times w,3,1)$
base blocks from Lemma \ref{w^11}, by Construction \ref{fill2}, we
obtain a $w$-cyclic $3$-GDP of type $(vw)^{11}$ with $J^{*}(11\times
v\times w,3,1)$ base blocks.

When $u\geq23$, by Lemma \ref{vw^{u,11}}, there exists a $w$-cyclic
$3$-IGDD of type $(vw)^{(u,11)}$. Filling in the hole with a
$w$-cyclic $3$-GDP of type $(vw)^{11}$ with $J^{*}(11\times v\times
w,3,1)$ base blocks, by Construction \ref{fill3}, we obtain a
$w$-cyclic $3$-GDP of type $(vw)^{u}$ with $J^{*}(u\times v\times
w,3,1)$ base blocks.\qed

\subsection{The case of $u\equiv2\hspace{-0.3cm}\pmod{6}$}

We first deal with the cases of $w\equiv0,1\hspace{-0.3cm}\pmod{6}$.

\begin{lemma}\label{12t+2 w=6mod12}
Suppose $u\equiv2\hspace{-0.3cm}\pmod{12}$ and $u\geq14$. Then there
exists a $w$-cyclic $3$-GDP of type $(vw)^{u}$ with $J^\ast(u\times
v\times w,3,1)$ base blocks for any
$v\equiv1\hspace{-0.3cm}\pmod{2}$ and $w\equiv6\pmod{12}$.
\end{lemma}

\Proof When $w=6$, start with a $6$-cyclic $3$-IGDD of type
$6^{(u,6)}$ from Lemma \ref{6^{u,2}}. Fill in the hole with a
$3$-SCGDP of type $6^{6}$ with $J^\ast(6\times 1\times 6,3,1)$ base
blocks, which exists by Corollary \ref{w-cyclic 3gdp}. Then we
obtain a $3$-SCGDP of type $6^{u}$ with $J^\ast(u\times 1\times
6,3,1)$ base blocks. Further apply Construction \ref{fill2} with a
$6$-cyclic $3$-IGDD of type $(6v,6)^{u}$ for any
$v\equiv1\hspace{-0.3cm}\pmod{2}$ and $v\geq3$, which can be found
in Lemma \ref{(6v,6)^u,u=12t+2}, then we obtain a $6$-cyclic $3$-GDP
of type $(6v)^{u}$ with $J^\ast(u\times v\times 6,3,1)$ base blocks.

When $w\geq18$, by Lemma \ref{(v,w,1)-cyclic hgdd}, there is a
$(v,w,1)$-cyclic $3$-HGDD of type $(u,(6v)^{w/6})$. Fill in the
holes with above $6$-cyclic $3$-GDP of type $(6v)^{u}$ with
$J^\ast(u\times v\times 6,3,1)$. We then obtain the required design
by Construction \ref{fill1}.\qed

\begin{lemma}\label{6t+2 w=6t+1}
Suppose $u\equiv2\hspace{-0.3cm}\pmod{6}$ and $u\geq8$. Then there
exists a $w$-cyclic $3$-GDP of type $(vw)^{u}$ with $J^\ast(u\times
v\times w,3,1)$ base blocks for any
$v\not\equiv0\hspace{-0.3cm}\pmod{6}$ and
$w\equiv1\hspace{-0.3cm}\pmod{6}$.
\end{lemma}

\Proof When $w=1$, an optimal $1$-cyclic $3$-GDP of type $v^{u}$
with $J^\ast(u\times v\times 1,3,1)$ base blocks can be found in
Corollary \ref{w-cyclic 3gdp}.

When $w\geq7$, by Lemma \ref{(v,w,1)-cyclic hgdd}, there exists a
$(v,w,1)$-cyclic $3$-HGDD of type $(u,v^{w})$. Then apply
Construction \ref{fill1} with above $1$-cyclic $3$-GDP of type
$v^{u}$ to obtain the required design. \qed

We next deal with the case of $w\equiv2\hspace{-0.3cm}\pmod{6}$.

\begin{lemma}\label{w=12t+2,v=6t+3}
Let $u\equiv2\hspace{-0.3cm}\pmod{12}$ and $u\geq14$. Then there
exists a $w$-cyclic $3$-GDP of type $(vw)^{u}$ with $J^{*}(u\times
v\times w,3,1)$ base blocks for any
$v\equiv3\hspace{-0.3cm}\pmod{6}$ and $w\equiv2\pmod{12}$.
\end{lemma}

\Proof  $(1)$ $w=2$. When $v=3$, by Corollary \ref{2-cyclic
6^{u,6}}, there exists a $2$-cyclic $3$-IGDD of type $6^{(u,6)}$.
Fill in the hole with a $2$-cyclic $3$-GDP of type $6^{6}$ from
Corollary \ref{w-cyclic 3gdp}. By Construction \ref{fill3}, we
obtain a $2$-cyclic $3$-GDP of type $6^{u}$ with $J^\ast(u\times
3\times 2,3,1)$ base blocks.

When $v\equiv3\hspace{-0.3cm}\pmod{6}$ and $v\geq9$, by Corollary
\ref{2-cyclic (6v,6)^u}, there exists a $2$-cyclic $3$-IGDD of type
$(2v,6)^{u}$. Fill in the hole with above $2$-cyclic $3$-GDP of type
$6^{u}$ to obtain a $2$-cyclic $3$-GDP of type $(2v)^{u}$ with
$J^\ast(u\times v\times 2,3,1)$ base blocks.

$(2)$ $w\geq14$. By Lemma \ref{(v,w,1)-cyclic hgdd v=6t+3}, there
exists a $(v,w,1)$-cyclic $3$-HGDD of type $(u,(2v)^{w/2})$. Apply
Construction \ref{fill1} with a $2$-cyclic $3$-GDP of type $(2v)^u$
with $J^{*}(u\times v\times 2,3,1)$ base blocks to obtain the
required designs.\qed

\begin{lemma}\label{12t+8 w=2mod6}
Suppose $u\equiv2\hspace{-0.3cm}\pmod{6}$ and $u\geq8$. Then there
exists a $w$-cyclic $3$-GDP of type $(vw)^{u}$ with $J^\ast(u\times
v\times w,3,1)$ base blocks for any
$v\equiv1,2\hspace{-0.3cm}\pmod{3}$ and $w\equiv2\pmod{6}$.
\end{lemma}
\Proof $(1)$ $u=8$. When $v=1$, the required $3$-SCGDPs of type
$w^8$ with $J^\ast(8\times 1\times w,3,1)$ base blocks are
constructed directly in Lemmas C.1 and C.5 in Supporting
Information. When $v\geq2$, there exists a $w$-cyclic $3$-IGDD of
type $(vw,w)^{8}$ by Lemma \ref{(vw,w)^8}. Fill in the hole with
above $3$-SCGDP of type $w^8$ with $J^\ast(8\times 1\times w,3,1)$
base blocks to obtain the required design.

$(2)$ $u\equiv2\hspace{-0.3cm}\pmod{12}$ and
$v\equiv1,5\hspace{-0.3cm}\pmod{6}$. We first deal with the case of
$(v,w)=(1,2)$. When $u=14$, by Lemma B.2 in Supporting Information,
there exists a $2$-cyclic $3$-IGDD of type $2^{(14,2)}$, which is
also a $3$-SCGDP of type $2^{14}$ with $J^\ast(14\times 1\times
2,3,1)$ base blocks. When $u=26$, by Lemma B.5 in Supporting
Information, there exists a $2$-cyclic $3$-IGDD of type
$2^{(26,11)}$. Fill in the hole with a $3$-SCGDP of type $2^{11}$
with $J^\ast(11\times 1\times 2,3,1)$ base blocks from Corollary
\ref{w-cyclic 3gdp} to obtain a $3$-SCGDP of type $2^{26}$ with
$J^\ast(26\times 1\times 2,3,1)$ base blocks. When $u\geq38$, by
Lemma \ref{2^{u,t}}, there exists a $2$-cyclic $3$-IGDD of type
$2^{(u,14)}$. Fill in the hole with above $3$-SCGDP of type $2^{14}$
to obtain a $3$-SCGDP of type $2^{u}$ with $J^\ast(u\times 1\times
2,3,1)$ base blocks.

When $v\geq5$ and $w=2$, by Lemma \ref{(2v,2)^8}, there exists a
$2$-cyclic $3$-IGDD of type $(2v,2)^{u}$. Filling in the hole with a
$3$-SCGDP of type $2^{u}$ with $J^\ast(u\times 1\times 2,3,1)$ base
blocks, by Construction \ref{fill2}, we obtain a $2$-cyclic $3$-GDP
of type $(2v)^{u}$ with $J^\ast(u\times v\times 2,3,1)$ base blocks.

When $w\geq8$, a $(v,w,1)$-cyclic $3$-HGDD of type $(u,(2v)^{w/2})$
exists by Lemma \ref{(v,w,1)-cyclic hgdd}. Fill in the holes with
above $2$-cyclic $3$-GDP of type $(2v)^{u}$ with $J^\ast(u\times
v\times 2,3,1)$ base blocks. Then we obtain the required design by
Construction \ref{fill1}.

$(3)$ Other values. By Lemma \ref{vw^{u,5} u=6t+2}, there exists a
$w$-cyclic $3$-IGDD of type $(vw)^{(u,5)}$. Take a $w$-cyclic
$3$-GDP of type $(vw)^{5}$ with $J^\ast(5\times v\times w,3,1)$ base
blocks from Corollary \ref{w-cyclic 3gdp} to fill in the hole. We
then obtain the required $w$-cyclic $3$-GDP of type $(vw)^{u}$ with
$J^\ast(u\times v\times w,3,1)$ base blocks.\qed

Below we deal with the case of $w\equiv3\hspace{-0.3cm}\pmod{6}$.

\begin{lemma}\label{w=6t+3,v=6t+3}
Let $u\equiv2\hspace{-0.3cm}\pmod{6}$ and $u\geq8$. Then there
exists a $w$-cyclic $3$-GDP of type $(vw)^{u}$ with $J^\ast(u\times
v\times w,3,1)$ base blocks for any $v\equiv
w\equiv3\hspace{-0.3cm}\pmod{6}$.
\end{lemma}

\Proof By Lemma \ref{(v,w,1)-cyclic hgdd v=6t+3 II}, there exists a
$(v,w,1)$-cyclic $3$-HGDD of type $(u,v^{w})$. Filling in the hole
with a $1$-cyclic $3$-GDP of type $v^{u}$ with $J^\ast(u\times
v\times1,3,1)$ base blocks from  Corollary \ref{w-cyclic 3gdp}, we
then obtain the required design by Construction \ref{fill1}.\qed

\begin{lemma}\label{w=6t+3,v=6t+1}
Let $u\equiv2\hspace{-0.3cm}\pmod{6}$ and $u\geq8$. Then there
exists a $w$-cyclic $3$-GDP of type $(vw)^{u}$ with $J^\ast(u\times
v\times w,3,1)$ base blocks for any
$v\equiv1,5\hspace{-0.3cm}\pmod{6}$ and
$w\equiv3\hspace{-0.3cm}\pmod{6}$.
\end{lemma}

\Proof $(1)$ $w=3$ and $u\in\{8,14\}$. For $v\in\{1,5\}$, Lemmas C.3
and C.4 in Supporting Information provide the required designs. For
$v\geq7$, let $v=6t+s$, where $s\in\{1,5\}$. By Lemma
\ref{(3v,3)^u}, there exists a $3$-cyclic $3$-IGDP of type
$(3v,3s)^u$, which has $u(v-s)(3(u-1)(v+s)-1)/6$ base blocks.
Filling in the hole with above $3$-cyclic $3$-GDP of type $(3s)^{u}$
to obtain the required designs.

$(2)$ $w=3$ and $u\geq20$. By Lemma \ref{w-cyclic 3-igdp}, there
exists a $3$-cyclic $3$-IGDP of type $(3v)^{(u,8)}$ with
$v(u-8)(3(u+7)v-1)/6$ base blocks. Fill in the hole with a
$3$-cyclic $3$-GDP of type $(3v)^{8}$ with $J^\ast(8\times v\times
3,3,1)$ base blocks. By Construction \ref{fill3}, we obtain a
$3$-cyclic $3$-GDP of type $(3v)^{u}$ with $J^\ast(u\times v\times
3,3,1)$ base blocks.

$(3)$ $w\geq9$. By Lemma \ref{(v,w,1)-cyclic hgdd}, there exists a
$(v,w,1)$-cyclic $3$-HGDD of type $(u,(3v)^{w/3})$. Filling in the
hole with above $3$-cyclic $3$-GDP of type $(3v)^{u}$ with
$J^\ast(u\times v\times3,3,1)$ base blocks, by Construction
\ref{fill1}, we obtain the required design. \qed

For dealing with the case of $w\equiv4\hspace{-0.3cm}\pmod{6}$, we
need to construct more $3$-SCGDP$^*$s.

\begin{lemma}\label{*gdp}
There exists a $3$-SCGDP$^*$ of type $2^{(u,5)}$ with
$(u(u-1)-14)/3$ base blocks for any
$u\equiv2\hspace{-0.3cm}\pmod{12}$ and $u\geq14$.
\end{lemma}

\Proof When $u=14$, we construct the required design in Lemma C.2 in
Supporting Information. When $u=26$, by Lemma B.5 in Supporting
Information, there exists a $2$-cyclic $3$-IGDD of type
$2^{(26,11)}$. Take a $3$-SCGDP$^*$ of type $2^{(11,5)}$ with $32$
base blocks from Example \ref{small
*gdp}. Then fill in the hole to obtain a $3$-SCGDP$^*$ of
type $2^{(26,5)}$ with $212$ base blocks. Note that the $2$-cyclic
$3$-IGDD of type $2^{(26,11)}$ has $180$ base blocks.

When $u\geq38$, by Lemma \ref{2^{u,t}}, there exists a $2$-cyclic
$3$-IGDD of type $2^{(u,14)}$. Filling in the hole with above
$3$-SCGDP$^*$ of type $2^{(14,5)}$ with $56$ base blocks, then we
obtain a $3$-SCGDP$^*$ of type $2^{(u,5)}$ with $(u(u-1)-14)/3$ base
blocks. Note that the $2$-cyclic $3$-IGDD of type $2^{(u,14)}$ has
$(u(u-1)-182)/3$ base blocks. \qed

\begin{lemma}\label{12t+2 w=4mod12}
Suppose $u\equiv2\hspace{-0.3cm}\pmod{6}$ and $u\geq8$. Then there
exists a $w$-cyclic $3$-GDP of type $(vw)^{u}$ with $J^\ast(u\times
v\times w,3,1)$ base blocks for any
$v\equiv1,2\hspace{-0.3cm}\pmod{3}$ and $w\equiv4\pmod{12}$.
\end{lemma}

\Proof When $w=4$, by Lemma \ref{4^{u,2}}, there exists a $4$-cyclic
$3$-IGDD of type $4^{(u,2)}$, which is also a $3$-SCGDP of type
$4^{u}$ with $J^\ast(u\times 1\times 4,3,1)$ base blocks. For
$v\geq2$, by Lemma \ref{(8,4)^u u=6t+2}, there exists a $4$-cyclic
$3$-IGDD of type $(4v,4)^{u}$. Filling in the hole with above
$3$-SCGDP of type $4^{u}$, by Construction \ref{fill2}, we obtain a
$4$-cyclic $3$-GDP of type $(4v)^{u}$ with $J^\ast(u\times v\times
4,3,1)$ base blocks.

For $w\geq16$, by Lemma \ref{(v,w,1)-cyclic hgdd}, there exists a
$(v,w,1)$-cyclic $3$-HGDD of type $(u,(4v)^{w/4})$. Filling in the
holes with a $4$-cyclic $3$-GDP of type $(4v)^{u}$ with
$J^\ast(u\times v\times 4,3,1)$ base blocks, by Construction
\ref{fill1}, we obtain a $w$-cyclic $3$-GDP of type $(vw)^{u}$ with
$J^\ast(u\times v\times w,3,1)$ base blocks.\qed

\begin{lemma}\label{u=6t+2,w=12t+10}
Let $u\equiv2\hspace{-0.3cm}\pmod{6}$ and $u\geq8$. Then there
exists a $w$-cyclic $3$-GDP of type $(vw)^{u}$ with $J^{*}(u\times
v\times w,3,1)$ base blocks for any
$v\equiv1,2\hspace{-0.3cm}\pmod{3}$ and $w\equiv10\pmod{12}$.
\end{lemma}

\Proof $(1)$ $u=8$. When $v=1$, there exists a $3$-SCGDP of type
$w^{8}$ with $J^{*}(8\times 1\times w,3,1)$ base blocks by Lemma C.6
in Supporting Information. When $v\geq2$, by Lemma \ref{(vw,w)^8},
there exists a $w$-cyclic $3$-IGDD of type $(vw,w)^{8}$. Filling in
the hole with above $3$-SCGDP of type $w^{8}$, we obtain a
$w$-cyclic $3$-GDP of type $(vw)^{8}$ with $J^{*}(8\times v\times
w,3,1)$ base blocks.

$(2)$ $u\equiv2\hspace{-0.3cm}\pmod{12}$ and
$v\equiv1,5\hspace{-0.3cm}\pmod{6}$. For $v=1$, the proof is similar
to that of Lemma \ref{w^11}. Here we start from a
PBD$(u,\{3,4,5^*\})$. Give weight $w$ to each point of this PBD. For
each block of size $k$, $k\in\{3,4\}$, construct a $3$-SCHGDD of
type $(k,2^{w/2})$. For the unique block of size $5$, construct a
$3$-SCHGDD of type $(5,1^{w})$. Note that the required $3$-SCHGDDs
of type $(k,2^{w/2})$ for $k\in\{3,4\}$ and $3$-SCGDP$^*$s of type
$2^{(u,5)}$ with $(u(u-1)-14)/3$ base blocks exist by Theorem
\ref{schgdd} and Lemma \ref{*gdp}, respectively.

When $v\geq5$, start with a $w$-cyclic $3$-IGDD of type $(vw,w)^u$
from Lemma \ref{(vw,sw)^u}, and fill in the hole with above
$3$-SCGDP of type $w^{u}$ with $J^\ast(u\times 1\times w,3,1)$ base
blocks. Then we obtain a $w$-cyclic $3$-GDP of type $(vw)^{u}$ with
$J^\ast(u\times v\times w,3,1)$ base blocks.

$(3)$ Other values. By Lemma \ref{vw^{u,5} u=6t+2}, there exists a
$w$-cyclic $3$-IGDD of type $(vw)^{(u,5)}$. Filling in the hole with
a $w$-cyclic $3$-GDP of type $(vw)^{5}$ with $J^{*}(5\times v\times
w,3,1)$ base blocks from Corollary \ref{w-cyclic 3gdp}, we obtain a
$w$-cyclic $3$-GDP of type $(vw)^{u}$ with $J^{*}(u\times v\times
w,3,1)$ base blocks. \qed

\begin{lemma}\label{w=12t+10,v=6t+3}
Let $u\equiv2\hspace{-0.3cm}\pmod{12}$ and $u\geq14$. Then there
exists a $w$-cyclic $3$-GDP of type $(vw)^{u}$ with $J^{*}(u\times
v\times w,3,1)$ base blocks for any
$v\equiv3\hspace{-0.3cm}\pmod{6}$ and $w\equiv10\pmod{12}$.
\end{lemma}

\Proof  By Lemma \ref{(v,w,1)-cyclic hgdd v=6t+3}, there exists a
$(v,w,1)$-cyclic $3$-HGDD of type $(u,(2v)^{w/2})$. Apply
Construction \ref{fill1} with a $2$-cyclic $3$-GDP of type $(2v)^u$
with $J^{*}(u\times v\times 2,3,1)$ base blocks, which exits by
Lemma \ref{w=12t+2,v=6t+3}. Then we obtain the required $w$-cyclic
$3$-GDP of type $(vw)^{u}$ with $J^{*}(u\times v\times w,3,1)$ base
blocks.\qed

Finally, we deal with the case of $w\equiv5\hspace{-0.3cm}\pmod{6}$.

\begin{lemma}\label{w=6t+5,v=6t+2,4}
Let $u\equiv2\hspace{-0.3cm}\pmod{6}$ and $u\geq8$. Then there
exists a $w$-cyclic $3$-GDP of type $(vw)^{u}$ with $J^\ast(u\times
v\times w,3,1)$ base blocks for any
$v\equiv2,4\hspace{-0.3cm}\pmod{6}$ and
$w\equiv5\hspace{-0.3cm}\pmod{6}$.
\end{lemma}

\Proof By Lemma \ref{gtw1}, there exists a $3$-GDD of type
$3^{2(u-2)/3}5^1$, which is also a PBD($2u+1,\{3,5^*\}$). Delete one
point which not belonging to the block of size $5$, we obtain a
$\{3,5^*\}$-GDD of type $2^u$. Give weight $vw/2$ and input
$w$-cyclic $3$-GDDs of type $(vw/2)^3$ and a $w$-cyclic $3$-GDP of
type $(vw/2)^5$ with $J^\ast(5\times v/2\times w,3,1)$ base blocks
from Corollary \ref{w-cyclic 3gdp}. Then we obtain a $w$-cyclic
$3$-GDP of type $(vw)^{u}$ with $J^\ast(u\times v\times w,3,1)$ base
blocks.\qed

\begin{lemma}\label{w=6t+5,v=6t+5}
Let $u\equiv2\hspace{-0.3cm}\pmod{6}$ and $u\geq8$. Then there
exists a $w$-cyclic $3$-GDP of type $(vw)^{u}$ with $J^{*}(u\times
v\times w,3,1)$ base blocks for any
$v\equiv5\hspace{-0.3cm}\pmod{6}$ and
$w\equiv5\hspace{-0.3cm}\pmod{6}$.
\end{lemma}

\Proof By Theorem \ref{schgdd}, there exists a $3$-SCHGDD of type
$(u,1^{vw})$. Fill in the holes with a $1$-cyclic $3$-GDP of type
$1^u$ with $u(u-2)/6$ blocks, which is also a $3$-GDD of type
$2^{u/2}$ and exists by Lemma \ref{gtw1}. We then obtain a $3$-SCGDP
of type $(vw)^u$ with $J^{*}(u\times 1\times vw,3,1)$ base blocks.
Then, by Lemma \ref{scgdp to w-cyclic gdp}, we obtain a $w$-cyclic
$3$-GDP of type $(vw)^u$ with $vJ^{*}(u\times 1\times
vw,3,1)$=$J^{*}(u\times v\times w,3,1)$ base blocks.\qed

\begin{lemma}\label{w=6t+5,v=6t+3}
Let $u\equiv2\hspace{-0.3cm}\pmod{6}$ and $u\geq8$. Then there
exists a $w$-cyclic $3$-GDP of type $(vw)^{u}$ with $J^{*}(u\times
v\times w,3,1)$ base blocks for any
$v\equiv3\hspace{-0.3cm}\pmod{6}$ and
$w\equiv5\hspace{-0.3cm}\pmod{6}$.
\end{lemma}

\Proof By Lemma \ref{(v,w,1)-cyclic hgdd v=6t+3}, there exists a
$(v,w,1)$-cyclic $3$-HGDD of type $(u,v^{w})$. Apply Construction
\ref{fill1} with a $1$-cyclic $3$-GDP of type $v^u$ with
$J^{*}(u\times v\times 1,3,1)$ base blocks, which exists by
Corollary \ref{w-cyclic 3gdp}. Then we obtain a $w$-cyclic $3$-GDP
of type $(vw)^{u}$ with $J^{*}(u\times v\times w,3,1)$ base
blocks.\qed

\begin{lemma}\label{w=6t+5,v=6t+1}
Let $u\equiv2\hspace{-0.3cm}\pmod{6}$ and $u\geq8$. Then there
exists a $w$-cyclic $3$-GDP of type $(vw)^{u}$ with $J^\ast(u\times
v\times w,3,1)$ base blocks for any
$v\equiv1\hspace{-0.3cm}\pmod{6}$ and
$w\equiv5\hspace{-0.3cm}\pmod{6}$.
\end{lemma}

\Proof $(1)$ $u\in\{8,14\}$. When $v=1$, Lemmas C.7 and C.8 in
Supporting Information provide the required designs. When $v\geq7$,
by Lemma \ref{w-cyclic igdp w=6t+5}, there exists a $w$-cyclic
$3$-IGDP of type $(vw,w)^u$, which has $u(v-1)((u-1)(v+1)w-1)/6$
base blocks. Filling in the hole using a $3$-SCGDP of type $w^{u}$
with $J^\ast(u\times 1\times w,3,1)$ base blocks, we then obtain the
required designs by Construction \ref{fill2}.

$(2)$ $u\geq20$. By Lemma \ref{w-cyclic 3-igdp}, there exists a
$w$-cyclic $3$-IGDP of type $(vw)^{(u,8)}$ which has
$v(u-8)((u+7)vw-1)/6$ base blocks. Filling in the hole with a
$w$-cyclic $3$-GDP of type $(vw)^{8}$ with $J^\ast(8\times v\times
w,3,1)$ base blocks, we obtain the required design by Construction
\ref{fill3}.\qed

Now, we are in the position to prove our main result of this
section.

\begin{theorem}\label{main result}
There exists a $w$-cyclic $3$-GDP of type $(vw)^{u}$ with
$J^\ast(u\times v\times w,3,1)$ base blocks for any positive
integers $v,w$ and $u\geq3$.
\end{theorem}

\Proof Combining Lemmas \ref{vw^{12t+11}}-\ref{w=6t+3,v=6t+1} and
\ref{12t+2 w=4mod12}-\ref{w=6t+5,v=6t+1}, and Corollary
\ref{w-cyclic 3gdp}, the conclusion then follows.\qed

\section{Concluding remarks}

By Theorems \ref{equi} and \ref{main result}, the size of an optimal
AM-OPP $3$-D $(u\times v\times w,3,1)$-OOC is finally determined.

\begin{theorem}\label{ooc}
There exists an optimal AM-OPP $3$-D $(u\times v\times w,3,1)$-OOC
with $J^\ast(u\times v\times w,3,1)$ codewords for any positive
integers $v,w$ and $u\geq3$.
\end{theorem}

In \cite{wy}, Wang and Yin proved that an SCHP$(2,k,uw)$ of type
$w^u$, which is in fact a $k$-SCGDP of type $w^u$, is equivalent to
an AM-OPPW $2$-D $(u\times w,k,1)$-OOC. Therefore the size of an
optimal AM-OPPW $2$-D $(u\times w,3,1)$-OOCs is also determined by
Theorem \ref{main result}.

\begin{theorem}\label{2-D OOc}
There exists an optimal AM-OPPW $2$-D $(u\times w,3,1)$-OOC with
$J^\ast(u\times 1\times w,3,1)$ codewords for any positive integers
$w$ and $u\geq3$.
\end{theorem}


\begin{thebibliography}{99}

\bibitem{abg}R. J. R. Abel, F. E. Bennett and M. Greig, PBD-closure, in: CRC
Handbook of Combinatorial Designs (C. J. Colbourn and J. H. Dinitz,
eds.), CRC Press, (2007), 247-255.

\bibitem{ab}R. J. R. Abel, M. Buratti, Some progress on $(v, 4,
1)$ difference families and optical orthogonal codes, J. Combin.
Theory (A), 106 (2004), 59-75.


\bibitem{be}S. Bitan, T. Etzion, Constructions for optimal constant weight
cyclically permutable codes and difference families, IEEE Trans.
Inf. Theory, 41 (1995),  77-87.

\bibitem{bl}E. J. Billington, C. C. Lindner, Maximum packings of uniform group divisible triple
systems, J. Combin. Des., 4 (1996), 397-404.

\bibitem{cfm}Y. Chang, R. Fuji-Hara and Y. Miao, Combinatorial constructions
of optimal optical orthogonal codes with weight 4, IEEE Trans. Inf.
Theory, 49 (2003), 1283-1292.

\bibitem{chr}C. J. Colbourn, D. G. Hoffman and R. S. Rees, A new class of
group-divisible designs with block size three, J. Combin. Theory
(A), 59 (1992), 73-89.

\bibitem{csw}F. R. K. Chung, J. A. Salehi and V. K. Wei, Optical orthogonal
codes: Design, analysis, and applications, IEEE Trans. Inf. Theory,
35 (1989), 595-604.

\bibitem{dc}S. Dai, Y. Chang and L. Wang, Combinatorial constructions of optimal three-dimensional optical
orthogonal codes with AM-OPPS/WP restriction, manuscript.

\bibitem{f}T. Feng, private communication.

\bibitem{fwc}T. Feng, X. Wang and Y. Chang, Semi-cyclic holey group divisible designs with block size three,
Des. Codes Cryptogr., 74 (2015), 301-324.

\bibitem{fww}T. Feng, X. Wang and R. Wei, Semicyclic holey group divisible designs and
applications to sampling designs and optical orthogonal codes, J.
Combin. Des., DOI: 10.1002/jcd.21417.

\bibitem{fm}R. Fuji-Hara, Y. Miao, Optical orthogonal codes: Their bounds
and new optimal constructions, IEEE Trans. Inf. Theory, 46 (2000),
2396-2406.

\bibitem{gm}G. Ge, Y.Miao, PBDs, Frames, and Resolvability, in: CRC
Handbook of Combinatorial Designs (C. J. Colbourn and J. H. Dinitz,
eds.), CRC Press, (2007), 261-265.

\bibitem{gy}G. Ge, J. Yin, Constructions for optimal $(v, 4, 1)$ optical
orthogonal codes, IEEE Trans. Inf. Theory, 47 (2001), 2998-3004.

\bibitem{h}H. Hanani, Balanced incomplete block designs and related designs,
Discr. Math., 11 (1975), 255-369.

\bibitem{j}S. M. Johnson, A new upper bound for error-correcting codes, IEEE Trans. Inf. Theory, 8 (1962) 203-207.

\bibitem{kyp}S. Kim, K. Yu and N. Park, A new family of space/wavelength/time spread
three-dimensional optical code for OCDMA networks, J. Lightwave
Technol., 18 (2000), 502-511.

\bibitem{mc1}S. Ma, Y. Chang, A new class of optimal optical orthogonal
codes with weight five, IEEE Trans. Inf. Theory, 50 (2004),
1848-1850.

\bibitem{mc2}S. Ma, Y. Chang,
Constructions of optimal optical orthogonal codes with weight five,
J. Combin. Des., 13 (2005), 54-69.

\bibitem{okeb}R. Omrani, G. Garg, P. V. Kumar, P. Elia and P. Bhambhani, Large families
of asymptotically optimal two-dimensional optical orthogonal codes,
IEEE Trans. Inf. Theory, 58 (2012), 1163-1185.


\bibitem{s}K. W. Shum, Optimal three-dimensional optical orthogonal codes of weight three, Des. Codes Cryptogr.,
75 (2015), 109-126.

\bibitem{wy}J. Wang, J. Yin, Two-dimensional optical orthogonal codes and
semicyclic group divisible designs, IEEE Trans. Inf. Theory, 56
(2010), 2177-2187.

\bibitem{wy1}J. Wang, J. Yin, Existence of holey $3$-GDDs of type $(u,g^tw^1)$, Discr.
Math., 202 (1999), 249-269.

\bibitem{wc}L. Wang, Y. Chang, Combinatorial constructions of optimal three-dimensional optical
orthogonal codes, IEEE Trans. Inf. Theory, 61 (2015), 671-687.

\bibitem{yk}G. C. Yang, W. C. Kwong, Performance comparison of
multiwavelength CDMA and WDMA+CDMA for fiber-optic networks, IEEE
Trans. Commun., 45 (1997), 1426-1436.

\bibitem{y}J. Yin, Some combinatorial constructions for optical orthogonal
codes, Discr. Math., 185 (1998), 201-219.

\bibitem{y1} J. Yin, Packing designs with equal-sized holes, J. Stat. Plan.
Infer., 94 (2001), 393-403.

\bibitem{zc}J. Zhang, Y. Chang, The spectrum of cyclic BSA$(v,3,\lambda;\alpha)$ with
$\alpha=2,3$,
J. Combin. Des., 13 (2005), 313-335.
\end{thebibliography}
\end{document}


\newcommand{\qed}{\hphantom{.}\hfill $\Box$\medbreak}
\newcommand{\Proof}{\noindent{\bf Proof \ }}

\newtheorem{theorem}{Theorem}[section]
\newtheorem{lemma}[theorem]{Lemma}
\newtheorem{corollary}[theorem]{Corollary}
\newtheorem{remark}[theorem]{Remark}
\newtheorem{example}[theorem]{Example}
\newtheorem{definition}[theorem]{Definition}
\newtheorem{construction}[theorem]{Construction}

\title{\large{\bf Supporting Information}}

\author
 {
       Lidong Wang, Yanxun Chang
  }
\date{}

\maketitle


\appendix

\section{$h$-cyclic $3$-IGDDs of type $(gh,th)^u$}

\begin{lemma}\label{small perfect igdd}
There exists an $h$-perfect $2h$-cyclic $3$-IGDD of type $(4h,2h)^8$
for $h\in\{1,2,4\}$.
\end{lemma}

\Proof Let $X=I\times I_2\times Z_{2h}$, $Y=I\times \{0\}\times
Z_{2h}$, and $\mathcal{G}=\{\{i\}\times I_2\times Z_{2h}:i\in I\}$,
where $I=Z_7\cup\{\infty\}$. All base blocks can be obtained by
developing following initial base blocks by $(+1,-,-)$ mod
$(7,-,-)$, where $\infty+1=\infty$.\vspace{0.2cm}

$h=1:$

\begin{center}
{ \small
\begin{tabular}{llll}
$\{(0,0,0),(1,1,0),(3,1,1)\}$, &$\{(0,0,0),(5,1,0),(2,1,0)\}$,
&$\{(0,0,0),(6,1,1),(\infty,1,1)\}$,\\
$\{(0,0,0),(6,1,0),(4,1,0)\}$, &$\{(0,0,0),(1,1,1),(2,1,1)\}$,
&$\{(0,0,0),(5,1,1),(\infty,1,0)\}$,\\
$\{(0,0,0),(4,1,1),(3,1,0)\}$, &$\{(1,1,0),(4,1,1),(\infty,0,0)\}$.
\end{tabular}}
\end{center}

$h=2:$

\begin{center}
{ \small
\begin{tabular}{llll}
$\{(0,0,0),(1,1,0),(3,1,1)\}$, &$\{(0,0,0),(5,1,2),(2,1,0)\}$,
&$\{(0,0,0),(4,1,1),(3,1,0)\}$,\\
$\{(0,0,0),(6,1,2),(2,1,3)\}$, &$\{(0,0,0),(1,1,3),(6,1,1)\}$,
&$\{(0,0,0),(5,1,1),(2,1,2)\}$,\\
$\{(0,0,0),(4,1,0),(3,1,2)\}$, &$\{(0,0,0),(1,1,2),(4,1,2)\}$,
&$\{(0,0,0),(3,1,3),(5,1,3)\}$,\\
$\{(0,0,0),(5,1,0),(6,1,0)\}$, &$\{(0,0,0),(4,1,3),(\infty,1,0)\}$,
&$\{(0,0,0),(1,1,1),(\infty,1,1)\}$,\\
$\{(0,0,0),(2,1,1),(\infty,1,3)\}$,
&$\{(0,0,0),(6,1,3),(\infty,1,2)\}$,
&$\{(1,1,0),(2,1,3),(\infty,0,0)\}$,\\
$\{(1,1,0),(6,1,1),(\infty,0,3)\}$.
\end{tabular}}
\end{center}

$h=4:$

\begin{center}
{ \small
\begin{tabular}{llll}
$\{(0,0,0),(1,1,0),(3,1,1)\}$, &$\{(0,0,0),(5,1,2),(2,1,4)\}$,
&$\{(0,0,0),(4,1,5),(3,1,0)\}$,\\
$\{(0,0,0),(6,1,6),(2,1,3)\}$, &$\{(0,0,0),(1,1,7),(6,1,5)\}$,
&$\{(0,0,0),(5,1,1),(2,1,2)\}$,\\
$\{(0,0,0),(4,1,4),(5,1,0)\}$, &$\{(0,0,0),(1,1,6),(3,1,6)\}$,
&$\{(0,0,0),(3,1,7),(2,1,1)\}$,\\
$\{(0,0,0),(4,1,3),(5,1,6)\}$, &$\{(0,0,0),(6,1,4),(2,1,0)\}$,
&$\{(0,0,0),(1,1,5),(6,1,2)\}$,\\
$\{(0,0,0),(5,1,7),(2,1,7)\}$, &$\{(0,0,0),(4,1,2),(5,1,4)\}$,
&$\{(0,0,0),(6,1,3),(1,1,1)\}$,\\
$\{(0,0,0),(1,1,4),(5,1,3)\}$, &$\{(0,0,0),(3,1,5),(2,1,5)\}$,
&$\{(0,0,0),(4,1,1),(6,1,0)\}$,\\
$\{(0,0,0),(1,1,3),(\infty,1,1)\}$, &$\{(0,0,0),(3,1,4),(6,1,7)\}$,
&$\{(0,0,0),(5,1,5),(4,1,6)\}$,\\
$\{(0,0,0),(4,1,0),(\infty,1,3)\}$,
&$\{(0,0,0),(6,1,1),(\infty,1,5)\}$,
&$\{(0,0,0),(1,1,2),(\infty,1,2)\}$,\\
$\{(0,0,0),(3,1,3),(\infty,1,0)\}$,
&$\{(0,0,0),(2,1,6),(\infty,1,7)\}$,
&$\{(0,0,0),(4,1,7),(\infty,1,6)\}$,\\
$\{(0,0,0),(3,1,2),(\infty,1,4)\}$,
&$\{(1,1,0),(6,1,4),(\infty,0,6)\}$,
&$\{(1,1,0),(2,1,1),(\infty,0,4)\}$,\\
$\{(1,1,0),(4,1,2),(\infty,0,1)\}$,
&$\{(1,1,0),(6,1,3),(\infty,0,0)\}$.
\end{tabular}}
\end{center}

\begin{lemma}\label{(8,4)^6}
There exists a $4$-cyclic $3$-IGDD of type $(8,4)^u$ for any
$u\in\{4,6,14\}$.
\end{lemma}

\Proof Let $X=I\times I_2\times Z_{4}$, $Y=I\times \{0\}\times
Z_{4}$, and $\mathcal{G}=\{\{i\}\times I_2\times Z_{4}:i\in I\}$,
where $I=Z_{u-1}\cup\{\infty\}$. All base blocks can be obtained by
developing following initial base blocks by $(+1,-,-)$ mod
$(u-1,-,-)$, where $\infty+1=\infty$.\vspace{0.2cm}

$u=4:$

\begin{center}
{ \small
\begin{tabular}{llll}
$\{(0,0,0),(1,1,0),(2,1,2)\}$, &$\{(0,0,0),(1,1,3),(2,1,3)\}$,
&$\{(0,0,0),(2,1,1),(\infty,1,0)\}$,\\
$\{(0,0,0),(1,1,2),(\infty,1,2)\}$,
&$\{(0,0,0),(2,1,0),(\infty,1,1)\}$,
&$\{(0,0,0),(1,1,1),(\infty,1,3)\}$,\\
$\{(1,1,0),(0,1,1),(\infty,0,0)\}$,
&$\{(1,1,0),(2,1,1),(\infty,0,2)\}$.
\end{tabular}}
\end{center}

$u=6:$

\begin{center}
{ \small
\begin{tabular}{llll}
$\{(0,0,0),(1,1,0),(3,1,1)\}$, &$\{(0,0,0),(2,1,3),(1,1,1)\}$,
&$\{(0,0,0),(4,1,0),(2,1,0)\}$,\\
$\{(0,0,0),(3,1,2),(4,1,1)\}$, &$\{(0,0,0),(1,1,2),(3,1,0)\}$,
&$\{(0,0,0),(3,1,3),(4,1,3)\}$,\\
$\{(0,0,0),(2,1,1),(\infty,1,0)\}$,
&$\{(0,0,0),(4,1,2),(\infty,1,2)\}$,
&$\{(0,0,0),(1,1,3),(\infty,1,1)\}$,\\
$\{(0,0,0),(2,1,2),(\infty,1,3)\}$,
&$\{(1,1,0),(0,1,3),(\infty,0,0)\}$,
&$\{(1,1,0),(4,1,1),(\infty,0,3)\}$.
\end{tabular}}
\end{center}

$u=14:$

\begin{center}
{ \small
\begin{tabular}{llll}
$\{(0,0,0),(1,1,0),(3,1,1)\}$, &$\{(0,0,0),(5,1,2),(8,1,3)\}$,
&$\{(0,0,0),(7,1,3),(\infty,1,1)\}$,\\
$\{(0,0,0),(9,1,0),(8,1,0)\}$, &$\{(0,0,0),(11,1,1),(1,1,3)\}$,
&$\{(0,0,0),(2,1,3),(7,1,0)\}$,\\
$\{(0,0,0),(4,1,0),(9,1,2)\}$, &$\{(0,0,0),(6,1,1),(2,1,2)\}$,
&$\{(0,0,0),(8,1,2),(12,1,3)\}$,\\
$\{(0,0,0),(10,1,3),(11,1,0)\}$, &$\{(0,0,0),(12,1,0),(4,1,3)\}$,
&$\{(0,0,0),(1,1,1),(12,1,1)\}$,\\
$\{(0,0,0),(3,1,2),(\infty,1,2)\}$, &$\{(0,0,0),(5,1,3),(4,1,1)\}$,
&$\{(0,0,0),(9,1,1),(\infty,1,0)\}$,\\
$\{(0,0,0),(11,1,2),(4,1,2)\}$, &$\{(0,0,0),(2,1,0),(6,1,0)\}$,
&$\{(0,0,0),(6,1,2),(8,1,1)\}$,\\
$\{(0,0,0),(10,1,0),(7,1,1)\}$, &$\{(0,0,0),(1,1,2),(3,1,0)\}$,
&$\{(0,0,0),(3,1,3),(11,1,3)\}$,\\
$\{(0,0,0),(5,1,0),(12,1,2)\}$, &$\{(0,0,0),(2,1,1),(5,1,1)\}$,
&$\{(0,0,0),(6,1,3),(10,1,1)\}$,\\
$\{(0,0,0),(7,1,2),(\infty,1,3)\}$, &$\{(0,0,0),(9,1,3),(10,1,2)\}$,
&$\{(1,1,0),(7,1,3),(\infty,0,0)\}$,\\
$\{(1,1,0),(8,1,3),(\infty,0,2)\}$.
\end{tabular}}
\end{center}

\begin{lemma}\label{{3,u-2}-IGDD}
Let $s\in\{1,5\}$ and $u\in\{8,14\}$. Then there exists a
$\{3,u-2\}$-IGDD of type $(6+s,s)^u$, where the blocks of size $u-2$
form a partition of the points outside the hole.
\end{lemma}

\Proof When $s=1$, let $X=Z_u\times I$, $Y=Z_u\times \{\infty\}$ and
$\mathcal{G}=\{\{i\}\times I:i\in Z_u\}$, where
$I=I_{6}\cup\{\infty\}$. When $s=5$, let $X=Z_u\times I$,
$Y=Z_u\times \{a,b,c,d,e\}$ and $\mathcal{G}=\{\{i\}\times I:i\in
Z_u\}$, where $I=I_{6}\cup\{a,b,c,d,e\}$. All blocks can be obtained
by developing base blocks by $(+1,-)$ mod $(u,-)$. Note that the
lengths of orbits generated by the blocks of size $u-2$ are
$u/2$.\vspace{0.1cm}

$(1)$ Base blocks of size $u-2$.\vspace{0.1cm}

$u=8:$\vspace{0.1cm}

\hspace{1cm}{\small$\{(0,0),(1,2),(2,4),(4,0),(5,2),(6,4)\},
\{(0,1),(1,3),(2,5),(4,1),(5,3),(6,5)\}.$}\vspace{0.1cm}

$u=14:$\vspace{0.1cm}

\hspace{1cm}{\small$\{(0,0),(1,1),(2,2),(3,3),(4,4),(5,5),(7,0),(8,1),(9,2),(10,3),(11,4),(12,5)\}.$}\vspace{0.1cm}

$(2)$ Base blocks of size $3$.\vspace{0.1cm}

$(u,s)=(8,1):$

\begin{center}
{ \small
\begin{tabular}{llll}
$\{(0,0),(1,0),(2,1)\}$, &$\{(0,0),(1,3),(2,0)\}$,
&$\{(0,0),(1,4),(2,2)\}$,
&$\{(0,0),(1,5),(2,3)\}$,\\
$\{(0,0),(2,5),(3,0)\}$, &$\{(0,0),(3,1),(4,1)\}$,
&$\{(0,0),(3,2),(4,2)\}$,
&$\{(0,0),(3,3),(4,3)\}$,\\
$\{(0,0),(3,4),(4,4)\}$, &$\{(0,0),(3,5),(4,5)\}$,
&$\{(0,0),(5,1),(6,2)\}$,
&$\{(0,0),(5,3),(6,1)\}$,\\
$\{(0,0),(5,4),(1,\infty)\}$, &$\{(0,0),(5,5),(2,\infty)\}$,
&$\{(0,0),(6,3),(3,\infty)\}$,
&$\{(0,0),(6,5),(7,\infty)\}$,\\
$\{(0,0),(7,1),(4,\infty)\}$, &$\{(0,0),(7,2),(5,\infty)\}$,
&$\{(0,0),(7,4),(6,\infty)\}$,
&$\{(0,1),(1,4),(2,1)\}$,\\
$\{(0,1),(1,5),(2,2)\}$, &$\{(0,1),(2,3),(3,2)\}$,
&$\{(0,1),(2,4),(3,3)\}$,
&$\{(0,1),(3,1),(6,4)\}$,\\
$\{(0,1),(3,5),(4,4)\}$, &$\{(0,1),(4,2),(5,5)\}$,
&$\{(0,1),(4,3),(5,4)\}$,
&$\{(0,1),(4,5),(2,\infty)\}$,\\
$\{(0,1),(5,2),(1,\infty)\}$, &$\{(0,1),(6,2),(7,\infty)\}$,
&$\{(0,1),(6,3),(4,\infty)\}$,
&$\{(0,1),(7,2),(6,\infty)\}$,\\
$\{(0,1),(7,5),(3,\infty)\}$, &$\{(0,2),(1,3),(3,2)\}$,
&$\{(0,2),(2,2),(4,3)\}$,
&$\{(0,2),(2,4),(4,4)\}$,\\
$\{(0,2),(2,5),(5,5)\}$, &$\{(0,2),(3,3),(5,3)\}$,
&$\{(0,2),(3,4),(5,\infty)\}$,
&$\{(0,2),(3,5),(2,\infty)\}$,\\
$\{(0,2),(4,5),(6,5)\}$, &$\{(0,2),(6,4),(3,\infty)\}$,
&$\{(0,3),(2,4),(3,\infty)\}$,
&$\{(0,3),(2,5),(6,4)\}$,\\
$\{(0,3),(3,3),(4,\infty)\}$, &$\{(0,3),(3,4),(6,5)\}$,
&$\{(0,3),(3,5),(5,4)\}$,
&$\{(0,3),(4,4),(2,\infty)\}$,\\
$\{(0,3),(4,5),(7,\infty)\}$, &$\{(0,4),(1,5),(3,\infty)\}$,
&$\{(0,4),(2,5),(5,4)\}$.
\end{tabular}}
\end{center}

$(u,s)=(8,5):$

\begin{center}
{ \small
\begin{tabular}{llll}
$\{(1,0),(2,0),(0,a)\}$, &$\{(1,1),(2,1),(0,a)\}$,
&$\{(1,2),(2,2),(0,a)\}$,
&$\{(1,3),(2,3),(0,a)\}$,\\
$\{(1,4),(2,4),(0,a)\}$, &$\{(1,5),(2,5),(0,a)\}$,
&$\{(3,0),(4,3),(0,a)\}$,
&$\{(3,1),(4,0),(0,a)\}$,\\
$\{(3,2),(4,1),(0,a)\}$, &$\{(3,3),(4,2),(0,a)\}$,
&$\{(3,4),(5,1),(0,a)\}$,
&$\{(3,5),(4,4),(0,a)\}$,\\
$\{(4,5),(6,0),(0,a)\}$, &$\{(5,0),(7,5),(0,a)\}$,
&$\{(5,2),(7,3),(0,a)\}$,
&$\{(5,3),(6,1),(0,a)\}$,\\
$\{(5,4),(6,2),(0,a)\}$, &$\{(5,5),(7,4),(0,a)\}$,
&$\{(6,3),(7,0),(0,a)\}$,
&$\{(6,4),(7,1),(0,a)\}$,\\
$\{(6,5),(7,2),(0,a)\}$, &$\{(1,0),(2,4),(0,b)\}$,
&$\{(1,1),(2,5),(0,b)\}$,
&$\{(1,2),(2,0),(0,b)\}$,\\
$\{(1,3),(3,0),(0,b)\}$, &$\{(1,4),(2,3),(0,b)\}$,
&$\{(1,5),(2,1),(0,b)\}$,
&$\{(2,2),(3,3),(0,b)\}$,\\
$\{(3,1),(4,4),(0,b)\}$, &$\{(3,2),(4,5),(0,b)\}$,
&$\{(3,4),(6,3),(0,b)\}$,
&$\{(3,5),(6,2),(0,b)\}$,\\
$\{(4,0),(7,0),(0,b)\}$, &$\{(4,1),(6,0),(0,b)\}$,
&$\{(4,2),(6,1),(0,b)\}$,
&$\{(4,3),(6,4),(0,b)\}$,\\
$\{(5,0),(6,5),(0,b)\}$, &$\{(5,1),(7,1),(0,b)\}$,
&$\{(5,2),(7,2),(0,b)\}$,
&$\{(5,3),(7,3),(0,b)\}$,\\
$\{(5,4),(7,4),(0,b)\}$, &$\{(5,5),(7,5),(0,b)\}$,
&$\{(1,0),(3,0),(0,c)\}$,
&$\{(1,1),(3,2),(0,c)\}$,\\
$\{(1,2),(3,4),(0,c)\}$, &$\{(1,3),(3,1),(0,c)\}$,
&$\{(1,4),(2,0),(0,c)\}$,
&$\{(1,5),(2,3),(0,c)\}$,\\
$\{(2,1),(4,3),(0,c)\}$, &$\{(2,2),(6,1),(0,c)\}$,
&$\{(2,4),(4,2),(0,c)\}$,
&$\{(2,5),(5,0),(0,c)\}$,\\
$\{(3,3),(6,2),(0,c)\}$, &$\{(3,5),(6,4),(0,c)\}$,
&$\{(4,0),(6,3),(0,c)\}$,
&$\{(4,1),(7,0),(0,c)\}$,\\
$\{(4,4),(6,5),(0,c)\}$, &$\{(4,5),(7,1),(0,c)\}$,
&$\{(5,1),(7,4),(0,c)\}$,
&$\{(5,2),(7,5),(0,c)\}$,\\
$\{(5,3),(7,2),(0,c)\}$, &$\{(5,4),(7,3),(0,c)\}$,
&$\{(5,5),(6,0),(0,c)\}$,
&$\{(1,0),(3,1),(0,d)\}$,\\
$\{(1,1),(4,1),(0,d)\}$, &$\{(1,2),(3,0),(0,d)\}$,
&$\{(1,3),(5,0),(0,d)\}$,
&$\{(1,4),(5,1),(0,d)\}$,\\
$\{(1,5),(6,0),(0,d)\}$, &$\{(2,0),(5,2),(0,d)\}$,
&$\{(2,1),(5,3),(0,d)\}$,
&$\{(2,2),(5,4),(0,d)\}$,\\
$\{(2,3),(5,5),(0,d)\}$, &$\{(2,4),(6,2),(0,d)\}$,
&$\{(2,5),(6,1),(0,d)\}$,
&$\{(3,2),(6,3),(0,d)\}$,\\
$\{(3,3),(6,4),(0,d)\}$, &$\{(3,4),(6,5),(0,d)\}$,
&$\{(3,5),(7,0),(0,d)\}$,
&$\{(4,0),(7,1),(0,d)\}$,\\
$\{(4,2),(7,2),(0,d)\}$, &$\{(4,3),(7,3),(0,d)\}$,
&$\{(4,4),(7,4),(0,d)\}$,
&$\{(4,5),(7,5),(0,d)\}$,\\
$\{(1,0),(4,3),(0,e)\}$, &$\{(1,1),(5,3),(0,e)\}$,
&$\{(1,2),(5,5),(0,e)\}$,
&$\{(1,3),(4,0),(0,e)\}$,\\
$\{(1,4),(6,0),(0,e)\}$, &$\{(1,5),(5,4),(0,e)\}$,
&$\{(2,0),(6,1),(0,e)\}$,
&$\{(2,1),(5,2),(0,e)\}$,\\
$\{(2,2),(5,1),(0,e)\}$, &$\{(2,3),(6,2),(0,e)\}$,
&$\{(2,4),(5,0),(0,e)\}$,
&$\{(2,5),(4,2),(0,e)\}$,\\
$\{(3,0),(7,2),(0,e)\}$, &$\{(3,1),(6,4),(0,e)\}$,
&$\{(3,2),(6,5),(0,e)\}$,
&$\{(3,3),(7,4),(0,e)\}$,\\
$\{(3,4),(7,0),(0,e)\}$, &$\{(3,5),(7,3),(0,e)\}$,
&$\{(4,1),(7,5),(0,e)\}$,
&$\{(4,4),(7,1),(0,e)\}$,\\
$\{(4,5),(6,3),(0,e)\}$, &$\{(0,0),(1,1),(2,2)\}$,
&$\{(1,3),(2,4),(3,5)\}$.
\end{tabular}}
\end{center}

$(u,s)=(14,1):$

\begin{center}
{ \small
\begin{tabular}{llll}
$\{(0,0),(1,0),(2,3)\}$, &$\{(0,0),(1,2),(8,3)\}$,
&$\{(0,0),(1,4),(8,5)\}$,
&$\{(0,0),(1,5),(12,1)\}$,\\
$\{(0,0),(2,0),(8,4)\}$, &$\{(0,0),(2,1),(12,2)\}$,
&$\{(0,0),(2,4),(10,1)\}$,
&$\{(0,0),(2,5),(4,5)\}$,\\
$\{(0,0),(3,0),(12,3)\}$, &$\{(0,0),(3,1),(5,1)\}$,
&$\{(0,0),(3,2),(13,4)\}$,
&$\{(0,0),(3,4),(7,3)\}$,\\
$\{(0,0),(3,5),(7,1)\}$, &$\{(0,0),(4,0),(13,1)\}$,
&$\{(0,0),(4,1),(10,\infty)\}$,
&$\{(0,0),(4,2),(5,\infty)\}$,\\
$\{(0,0),(4,3),(5,0)\}$, &$\{(0,0),(5,2),(4,\infty)\}$,
&$\{(0,0),(5,3),(9,4)\}$,
&$\{(0,0),(5,4),(13,2)\}$,\\
$\{(0,0),(6,0),(3,\infty)\}$, &$\{(0,0),(6,1),(7,5)\}$,
&$\{(0,0),(6,2),(12,\infty)\}$,
&$\{(0,0),(6,3),(1,\infty)\}$,\\
$\{(0,0),(6,5),(10,4)\}$, &$\{(0,0),(7,2),(13,5)\}$,
&$\{(0,0),(7,4),(2,\infty)\}$,
&$\{(0,0),(8,2),(10,5)\}$,\\
$\{(0,0),(9,5),(6,\infty)\}$, &$\{(0,0),(10,2),(11,1)\}$,
&$\{(0,0),(11,2),(9,\infty)\}$,
&$\{(0,0),(11,3),(8,\infty)\}$,\\
$\{(0,0),(11,5),(7,\infty)\}$, &$\{(0,0),(12,4),(13,\infty)\}$,
&$\{(0,1),(1,1),(11,3)\}$,
&$\{(0,1),(1,3),(7,3)\}$,\\
$\{(0,1),(1,4),(7,4)\}$, &$\{(0,1),(2,2),(9,4)\}$,
&$\{(0,1),(2,4),(5,2)\}$,
&$\{(0,1),(2,5),(7,2)\}$,\\
$\{(0,1),(3,1),(8,1)\}$, &$\{(0,1),(3,2),(5,\infty)\}$,
&$\{(0,1),(3,3),(4,2)\}$,
&$\{(0,1),(4,1),(12,3)\}$,\\
$\{(0,1),(4,3),(13,3)\}$, &$\{(0,1),(4,4),(5,3)\}$,
&$\{(0,1),(5,4),(7,5)\}$,
&$\{(0,1),(5,5),(11,2)\}$,\\
$\{(0,1),(6,2),(13,5)\}$, &$\{(0,1),(6,3),(4,\infty)\}$,
&$\{(0,1),(6,5),(7,\infty)\}$,
&$\{(0,1),(8,4),(10,\infty)\}$,\\
$\{(0,1),(8,5),(13,\infty)\}$, &$\{(0,1),(9,2),(3,\infty)\}$,
&$\{(0,1),(9,5),(11,\infty)\}$,
&$\{(0,1),(11,4),(9,\infty)\}$,\\
$\{(0,1),(12,2),(2,\infty)\}$, &$\{(0,1),(12,4),(8,\infty)\}$,
&$\{(0,1),(12,5),(1,\infty)\}$,
&$\{(0,1),(13,4),(12,\infty)\}$,\\
$\{(0,2),(1,2),(10,2)\}$, &$\{(0,2),(1,4),(4,3)\}$,
&$\{(0,2),(1,5),(11,2)\}$,
&$\{(0,2),(2,2),(13,5)\}$,\\
$\{(0,2),(2,3),(12,3)\}$, &$\{(0,2),(3,3),(10,\infty)\}$,
&$\{(0,2),(3,4),(12,5)\}$,
&$\{(0,2),(4,4),(13,4)\}$,\\
$\{(0,2),(5,3),(12,4)\}$, &$\{(0,2),(5,4),(9,\infty)\}$,
&$\{(0,2),(5,5),(9,3)\}$,
&$\{(0,2),(6,2),(3,\infty)\}$,\\
$\{(0,2),(6,3),(7,\infty)\}$, &$\{(0,2),(8,4),(5,\infty)\}$,
&$\{(0,2),(10,3),(11,3)\}$,
&$\{(0,3),(1,5),(11,5)\}$,\\
$\{(0,3),(2,3),(7,5)\}$, &$\{(0,3),(2,4),(5,\infty)\}$,
&$\{(0,3),(3,3),(6,\infty)\}$,
&$\{(0,3),(3,4),(8,5)\}$,\\
$\{(0,3),(3,5),(6,5)\}$, &$\{(0,3),(4,5),(2,\infty)\}$,
&$\{(0,3),(5,4),(10,\infty)\}$,
&$\{(0,3),(6,4),(13,\infty)\}$,\\
$\{(0,3),(9,4),(12,5)\}$, &$\{(0,3),(12,4),(4,\infty)\}$,
&$\{(0,3),(13,5),(8,\infty)\}$,
&$\{(0,4),(1,4),(11,4)\}$,\\
$\{(0,4),(2,4),(13,5)\}$, &$\{(0,4),(4,5),(8,\infty)\}$,
&$\{(0,4),(6,5),(12,5)\}$,
&$\{(0,5),(1,5),(7,\infty)\}$,\\
$\{(0,5),(5,5),(13,\infty)\}$.
\end{tabular}}
\end{center}

$(u,s)=(14,5):$

\begin{center}
{ \small
\begin{tabular}{llll}
$\{(1,0),(2,0),(0,a)\}$, &$\{(1,1),(8,0),(0,a)\}$,
&$\{(1,2),(13,1),(0,a)\}$,
&$\{(1,3),(7,1),(0,a)\}$,\\
$\{(1,4),(8,2),(0,a)\}$, &$\{(1,5),(4,4),(0,a)\}$,
&$\{(2,1),(5,0),(0,a)\}$,
&$\{(2,2),(11,2),(0,a)\}$,\\
$\{(2,3),(9,5),(0,a)\}$, &$\{(2,4),(6,2),(0,a)\}$,
&$\{(2,5),(3,4),(0,a)\}$,
&$\{(3,0),(5,1),(0,a)\}$,\\
$\{(3,1),(7,4),(0,a)\}$, &$\{(3,2),(11,5),(0,a)\}$,
&$\{(3,3),(12,4),(0,a)\}$,
&$\{(3,5),(6,3),(0,a)\}$,\\
$\{(4,0),(10,4),(0,a)\}$, &$\{(4,1),(13,0),(0,a)\}$,
&$\{(4,2),(11,0),(0,a)\}$,
&$\{(4,3),(11,1),(0,a)\}$,\\
$\{(4,5),(10,3),(0,a)\}$, &$\{(5,2),(13,2),(0,a)\}$,
&$\{(5,3),(10,5),(0,a)\}$,
&$\{(5,4),(9,3),(0,a)\}$,\\
$\{(5,5),(6,5),(0,a)\}$, &$\{(6,0),(13,5),(0,a)\}$,
&$\{(6,1),(9,1),(0,a)\}$,
&$\{(6,4),(8,3),(0,a)\}$,\\
$\{(7,0),(11,3),(0,a)\}$, &$\{(7,2),(13,3),(0,a)\}$,
&$\{(7,3),(10,2),(0,a)\}$,
&$\{(7,5),(12,0),(0,a)\}$,\\
$\{(8,1),(12,1),(0,a)\}$, &$\{(8,4),(10,1),(0,a)\}$,
&$\{(8,5),(12,3),(0,a)\}$,
&$\{(9,0),(12,2),(0,a)\}$,\\
$\{(9,2),(13,4),(0,a)\}$, &$\{(9,4),(12,5),(0,a)\}$,
&$\{(10,0),(11,4),(0,a)\}$,
&$\{(1,0),(4,4),(0,b)\}$,\\
$\{(1,1),(5,0),(0,b)\}$, &$\{(1,2),(8,5),(0,b)\}$,
&$\{(1,3),(11,2),(0,b)\}$,
&$\{(1,4),(11,3),(0,b)\}$,\\
$\{(1,5),(11,4),(0,b)\}$, &$\{(2,0),(13,5),(0,b)\}$,
&$\{(2,1),(5,5),(0,b)\}$,
&$\{(2,2),(5,2),(0,b)\}$,\\
$\{(2,3),(3,0),(0,b)\}$, &$\{(2,4),(4,2),(0,b)\}$,
&$\{(2,5),(12,0),(0,b)\}$,
&$\{(3,1),(4,1),(0,b)\}$,\\
$\{(3,2),(5,1),(0,b)\}$, &$\{(3,3),(9,0),(0,b)\}$,
&$\{(3,4),(6,4),(0,b)\}$,&$\{(3,5),(5,4),(0,b)\}$,\\
$\{(4,0),(9,2),(0,b)\}$, &$\{(4,3),(12,1),(0,b)\}$,
&$\{(4,5),(6,1),(0,b)\}$,
&$\{(5,3),(6,2),(0,b)\}$,\\
$\{(6,0),(7,3),(0,b)\}$, &$\{(6,3),(9,1),(0,b)\}$,
&$\{(6,5),(12,5),(0,b)\}$,
&$\{(7,0),(12,4),(0,b)\}$,\\
$\{(7,1),(9,4),(0,b)\}$, &$\{(7,2),(12,3),(0,b)\}$,
&$\{(7,4),(13,4),(0,b)\}$,
&$\{(7,5),(10,5),(0,b)\}$,\\
$\{(8,0),(12,2),(0,b)\}$, &$\{(8,1),(9,3),(0,b)\}$,
&$\{(8,2),(11,1),(0,b)\}$,
&$\{(8,3),(11,0),(0,b)\}$,\\
$\{(8,4),(13,1),(0,b)\}$, &$\{(9,5),(10,3),(0,b)\}$,
&$\{(10,0),(11,5),(0,b)\}$,
&$\{(10,1),(13,2),(0,b)\}$,\\
$\{(10,2),(13,0),(0,b)\}$, &$\{(10,4),(13,3),(0,b)\}$,
&$\{(1,0),(13,0),(0,c)\}$,
&$\{(1,1),(3,1),(0,c)\}$,\\
$\{(1,2),(5,1),(0,c)\}$, &$\{(1,3),(10,1),(0,c)\}$,
&$\{(1,4),(12,3),(0,c)\}$,
&$\{(1,5),(13,2),(0,c)\}$,\\
$\{(2,0),(8,1),(0,c)\}$, &$\{(2,1),(6,2),(0,c)\}$,
&$\{(2,2),(6,5),(0,c)\}$,
&$\{(2,3),(13,1),(0,c)\}$,\\
\end{tabular}}
\end{center}

\begin{center}
{ \small
\begin{tabular}{llll}
$\{(2,4),(12,5),(0,c)\}$, &$\{(2,5),(13,3),(0,c)\}$,
&$\{(3,0),(11,5),(0,c)\}$,
&$\{(3,2),(7,0),(0,c)\}$,\\
$\{(3,3),(7,1),(0,c)\}$, &$\{(3,4),(8,4),(0,c)\}$,
&$\{(3,5),(13,5),(0,c)\}$,
&$\{(4,0),(10,5),(0,c)\}$,\\
$\{(4,1),(8,3),(0,c)\}$, &$\{(4,2),(11,3),(0,c)\}$,
&$\{(4,3),(9,4),(0,c)\}$,
&$\{(4,4),(10,0),(0,c)\}$,\\
$\{(4,5),(12,1),(0,c)\}$, &$\{(5,0),(11,2),(0,c)\}$,
&$\{(5,2),(9,2),(0,c)\}$,
&$\{(5,3),(9,3),(0,c)\}$,\\
$\{(5,4),(12,0),(0,c)\}$, &$\{(5,5),(10,4),(0,c)\}$,
&$\{(6,0),(9,5),(0,c)\}$,
&$\{(6,1),(13,4),(0,c)\}$,\\
$\{(6,3),(12,4),(0,c)\}$, &$\{(6,4),(9,1),(0,c)\}$,
&$\{(7,2),(8,5),(0,c)\}$,
&$\{(7,3),(9,0),(0,c)\}$,\\
$\{(7,4),(11,4),(0,c)\}$, &$\{(7,5),(11,1),(0,c)\}$,
&$\{(8,0),(11,0),(0,c)\}$,
&$\{(8,2),(10,3),(0,c)\}$,\\
$\{(10,2),(12,2),(0,c)\}$, &$\{(1,0),(7,0),(0,d)\}$,
&$\{(1,1),(2,5),(0,d)\}$,
&$\{(1,2),(12,4),(0,d)\}$,\\
$\{(1,3),(4,3),(0,d)\}$, &$\{(1,4),(3,0),(0,d)\}$,
&$\{(1,5),(13,4),(0,d)\}$,
&$\{(2,0),(7,3),(0,d)\}$,\\
$\{(2,1),(7,2),(0,d)\}$, &$\{(2,2),(10,1),(0,d)\}$,
&$\{(2,3),(9,4),(0,d)\}$,
&$\{(2,4),(8,5),(0,d)\}$,\\
$\{(3,1),(8,0),(0,d)\}$, &$\{(3,2),(12,3),(0,d)\}$,
&$\{(3,3),(5,4),(0,d)\}$,
&$\{(3,4),(12,2),(0,d)\}$,\\
$\{(3,5),(11,2),(0,d)\}$, &$\{(4,0),(13,0),(0,d)\}$,
&$\{(4,1),(13,5),(0,d)\}$,
&$\{(4,2),(5,1),(0,d)\}$,\\
$\{(4,4),(9,0),(0,d)\}$, &$\{(4,5),(11,4),(0,d)\}$,
&$\{(5,0),(8,1),(0,d)\}$,
&$\{(5,2),(8,4),(0,d)\}$,\\
$\{(5,3),(10,3),(0,d)\}$, &$\{(5,5),(11,1),(0,d)\}$,
&$\{(6,0),(8,3),(0,d)\}$,
&$\{(6,1),(13,2),(0,d)\}$,\\
$\{(6,2),(12,0),(0,d)\}$, &$\{(6,3),(12,5),(0,d)\}$,
&$\{(6,4),(12,1),(0,d)\}$,
&$\{(6,5),(11,5),(0,d)\}$,\\
$\{(7,1),(13,1),(0,d)\}$, &$\{(7,4),(11,0),(0,d)\}$,
&$\{(7,5),(9,5),(0,d)\}$,
&$\{(8,2),(9,2),(0,d)\}$,\\
$\{(9,1),(10,0),(0,d)\}$, &$\{(9,3),(10,5),(0,d)\}$,
&$\{(10,2),(13,3),(0,d)\}$,
&$\{(10,4),(11,3),(0,d)\}$,\\
$\{(1,0),(13,1),(0,e)\}$, &$\{(1,1),(6,1),(0,e)\}$,
&$\{(1,2),(10,5),(0,e)\}$,
&$\{(1,3),(13,3),(0,e)\}$,\\
$\{(1,4),(7,2),(0,e)\}$, &$\{(1,5),(10,1),(0,e)\}$,
&$\{(2,0),(4,5),(0,e)\}$,
&$\{(2,1),(7,4),(0,e)\}$,\\
$\{(2,2),(3,0),(0,e)\}$, &$\{(2,3),(6,2),(0,e)\}$,
&$\{(2,4),(7,5),(0,e)\}$,
&$\{(2,5),(5,2),(0,e)\}$,\\
$\{(3,1),(13,0),(0,e)\}$, &$\{(3,2),(5,0),(0,e)\}$,
&$\{(3,3),(4,1),(0,e)\}$,
&$\{(3,4),(11,1),(0,e)\}$,\\
$\{(3,5),(7,0),(0,e)\}$, &$\{(4,0),(10,3),(0,e)\}$,
&$\{(4,2),(9,1),(0,e)\}$,
&$\{(4,3),(11,0),(0,e)\}$,\\
$\{(4,4),(6,4),(0,e)\}$, &$\{(5,1),(12,5),(0,e)\}$,
&$\{(5,3),(10,0),(0,e)\}$,
&$\{(5,4),(13,2),(0,e)\}$,\\
$\{(5,5),(6,0),(0,e)\}$, &$\{(6,3),(12,3),(0,e)\}$,
&$\{(6,5),(7,1),(0,e)\}$,
&$\{(7,3),(9,2),(0,e)\}$,\\
$\{(8,0),(12,0),(0,e)\}$, &$\{(8,1),(9,4),(0,e)\}$,
&$\{(8,2),(13,5),(0,e)\}$,
&$\{(8,3),(9,3),(0,e)\}$,\\
$\{(8,4),(9,0),(0,e)\}$, &$\{(8,5),(10,2),(0,e)\}$,
&$\{(9,5),(11,3),(0,e)\}$,
&$\{(10,4),(11,2),(0,e)\}$,\\
$\{(11,4),(12,1),(0,e)\}$, &$\{(11,5),(12,2),(0,e)\}$,
&$\{(12,4),(13,4),(0,e)\}$, &$\{(0,0),(1,2),(2,4)\}$,\\
$\{(1,3),(3,1),(5,5)\}$.
\end{tabular}}
\end{center}

\section{$h$-cyclic $3$-IGDDs of type $gh^{(u,t)}$}

\begin{lemma}\label{2^{12t+8,5} for small t}
Let $u\in\{20,32,44,56\}$. Then there exists a $2$-cyclic $3$-IGDD
of type $2^{(u,5)}$.
\end{lemma}

\Proof Suppose $I=Z_{u-5}\cup\{a,b,c,d,e\}$. Let $X=I\times Z_{2}$,
$\mathcal{G}=\{\{i\}\times Z_2:i\in I\}$, and $Y=\{a,b,c,d,e\}\times
Z_{2}$. Base blocks are listed below. Developing these base blocks
by $+(i,j)$ mod $(u-5,2)$ gives all blocks, where $(i,j)\in
Z_{u-5}\times Z_2$ and $a,b,c,d,e$ are fixed points. It is readily
checked that each design is isomorphism to a $2$-cyclic $3$-IGDD of
type $2^{(u,5)}$.

$u=20:$

\begin{multicols}{4}

{\small\hspace{-0.6cm}$\{(0,0),(1,0),(3,0)\}$,
$\{(0,0),(4,0),(9,0)\}$, $\{(0,0),(7,0),(1,1)\}$,
$\{(0,0),(2,1),(a,0)\}$, $\{(0,0),(3,1),(b,0)\}$,
$\{(0,0),(4,1),(c,0)\}$, $\{(0,0),(5,1),(d,0)\}$,
$\{(0,0),(7,1),(e,0)\}$.}
\end{multicols}

$u=32:$

\begin{multicols}{4}

{\small\hspace{-0.6cm}$\{(0,0),(1,0),(3,0)\}$,
$\{(0,0),(4,0),(9,0)\}$, $\{(0,0),(6,0),(13,0)\}$,
$\{(0,0),(8,0),(1,1)\}$, $\{(0,0),(10,0),(2,1)\}$,
$\{(0,0),(11,0),(5,1)\}$, $\{(0,0),(12,0),(3,1)\}$,
$\{(0,0),(4,1),(a,0)\}$, $\{(0,0),(10,1),(b,0)\}$,
$\{(0,0),(11,1),(c,0)\}$, $\{(0,0),(12,1),(d,0)\}$,
$\{(0,0),(13,1),(e,0)\}$.}
\end{multicols}

$u=44:$

\begin{multicols}{4}
{\small\hspace{-0.6cm}$\{(0,0),(1,0),(3,0)\}$,
$\{(0,0),(4,0),(9,0)\}$, $\{(0,0),(6,0),(13,0)\}$,
$\{(0,0),(8,0),(18,0)\}$, $\{(0,0),(11,0),(1,1)\}$,
$\{(0,0),(12,0),(3,1)\}$, $\{(0,0),(14,0),(2,1)\}$,
$\{(0,0),(15,0),(4,1)\}$, $\{(0,0),(16,0),(21,1)\}$,
$\{(0,0),(17,0),(23,1)\}$, $\{(0,0),(19,0),(26,1)\}$,
$\{(0,0),(8,1),(a,0)\}$, $\{(0,0),(14,1),(b,0)\}$,
$\{(0,0),(15,1),(c,0)\}$, $\{(0,0),(17,1),(d,0)\}$,
$\{(0,0),(19,1),(e,0)\}$.}
\end{multicols}

$u=56:$

\begin{multicols}{4}
{\small\hspace{-0.6cm}$\{(0,0),(4,0),(46,0)\}$,
$\{(0,0),(8,0),(25,1)\}$, $\{(0,0),(6,0),(17,0)\}$,
$\{(0,0),(7,1),(25,0)\}$, $\{(0,0),(14,0),(32,0)\}$,
$\{(0,0),(10,0),(3,0)\}$, $\{(0,0),(11,1),(15,0)\}$,
$\{(0,0),(12,0),(27,1)\}$, $\{(0,0),(13,1),(50,1)\}$,
$\{(0,0),(1,1),(3,1)\}$, $\{(0,0),(19,0),(b,0)\}$,
$\{(0,0),(20,0),(22,1)\}$, $\{(0,0),(21,1),(9,0)\}$,
$\{(0,0),(22,0),(16,1)\}$, $\{(0,0),(23,1),(10,1)\}$,
$\{(0,0),(5,0),(a,0)\}$, $\{(0,0),(16,0),(43,0)\}$,
$\{(0,0),(20,1),(e,0)\}$, $\{(0,0),(21,0),(d,1)\}$,
$\{(0,0),(23,0),(c,1)\}$.}
\end{multicols}

\begin{lemma}\label{2^{14,2}}
There exists a $2$-cyclic $3$-IGDD of type $2^{(14,2)}$.
\end{lemma}

\Proof Let $X=(Z_{6}\cup\{\infty\})\times Z_{4}$,
$\mathcal{G}=\{\{i\}\times\{j,2+j\}:(i,j)\in Z_{6}\times
Z_2\}\cup\{\{\infty\}\times\{l,l+2\}:l\in Z_2\}$, and
$Y=\{\infty\}\times Z_{4}$. We list $10$ base blocks.\vspace{0.2cm}

\begin{center}
{ \small
\begin{tabular}{llll}
$\{(0,0),(1,0),(2,0)\}$, &$\{(0,0),(3,0),(0,1)\}$,
&$\{(0,0),(1,1),(3,1)\}$,
&$\{(0,0),(2,1),(1,2)\}$,\\
$\{(0,0),(4,1),(1,3)\}$, &$\{(0,0),(2,2),(\infty,0)\}$,
&$\{(0,0),(5,2),(\infty,1)\}$,
&$\{(0,0),(5,3),(\infty,3)\}$,\\
$\{(1,0),(1,1),(3,2)\}$, &$\{(1,0),(5,1),(\infty,2)\}$.
\end{tabular}}
\end{center}

Developing these base blocks by $+(2s,t)$ mod $(6,4)$ gives all
blocks, where $(s,t)\in Z_3\times Z_4$ and $\infty$ is a fixed
point. It is readily checked that each design is isomorphism to a
$2$-cyclic $3$-IGDD of type $2^{(u,2)}$.\qed

\begin{lemma}\label{4^{8,2}}
Let $u\in\{8,14\}$. Then there exists a $4$-cyclic $3$-IGDD of type
$4^{(u,2)}$.
\end{lemma}

\Proof Let $X=(Z_{(u-2)/2}\cup\{\infty\})\times Z_{8}$,
$\mathcal{G}=\{\{i\}\times\{j,2+j,4+j,6+j\}:(i,j)\in
Z_{(u-2)/2}\times Z_2\}\cup\{\{\infty\}\times\{l,2+l,4+l,6+l\}:l\in
Z_2\}$, and $Y=\{\infty\}\times Z_{8}$. Developing the following
base blocks by $+(2s,t)$ mod $((u-2)/2,8)$ gives all blocks, where
$(s,t)\in Z_3\times Z_8$ and $\infty$ is a fixed point. It is
readily checked that each design is isomorphism to a $4$-cyclic
$3$-IGDD of type $4^{(u,2)}$.

$u=8:$

\begin{center}
{ \small
\begin{tabular}{lll}
$\{(0,0),(1,1),(0,3)\}$, &$\{(0,0),(1,4),(\infty,1)\}$,
&$\{(0,0),(2,5),(1,5)\}$,\\
$\{(0,0),(1,7),(\infty,3)\}$, &$\{(0,0),(0,1),(\infty,7)\}$,
&$\{(0,0),(1,2),(\infty,2)\}$.
\end{tabular}}
\end{center}

$u=14:$

\begin{center}
{ \small
\begin{tabular}{llll}
$\{(0,0),(1,0),(2,0)\}$, &$\{(0,0),(3,0),(0,1)\}$,
&$\{(0,0),(1,1),(3,1)\}$,
&$\{(0,0),(2,1),(1,2)\}$,\\
$\{(0,0),(4,1),(0,3)\}$, &$\{(0,0),(3,2),(1,3)\}$,
&$\{(0,0),(4,2),(3,4)\}$,
&$\{(0,0),(2,3),(1,6)\}$,\\
$\{(0,0),(3,3),(2,4)\}$, &$\{(0,0),(4,3),(\infty,0)\}$,
&$\{(0,0),(1,4),(1,5)\}$,
&$\{(0,0),(5,4),(\infty,1)\}$,\\
$\{(0,0),(3,5),(\infty,3)\}$, &$\{(0,0),(5,5),(\infty,4)\}$,
&$\{(0,0),(3,6),(\infty,6)\}$,
&$\{(0,0),(5,6),(\infty,7)\}$,\\
$\{(0,0),(5,7),(\infty,2)\}$, &$\{(1,0),(3,1),(5,4)\}$,
&$\{(1,0),(3,2),(1,5)\}$, &$\{(1,0),(5,2),(\infty,4)\}$.
\end{tabular}}
\end{center}

\begin{lemma}\label{6^{6,2}}
There exists a $6$-cyclic $3$-IGDD of type $6^{(6,2)}$.
\end{lemma}

\Proof Let $X=Z_{36}$, $\mathcal{G}=\{i,6+i,\ldots,30+i:0\leq i\leq
5\}$, $Y=\{x:x=0,1\hspace{-0.3cm}\pmod{6},x\in Z_{36}\}$. Developing
the following base blocks by $+6$ mod $36$ gives all blocks.

\begin{center}
{ \small
\begin{tabular}{lllllll}
$\{0,2,3\}$, &$\{0,4,5\}$, &$\{0,8,10\}$, &$\{0,9,11\}$,
&$\{0,14,17\}$, &$\{0,15,16\}$,
&$\{0,20,27\}$,\\
$\{0,21,26\}$, &$\{0,22,32\}$, &$\{0,23,33\}$, &$\{0,28,35\}$,
&$\{0,29,34\}$, &$\{1,2,10\}$,
&$\{1,3,11\}$,\\
$\{1,4,8\}$, &$\{1,5,26\}$, &$\{1,9,28\}$, &$\{1,14,33\}$,
&$\{1,15,22\}$, &$\{1,16,35\}$,
&$\{1,17,32\}$,\\
$\{1,20,29\}$, &$\{1,21,34\}$, &$\{1,23,27\}$, &$\{2,15,29\}$,
&$\{2,16,27\}$, &$\{2,22,35\}$, &$\{3,23,34\}$.
\end{tabular}}
\end{center}

\begin{lemma}\label{2^{26,11}}
There exists a $2$-cyclic $3$-IGDD of type $2^{(26,11)}$.
\end{lemma}

\Proof The required design is constructed on $I_{52}$ with group set
$\{\{i,20+i\}:0\leq i\leq 19\}\cup\{\{j,j+6\}:40\leq j\leq45\}$ and
hole set $\{x:x\equiv0\hspace{-0.3cm}\pmod4,x<40\}\cup\{y:y\geq
40\}$. The base blocks are listed as follows:
\begin{center}
{ \small
\begin{tabular}{llllll}
$\{0,1,3\}$, &$\{0,2,10\}$, &$\{0,5,37\}$, &$\{0,6,39\}$,
&$\{0,7,38\}$,
&$\{0,9,19\}$,\\
$\{0,11,25\}$, &$\{0,13,22\}$, &$\{0,14,29\}$, &$\{0,15,31\}$,
&$\{0,17,30\}$,
&$\{0,18,34\}$,\\
$\{0,21,26\}$, &$\{0,23,35\}$, &$\{0,27,33\}$, &$\{1,2,38\}$,
&$\{1,5,40\}$,
&$\{1,7,47\}$,\\
$\{1,13,42\}$, &$\{1,15,17\}$, &$\{1,18,51\}$, &$\{1,19,45\}$,
&$\{1,22,49\}$,
&$\{1,23,50\}$,\\
$\{1,30,43\}$, &$\{1,31,44\}$, &$\{1,34,41\}$, &$\{2,3,40\}$,
&$\{2,7,15\}$,
&$\{2,14,44\}$,\\
$\{2,19,46\}$, &$\{2,23,42\}$, &$\{2,27,47\}$, &$\{2,31,45\}$,
&$\{2,39,48\}$, &$\{3,7,43\}$.
\end{tabular}}
\end{center}

Let
$\alpha=\prod_{i=0}^{3}(i~\,4+i~\cdots~36+i)\prod_{i=0}^{6}(40+i~\,46+i)$
be a permutation on $I_{52}$. Let $G$ be the group generated by
$\alpha$. All blocks are obtained by developing the base blocks
under the action of $G$. \qed

\section{$w$-cyclic $3$-GDPs of type $(vw)^u$}

\begin{lemma}\label{2^8}
There exists a $3$-SCGDP of type $2^{8}$ with $J^\ast(8\times
1\times 2,3,1)$ base blocks.
\end{lemma}

\Proof Let $X=I_8\times Z_2$, $\mathcal{G}=\{\{i\}\times Z_2:i\in
I_8\}$. Only the $J^\ast(8\times 1\times 2,3,1)$ base blocks are
listed below.\vspace{0.15cm}

\begin{center}
{ \small
\begin{tabular}{llll}
$\{(0,0),(1,0),(2,0)\}$, &$\{(0,0),(3,0),(4,0)\}$,
&$\{(0,0),(5,0),(6,0)\}$,
&$\{(0,0),(7,0),(1,1)\}$,\\

$\{(0,0),(2,1),(3,1)\}$, &$\{(0,0),(4,1),(5,1)\}$,
&$\{(0,0),(6,1),(7,1)\}$, &$\{(1,0),(3,0),(5,0)\}$,\\

$\{(1,0),(4,0),(6,0)\}$, &$\{(1,0),(7,0),(4,1)\}$,
&$\{(1,0),(2,1),(5,1)\}$, &$\{(1,0),(3,1),(6,1)\}$,\\

$\{(2,0),(4,0),(3,1)\}$, &$\{(2,0),(6,0),(4,1)\}$,
&$\{(2,0),(7,0),(6,1)\}$, &$\{(2,0),(5,1),(7,1)\}$,\\

$\{(3,0),(7,0),(5,1)\}$.
\end{tabular}}
\end{center}

\begin{lemma}\label{small *gdp}
There exists a $3$-SCGDP$^*$ of type $2^{(u,5)}$ with
$(u(u-1)-14)/3$ base blocks for any $u\in\{11,14\}$.
\end{lemma}

\Proof Let $X=I_{u}\times Z_2$, $\mathcal{G}=\{\{i\}\times Z_2:i\in
I_{u}\}$. Only base blocks are listed below.

$u=11:$
\begin{center}
{ \small
\begin{tabular}{llll}
$\{(0,0),(1,0),(2,0)\}$, &$\{(0,0),(3,0),(4,0)\}$,
&$\{(0,0),(5,0),(6,0)\}$,
&$\{(0,0),(7,0),(8,0)\}$,\\
$\{(0,0),(9,0),(10,0)\}$, &$\{(0,0),(5,1),(7,1)\}$,
&$\{(0,0),(6,1),(9,1)\}$,
&$\{(0,0),(8,1),(10,1)\}$,\\
$\{(1,0),(3,0),(5,0)\}$, &$\{(1,0),(4,0),(6,0)\}$,
&$\{(1,0),(7,0),(9,0)\}$,
&$\{(1,0),(8,0),(5,1)\}$,\\
$\{(1,0),(10,0),(6,1)\}$, &$\{(1,0),(7,1),(10,1)\}$,
&$\{(1,0),(8,1),(9,1)\}$,
&$\{(2,0),(3,0),(7,0)\}$,\\
$\{(2,0),(4,0),(8,0)\}$, &$\{(2,0),(5,0),(9,0)\}$,
&$\{(2,0),(6,0),(7,1)\}$,
&$\{(2,0),(10,0),(9,1)\}$,\\
$\{(2,0),(5,1),(8,1)\}$, &$\{(2,0),(6,1),(10,1)\}$,
&$\{(3,0),(6,0),(8,1)\}$,
&$\{(3,0),(8,0),(9,1)\}$,\\
$\{(3,0),(9,0),(6,1)\}$, &$\{(3,0),(10,0),(7,1)\}$,
&$\{(3,0),(5,1),(10,1)\}$,
&$\{(4,0),(5,0),(10,1)\}$,\\
$\{(4,0),(7,0),(9,1)\}$, &$\{(4,0),(9,0),(5,1)\}$,
&$\{(4,0),(10,0),(8,1)\}$, &$\{(4,0),(6,1),(7,1)\}$.
\end{tabular}}
\end{center}

$u=14:$

\begin{center}
{ \small
\begin{tabular}{llll}
$\{(0,0),(1,0),(2,0)\}$, &$\{(0,0),(3,0),(5,0)\}$,
&$\{(0,0),(4,0),(6,0)\}$,
&$\{(0,0),(7,0),(9,0)\}$,\\
$\{(0,0),(8,0),(10,0)\}$, &$\{(0,0),(11,0),(13,0)\}$,
&$\{(0,0),(12,0),(5,1)\}$,
&$\{(0,0),(6,1),(8,1)\}$,\\
$\{(0,0),(7,1),(12,1)\}$, &$\{(0,0),(9,1),(11,1)\}$,
&$\{(0,0),(10,1),(13,1)\}$,
&$\{(1,0),(3,0),(6,0)\}$,\\
$\{(1,0),(4,0),(7,0)\}$, &$\{(1,0),(5,0),(8,0)\}$,
&$\{(1,0),(9,0),(12,0)\}$,
&$\{(1,0),(10,0),(5,1)\}$,\\
$\{(1,0),(11,0),(6,1)\}$, &$\{(1,0),(13,0),(7,1)\}$,
&$\{(1,0),(8,1),(11,1)\}$,
&$\{(1,0),(9,1),(13,1)\}$,\\
$\{(1,0),(10,1),(12,1)\}$, &$\{(2,0),(3,0),(7,0)\}$,
&$\{(2,0),(4,0),(8,0)\}$,
&$\{(2,0),(5,0),(9,0)\}$,\\
$\{(2,0),(6,0),(10,0)\}$, &$\{(2,0),(11,0),(5,1)\}$,
&$\{(2,0),(12,0),(8,1)\}$,
&$\{(2,0),(13,0),(9,1)\}$,\\
$\{(2,0),(6,1),(11,1)\}$, &$\{(2,0),(7,1),(10,1)\}$,
&$\{(2,0),(12,1),(13,1)\}$,
&$\{(3,0),(4,0),(9,0)\}$,\\
$\{(3,0),(8,0),(12,0)\}$, &$\{(3,0),(10,0),(11,0)\}$,
&$\{(3,0),(13,0),(8,1)\}$,
&$\{(3,0),(5,1),(6,1)\}$,\\
$\{(3,0),(7,1),(13,1)\}$, &$\{(3,0),(9,1),(10,1)\}$,
&$\{(3,0),(11,1),(12,1)\}$,
&$\{(4,0),(5,0),(8,1)\}$,\\
$\{(4,0),(10,0),(9,1)\}$, &$\{(4,0),(11,0),(10,1)\}$,
&$\{(4,0),(12,0),(11,1)\}$,
&$\{(4,0),(13,0),(12,1)\}$,\\
$\{(4,0),(5,1),(7,1)\}$, &$\{(4,0),(6,1),(13,1)\}$,
&$\{(5,0),(10,0),(13,1)\}$,
&$\{(5,0),(11,0),(7,1)\}$,\\
$\{(5,0),(12,0),(9,1)\}$, &$\{(5,0),(13,0),(6,1)\}$,
&$\{(6,0),(7,0),(12,1)\}$,
&$\{(6,0),(9,0),(7,1)\}$,\\
$\{(6,0),(12,0),(10,1)\}$, &$\{(6,0),(8,1),(9,1)\}$,
&$\{(7,0),(8,0),(10,1)\}$, &$\{(7,0),(11,0),(8,1)\}$.
\end{tabular}}
\end{center}

\begin{lemma}\label{3^8}
There exists a $3$-SCGDP of type $3^{u}$ with $J^\ast(u\times
1\times 3,3,1)$ base blocks for any $u\in\{8,14\}$.
\end{lemma}

\Proof Let $X=I\times Z_{3}$ and $\mathcal{G}=\{\{i\}\times Z_3:i\in
I\}$, where $I=Z_{u-2}\cup\{a,b\}$. The total $J^\ast(u\times
1\times3,3,1)=(3u^2-4u-4)/6$ base blocks can be obtained by two
parts.\vspace{0.1cm}

Part I: Developing {\small$\{(0,0),((u-2)/3,0),(2(u-2)/3,0)\}$} by
$+(1,-)$ mod $(u-2,-)$ can generate $(u-2)/3$ base
blocks.\vspace{0.1cm}

Part II: Developing the following $u$ initial base blocks by
$+(2,-)$ mod $(u-2,-)$ can generate $u(u-2)/2$ base blocks. Here
$a,b$ are fixed points. \vspace{0.1cm}

$u=8:$

\begin{center}
{ \small
\begin{tabular}{llll}
$\{(0,0),(1,0),(2,1)\}$, &$\{(0,0),(5,0),(1,2)\}$,
&$\{(0,0),(1,1),(a,0)\}$,
&$\{(0,0),(3,1),(a,1)\}$,\\
$\{(0,0),(4,1),(b,1)\}$, &$\{(0,0),(5,1),(a,2)\}$,
&$\{(0,0),(3,2),(b,2)\}$, &$\{(1,0),(3,1),(b,2)\}$.
\end{tabular}}
\end{center}

$u=14:$

\begin{center}
{ \small
\begin{tabular}{llll}
$\{(0,0),(1,0),(2,0)\}$, &$\{(0,0),(3,0),(5,0)\}$,
&$\{(0,0),(7,0),(1,1)\}$,
&$\{(0,0),(9,0),(2,1)\}$,\\
$\{(0,0),(3,1),(1,2)\}$, &$\{(0,0),(4,1),(3,2)\}$,
&$\{(0,0),(5,1),(6,2)\}$,
&$\{(0,0),(7,1),(9,2)\}$,\\
$\{(0,0),(8,1),(a,0)\}$, &$\{(0,0),(9,1),(b,0)\}$,
&$\{(0,0),(10,1),(b,2)\}$,
&$\{(0,0),(5,2),(a,1)\}$,\\
$\{(1,0),(5,1),(a,1)\}$, &$\{(1,0),(9,1),(b,1)\}$.
\end{tabular}}
\end{center}

\begin{lemma}\label{15^8}
There exists a $3$-cyclic $3$-GDP of type $15^{u}$ with
$J^\ast(u\times 5\times 3,3,1)$ base blocks for any $u\in\{8,14\}$.
\end{lemma}

\Proof We construct the required designs directly. Let $X=I\times
I_5\times Z_3$ and $\mathcal{G}=\{\{i\}\times I_5\times Z_3:i\in
I\}$, where $I=Z_{u-1}\cup\{\infty\}$. The $J^\ast(u\times 5\times
3,3,1)=(75u^2-80u-2)/6$ base blocks are divide into two parts. The
first part consists of $(4u^2-5u-6)/6$ base blocks:

\begin{center}
$\{(2i,0,0),(2i+1,0,0),(\infty,0,0)\}$, $1\leq i\leq(u-2)/2$,

$\{(a_j,l,0),(b_j,l,0),(c_j,l,0)\}$, $1\leq l\leq4$, $1\leq j\leq
u(u-2)/6$.
\end{center}

\hspace{-0.57cm}Here, $\{\{a_j,b_j,c_j)\}:1\leq j\leq u(u-2)/6\}$
are the blocks of a $3$-GDD of type $2^{u/2}$ on $I$.

The second part consists of $(u-1)(71u-4)/6$ base blocks, which can
be obtained by developing $(71u-4)/6$ initial base blocks by
$(+1,-,-)$ mod $(u-1,-,-)$, where $\infty+1=\infty$. We list these
initial base blocks below.

$u=8:$

\begin{center}
{ \small
\begin{tabular}{lll}
$\{(0,0,0),(2,0,0),(1,1,0)\}$, &$\{(0,0,0),(3,0,0),(4,0,1)\}$,
&$\{(0,0,0),(2,1,0),(5,4,2)\}$,\\
$\{(0,0,0),(3,1,0),(1,3,2)\}$, &$\{(0,0,0),(4,1,0),(1,4,2)\}$,
&$\{(0,0,0),(5,1,0),(6,2,2)\}$,\\
$\{(0,0,0),(\infty,1,0),(3,2,0)\}$,
&$\{(0,0,0),(1,2,0),(\infty,1,2)\}$,
&$\{(0,0,0),(2,2,0),(1,2,2)\}$,\\
$\{(0,0,0),(4,2,0),(2,3,2)\}$, &$\{(0,0,0),(5,2,0),(\infty,0,2)\}$,
&$\{(0,0,0),(6,2,0),(1,1,2)\}$,\\
$\{(0,0,0),(\infty,2,0),(1,3,0)\}$, &$\{(0,0,0),(2,3,0),(5,1,1)\}$,
&$\{(0,0,0),(3,3,0),(1,4,0)\}$,\\
$\{(0,0,0),(4,3,0),(5,0,1)\}$, &$\{(0,0,0),(5,3,0),(3,1,2)\}$,
&$\{(0,0,0),(6,3,0),(2,2,1)\}$,\\
$\{(0,0,0),(\infty,3,0),(4,3,1)\}$, &$\{(0,0,0),(2,4,0),(6,4,2)\}$,
&$\{(0,0,0),(3,4,0),(2,2,2)\}$,\\
$\{(0,0,0),(4,4,0),(5,3,1)\}$, &$\{(0,0,0),(5,4,0),(2,1,2)\}$,
&$\{(0,0,0),(6,4,0),(4,2,1)\}$,\\
$\{(0,0,0),(\infty,4,0),(6,2,1)\}$, &$\{(0,0,0),(2,0,1),(5,1,2)\}$,
&$\{(0,0,0),(3,0,1),(4,1,2)\}$,\\
$\{(0,0,0),(6,0,1),(4,2,2)\}$, &$\{(0,0,0),(\infty,0,1),(3,2,1)\}$,
&$\{(0,0,0),(2,1,1),(3,4,1)\}$,\\
$\{(0,0,0),(4,1,1),(1,4,1)\}$, &$\{(0,0,0),(6,1,1),(4,3,2)\}$,
&$\{(0,0,0),(\infty,1,1),(6,4,1)\}$,\\
$\{(0,0,0),(1,2,1),(5,3,2)\}$, &$\{(0,0,0),(\infty,2,1),(5,2,2)\}$,
&$\{(0,0,0),(1,3,1),(\infty,3,2)\}$,\\
$\{(0,0,0),(2,3,1),(6,1,2)\}$, &$\{(0,0,0),(3,3,1),(\infty,2,2)\}$,
&$\{(0,0,0),(6,3,1),(3,3,2)\}$,\\
$\{(0,0,0),(\infty,3,1),(3,2,2)\}$,
&$\{(0,0,0),(2,4,1),(\infty,4,2)\}$,
&$\{(0,0,0),(4,4,1),(3,4,2)\}$,\\
$\{(0,0,0),(5,4,1),(2,4,2)\}$, &$\{(0,0,0),(\infty,4,1),(4,4,2)\}$,
&$\{(\infty,0,0),(0,1,0),(1,3,1)\}$,\\
$\{(\infty,0,0),(0,3,0),(6,3,2)\}$,
&$\{(\infty,0,0),(0,4,0),(5,4,2)\}$,
&$\{(\infty,0,0),(0,1,1),(2,2,2)\}$,\\
$\{(\infty,0,0),(0,4,1),(1,1,2)\}$, &$\{(0,1,0),(1,2,0),(4,3,1)\}$,
&$\{(0,1,0),(2,2,0),(\infty,4,1)\}$,\\
$\{(0,1,0),(3,2,0),(6,4,1)\}$, &$\{(0,1,0),(4,2,0),(2,2,2)\}$,
&$\{(0,1,0),(5,2,0),(1,2,1)\}$,\\
$\{(0,1,0),(6,2,0),(4,4,1)\}$, &$\{(0,1,0),(\infty,2,0),(4,1,1)\}$,
&$\{(0,1,0),(1,3,0),(6,2,1)\}$,\\
$\{(0,1,0),(2,3,0),(3,2,1)\}$, &$\{(0,1,0),(3,3,0),(5,4,2)\}$,
&$\{(0,1,0),(4,3,0),(6,4,0)\}$,\\
$\{(0,1,0),(5,3,0),(2,1,2)\}$, &$\{(0,1,0),(6,3,0),(2,4,2)\}$,
&$\{(0,1,0),(\infty,3,0),(4,2,2)\}$,\\
$\{(0,1,0),(2,4,0),(4,1,2)\}$, &$\{(0,1,0),(3,4,0),(6,3,2)\}$,
&$\{(0,1,0),(5,4,0),(3,2,2)\}$,\\
$\{(0,1,0),(\infty,4,0),(2,3,2)\}$,
&$\{(0,1,0),(1,1,1),(\infty,3,2)\}$,
&$\{(0,1,0),(2,1,1),(\infty,1,2)\}$,\\
$\{(0,1,0),(6,1,1),(1,4,2)\}$, &$\{(0,1,0),(4,2,1),(1,4,1)\}$,
&$\{(0,1,0),(\infty,2,1),(1,3,2)\}$,\\
$\{(0,1,0),(6,3,1),(5,2,2)\}$, &$\{(0,1,0),(6,2,2),(\infty,4,2)\}$,
&$\{(\infty,1,0),(0,3,0),(4,3,2)\}$,\\
$\{(\infty,1,0),(0,3,1),(5,4,2)\}$,
&$\{(\infty,1,0),(0,4,1),(1,2,2)\}$,
&$\{(0,2,0),(1,3,0),(5,4,0)\}$,\\
$\{(0,2,0),(2,3,0),(4,4,1)\}$, &$\{(0,2,0),(3,3,0),(6,4,1)\}$,
&$\{(0,2,0),(4,3,0),(3,2,2)\}$,\\
$\{(0,2,0),(5,3,0),(1,4,0)\}$, &$\{(0,2,0),(6,3,0),(1,2,2)\}$,
&$\{(0,2,0),(\infty,3,0),(2,4,0)\}$,\\
$\{(0,2,0),(3,4,0),(6,3,1)\}$, &$\{(0,2,0),(6,4,0),(5,2,1)\}$,
&$\{(0,2,0),(\infty,2,1),(5,4,2)\}$,\\
$\{(0,2,0),(2,3,1),(3,4,2)\}$, &$\{(0,2,0),(3,3,2),(4,4,2)\}$,
&$\{(\infty,2,0),(0,4,0),(2,4,2)\}$,\\
$\{(0,3,0),(6,4,0),(5,3,1)\}$, &$\{(0,3,0),(\infty,4,0),(2,3,1)\}$,
&$\{(0,3,0),(6,3,1),(5,4,2)\}$,\\
$\{(\infty,3,0),(0,4,1),(1,4,2)\}$.
\end{tabular}}
\end{center}

$u=14:$

\begin{center}
{ \small
\begin{tabular}{lll}
$\{(0,0,0),(2,0,0),(5,0,0)\}$, &$\{(0,0,0),(4,0,0),(3,3,0)\}$,
&$\{(0,0,0),(6,0,0),(7,4,2)\}$,\\
$\{(0,0,0),(1,1,0),(7,3,0)\}$, &$\{(0,0,0),(2,1,0),(1,4,0)\}$,
&$\{(0,0,0),(3,1,0),(6,0,1)\}$,\\
$\{(0,0,0),(4,1,0),(6,4,2)\}$, &$\{(0,0,0),(5,1,0),(3,1,2)\}$,
&$\{(0,0,0),(6,1,0),(7,3,2)\}$,\\
$\{(0,0,0),(7,1,0),(5,2,1)\}$, &$\{(0,0,0),(8,1,0),(1,4,1)\}$,
&$\{(0,0,0),(9,1,0),(2,4,0)\}$,\\
\end{tabular}}
\end{center}

\begin{center}
{ \small
\begin{tabular}{lll}
$\{(0,0,0),(10,1,0),(1,2,0)\}$, &$\{(0,0,0),(11,1,0),(10,4,2)\}$,
&$\{(0,0,0),(12,1,0),(11,3,0)\}$,\\
$\{(0,0,0),(\infty,1,0),(6,3,1)\}$, &$\{(0,0,0),(2,2,0),(3,2,1)\}$,
&$\{(0,0,0),(3,2,0),(6,1,1)\}$,\\
$\{(0,0,0),(4,2,0),(\infty,0,1)\}$, &$\{(0,0,0),(5,2,0),(8,0,2)\}$,
&$\{(0,0,0),(6,2,0),(2,1,2)\}$,\\
$\{(0,0,0),(7,2,0),(6,3,2)\}$, &$\{(0,0,0),(8,2,0),(10,3,1)\}$,
&$\{(0,0,0),(9,2,0),(8,4,0)\}$,\\
$\{(0,0,0),(10,2,0),(12,3,2)\}$, &$\{(0,0,0),(11,2,0),(9,4,1)\}$,
&$\{(0,0,0),(12,2,0),(4,3,1)\}$,\\
$\{(0,0,0),(\infty,2,0),(4,1,1)\}$, &$\{(0,0,0),(1,3,0),(8,1,1)\}$,
&$\{(0,0,0),(2,3,0),(\infty,4,0)\}$,\\
$\{(0,0,0),(4,3,0),(1,0,1)\}$, &$\{(0,0,0),(5,3,0),(8,3,2)\}$,
&$\{(0,0,0),(6,3,0),(3,4,1)\}$,\\
$\{(0,0,0),(8,3,0),(12,0,1)\}$, &$\{(0,0,0),(9,3,0),(4,0,2)\}$,
&$\{(0,0,0),(10,3,0),(1,1,1)\}$,\\
$\{(0,0,0),(\infty,3,0),(1,2,1)\}$, &$\{(0,0,0),(3,4,0),(2,1,1)\}$,
&$\{(0,0,0),(4,4,0),(5,2,2)\}$,\\
$\{(0,0,0),(5,4,0),(\infty,4,2)\}$, &$\{(0,0,0),(6,4,0),(3,4,2)\}$,
&$\{(0,0,0),(7,4,0),(2,4,2)\}$,\\
$\{(0,0,0),(9,4,0),(\infty,2,2)\}$,
&$\{(0,0,0),(10,4,0),(\infty,0,2)\}$,
&$\{(0,0,0),(11,4,0),(7,3,1)\}$,\\
$\{(0,0,0),(12,4,0),(1,3,2)\}$, &$\{(0,0,0),(2,0,1),(1,1,2)\}$,
&$\{(0,0,0),(3,0,1),(12,1,2)\}$,\\
$\{(0,0,0),(4,0,1),(6,2,2)\}$, &$\{(0,0,0),(7,0,1),(4,1,2)\}$,
&$\{(0,0,0),(8,0,1),(6,1,2)\}$,\\
$\{(0,0,0),(10,0,1),(3,2,2)\}$,
&$\{(0,0,0),(11,0,1),(\infty,1,2)\}$,
&$\{(0,0,0),(3,1,1),(4,2,1)\}$,\\
$\{(0,0,0),(5,1,1),(7,2,1)\}$, &$\{(0,0,0),(7,1,1),(\infty,2,1)\}$,
&$\{(0,0,0),(8,2,1),(9,4,2)\}$,\\
$\{(0,0,0),(9,2,1),(2,4,1)\}$, &$\{(0,0,0),(11,2,1),(8,1,2)\}$,
&$\{(0,0,0),(12,2,1),(8,4,1)\}$,\\
$\{(0,0,0),(1,3,1),(4,4,1)\}$, &$\{(0,0,0),(2,3,1),(12,2,2)\}$,
&$\{(0,0,0),(3,3,1),(4,4,2)\}$,\\
$\{(0,0,0),(8,3,1),(5,1,2)\}$, &$\{(0,0,0),(9,3,1),(10,3,2)\}$,
&$\{(0,0,0),(11,3,1),(5,3,2)\}$,\\
$\{(0,0,0),(12,3,1),(11,3,2)\}$,
&$\{(0,0,0),(\infty,3,1),(11,4,2)\}$,
&$\{(0,0,0),(5,4,1),(10,2,2)\}$,\\
$\{(0,0,0),(6,4,1),(9,1,2)\}$, &$\{(0,0,0),(7,4,1),(11,1,2)\}$,
&$\{(0,0,0),(10,4,1),(4,2,2)\}$,\\
$\{(0,0,0),(11,4,1),(\infty,3,2)\}$,
&$\{(0,0,0),(12,4,1),(11,2,2)\}$,
&$\{(0,0,0),(\infty,4,1),(7,2,2)\}$,\\
$\{(0,0,0),(7,1,2),(8,4,2)\}$, &$\{(0,0,0),(1,2,2),(5,4,2)\}$,
&$\{(0,0,0),(2,2,2),(12,4,2)\}$,\\
$\{(0,0,0),(8,2,2),(4,3,2)\}$, &$\{(0,0,0),(9,2,2),(2,3,2)\}$,
&$\{(\infty,0,0),(0,1,0),(8,4,2)\}$,\\
$\{(\infty,0,0),(0,2,0),(1,1,1)\}$,
&$\{(\infty,0,0),(0,3,0),(5,3,2)\}$,
&$\{(\infty,0,0),(0,4,0),(3,2,1)\}$,\\
$\{(\infty,0,0),(0,3,1),(1,1,2)\}$, &$\{(0,1,0),(3,2,0),(1,3,0)\}$,
&$\{(0,1,0),(5,2,0),(\infty,1,2)\}$,\\
$\{(0,1,0),(6,2,0),(4,4,0)\}$, &$\{(0,1,0),(7,2,0),(3,4,2)\}$,
&$\{(0,1,0),(8,2,0),(12,2,1)\}$,\\
$\{(0,1,0),(9,2,0),(10,3,2)\}$,
&$\{(0,1,0),(10,2,0),(\infty,3,0)\}$,
&$\{(0,1,0),(11,2,0),(4,1,1)\}$,\\
$\{(0,1,0),(12,2,0),(4,3,2)\}$, &$\{(0,1,0),(2,3,0),(8,1,2)\}$,
&$\{(0,1,0),(3,3,0),(2,4,0)\}$,\\
$\{(0,1,0),(4,3,0),(\infty,3,2)\}$, &$\{(0,1,0),(5,3,0),(1,4,1)\}$,
&$\{(0,1,0),(7,3,0),(9,1,1)\}$,\\
$\{(0,1,0),(8,3,0),(10,4,1)\}$, &$\{(0,1,0),(9,3,0),(6,1,2)\}$,
&$\{(0,1,0),(10,3,0),(12,3,1)\}$,\\
$\{(0,1,0),(11,3,0),(\infty,4,1)\}$, &$\{(0,1,0),(3,4,0),(1,1,2)\}$,
&$\{(0,1,0),(5,4,0),(2,1,2)\}$,\\
$\{(0,1,0),(7,4,0),(11,2,2)\}$, &$\{(0,1,0),(8,4,0),(2,2,2)\}$,
&$\{(0,1,0),(9,4,0),(4,3,1)\}$,\\
$\{(0,1,0),(10,4,0),(3,1,2)\}$, &$\{(0,1,0),(11,4,0),(8,4,1)\}$,
&$\{(0,1,0),(\infty,4,0),(8,3,1)\}$,\\
$\{(0,1,0),(1,1,1),(4,2,2)\}$, &$\{(0,1,0),(3,1,1),(5,2,2)\}$,
&$\{(0,1,0),(6,1,1),(8,3,2)\}$,\\
$\{(0,1,0),(8,1,1),(5,3,2)\}$, &$\{(0,1,0),(\infty,1,1),(1,3,1)\}$,
&$\{(0,1,0),(1,2,1),(8,2,2)\}$,\\
$\{(0,1,0),(5,2,1),(7,4,2)\}$, &$\{(0,1,0),(6,2,1),(11,4,2)\}$,
&$\{(0,1,0),(7,2,1),(9,2,2)\}$,\\
$\{(0,1,0),(8,2,1),(1,2,2)\}$, &$\{(0,1,0),(9,2,1),(6,2,2)\}$,
&$\{(0,1,0),(10,2,1),(4,4,2)\}$,\\
$\{(0,1,0),(\infty,2,1),(11,3,1)\}$, &$\{(0,1,0),(5,3,1),(9,4,1)\}$,
&$\{(0,1,0),(6,3,1),(2,3,2)\}$,\\
$\{(0,1,0),(9,3,1),(7,3,2)\}$, &$\{(0,1,0),(\infty,3,1),(5,4,1)\}$,
&$\{(0,1,0),(4,4,1),(5,4,2)\}$,\\
$\{(0,1,0),(11,4,1),(\infty,4,2)\}$,
&$\{(0,1,0),(12,4,1),(6,4,2)\}$,
&$\{(\infty,1,0),(0,2,0),(5,4,0)\}$,\\
$\{(\infty,1,0),(0,4,1),(11,4,2)\}$,
&$\{(\infty,1,0),(0,2,2),(5,3,2)\}$,
&$\{(0,2,0),(1,3,0),(3,2,1)\}$,\\
$\{(0,2,0),(2,3,0),(12,4,2)\}$, &$\{(0,2,0),(3,3,0),(6,3,1)\}$,
&$\{(0,2,0),(4,3,0),(11,2,1)\}$,\\
$\{(0,2,0),(7,3,0),(4,2,2)\}$, &$\{(0,2,0),(8,3,0),(2,4,2)\}$,
&$\{(0,2,0),(10,3,0),(3,4,1)\}$,\\
$\{(0,2,0),(12,3,0),(\infty,3,1)\}$,
&$\{(0,2,0),(1,4,0),(12,3,1)\}$,
&$\{(0,2,0),(2,4,0),(4,3,1)\}$,\\
$\{(0,2,0),(3,4,0),(5,2,1)\}$, &$\{(0,2,0),(7,4,0),(3,4,2)\}$,
&$\{(0,2,0),(8,4,0),(7,3,1)\}$,\\
$\{(0,2,0),(\infty,4,0),(5,2,2)\}$,
&$\{(0,2,0),(12,2,1),(10,3,2)\}$,
&$\{(0,2,0),(\infty,2,1),(9,3,2)\}$,\\
$\{(0,2,0),(1,3,1),(5,4,2)\}$, &$\{(0,2,0),(8,3,1),(7,4,2)\}$,
&$\{(0,2,0),(9,3,1),(\infty,2,2)\}$,\\
$\{(0,2,0),(10,3,1),(8,4,1)\}$, &$\{(0,2,0),(4,4,1),(8,3,2)\}$,
&$\{(0,2,0),(10,4,1),(4,3,2)\}$,\\
$\{(0,2,0),(7,3,2),(4,4,2)\}$, &$\{(\infty,2,0),(0,4,0),(1,4,2)\}$,
&$\{(0,3,0),(1,4,0),(8,4,2)\}$,\\
$\{(0,3,0),(2,4,0),(8,3,2)\}$, &$\{(0,3,0),(5,4,0),(6,3,1)\}$,
&$\{(0,3,0),(6,4,0),(8,4,1)\}$,\\
$\{(0,3,0),(7,4,0),(4,3,1)\}$, &$\{(0,3,0),(8,4,0),(3,4,1)\}$,
&$\{(0,3,0),(9,4,0),(5,4,1)\}$.
\end{tabular}}
\end{center}

\begin{lemma}\label{u=8 w=2mod6}
There exists a $3$-SCGDP of type $w^{8}$ with $J^\ast(8\times
1\times w,3,1)$ base blocks for any
$w\equiv2\hspace{-0.3cm}\pmod{6}$ and $w\geq8$.
\end{lemma}

\Proof Let $X=I\times Z_w$ and $\mathcal{G}=\{\{i\}\times Z_w:i\in
I\}$, where $I=Z_7\cup\{\infty\}$. We construct the required designs
directly.

$(1)$ $w=12s+2$ and $s\geq1$. All $112s+18$ are divided into two
parts. The first part consists of following $46$ base blocks:

\begin{center}
{ \small
\begin{tabular}{lll}
$\{(2,0),(4,1),(\infty,s)\}$, &$\{(1,0),(2,6s+1),(5,10s+1)\}$,
&$\{(2,0),(3,6s+1),(6,10s+1)\}$,\\

$\{(0,0),(5,s),(3,4s)\}$, &$\{(0,0),(5,4s+1),(6,10s+2)\}$,
&$\{(0,0),(6,10s+1),(1,10s+2)\}$,\\

$\{(0,0),(6,0),(\infty,s)\}$, &$\{(0,0),(1,4s+1),(2,9s+2)\}$,
&$\{(0,0),(2,4s+1),(1,6s+1)\}$,\\

$\{(0,0),(1,0),(2,0)\}$, &$\{(0,0),(5,6s+1),(6,8s+1)\}$,
&$\{(0,0),(6,6s+1),(5,8s+1)\}$,\\

$\{(1,0),(3,0),(\infty,s)\}$, &$\{(1,0),(6,4s+1),(4,8s+2)\}$,
&$\{(0,0),(4,10s+1),(5,12s+1)\}$,\\

$\{(1,0),(6,3s),(4,4s)\}$, &$\{(3,0),(4,4s+1),(6,8s+2)\}$,
&$\{(3,0),(4,5s+1),(5,6s+1)\}$,\\

$\{(4,0),(5,0),(\infty,s)\}$, &$\{(0,0),(1,2s+1),(3,8s+2)\}$,
&$\{(4,0),(5,5s+1),(6,6s+1)\}$,\\

$\{(0,0),(1,s),(6,2s)\}$, &$\{(4,0),(5,4s+1),(6,9s+2)\}$,
&$\{(1,0),(5,8s+2),(3,9s+2)\}$,\\

$\{(2,0),(3,0),(\infty,s-1)\}$, &$\{(1,0),(2,4s+1),(6,8s+1)\}$,
&$\{(1,0),(2,2s+1),(4,2s+1)\}$,\\

$\{(1,0),(5,4s),(3,8s+1)\}$, &$\{(2,0),(3,4s+1),(4,4s+1)\}$,
&$\{(2,0),(3,s),(4,11s+2)\}$,\\

$\{(2,0),(3,2s),(5,2s+1)\}$, &$\{(2,0),(5,8s+2),(6,8s+2)\}$,
&$\{(3,0),(4,2s),(5,4s+1)\}$,\\

$\{(3,0),(5,0),(4,6s+1)\}$, &$\{(3,0),(4,2s+1),(6,2s+1)\}$,
&$\{(4,0),(6,1),(5,10s+2)\}$,\\

$\{(0,0),(5,0),(\infty,s-1)\}$, &$\{(2,0),(3,5s+1),(4,6s+1)\}$,
&$\{(1,0),(2,s),(3,11s+2)\}$,\\

$\{(0,0),(4,4s),(2,8s+1)\}$, &$\{(0,0),(1,5s+1),(6,11s+2)\}$,
&$\{(1,0),(6,0),(\infty,s-1)\}$,\\

$\{(0,0),(5,3s),(6,7s+1)\}$, &$\{(0,0),(4,8s+2),(2,11s+2)\}$,
&$\{(0,0),(1,2s),(3,2s+1)\}$,\\

$\{(1,0),(2,2s),(3,4s+1)\}$.
\end{tabular}}
\end{center}

The second part consists of $112s-28$ base blocks. All base blocks
can be obtained by developing following $16s-4$ initial base blocks
by $(+1,-)$ mod $(7,-)$, where $\infty+1=\infty$.

\begin{center}
{ \small
\begin{tabular}{lll}
$\{(0,0),(3,6s+1),(\infty,8s+2)\}$,\\

$\{(0,0),(3,8s+1+2i),(\infty,4s+i)\}$, $i\in[0,s]$,\\

$\{(0,0),(3,4s+1+2i),(\infty,6s+1+i)\}$, $i\in[0,s-1]$,\\

$\{(0,0),(3,6s+2+2i),(\infty,5s+1+i)\}$, $i\in[0,s-1]$,\\

$\{(0,0),(3,8s+4+2i),(\infty,8s+3+i)\}$, $i\in[0,s-1]$,\\

$\{(0,0),(1,10s+1-i),(3,10s+3+i)\}$, $i\in[0,4s-1]$,\\

$\{(0,0),(2,8s+2+i),(4,6s+2+2i)\}$, $i\in[0,2s-1]\setminus\{s\}$,\\

$\{(0,0),(1,2s+1+i),(2,2i+1)\}$, $i\in[1,4s-1]\setminus\{2s,3s\}$,\\

$\{(0,0),(3,2s+1+2i),(\infty,3s+i)\}$, $i\in[1,s-1]$, (null if
$s=1$),\\

$\{(0,0),(3,6s+3+2i),(\infty,9s+3+i)\}$, $i\in[0,s-2]$, (null if
$s=1$).
\end{tabular}}
\end{center}

$(2)$ $w=12s+8$. All $112s+74$ are divided into two parts. The first
part consists of following $11$ base blocks:

\begin{center}
{ \small
\begin{tabular}{lll}
$\{(1,0),(6,0),(3,w/2)\}$, &$\{(0,0),(1,0),(2,0)\}$,
&$\{(0,0),(5,w/2),(6,w/2)\}$,\\

$\{(0,0),(5,0),(1,w/2)\}$, &$\{(0,0),(6,0),(2,w/2)\}$,
&$\{(0,0),(3,w/2),(4,w/2)\}$,\\

$\{(1,0),(3,0),(2,w/2)\}$, &$\{(4,0),(5,0),(6,w/2)\}$,
&$\{(1,0),(4,w/2),(6,w/2)\}$,\\

$\{(2,0),(3,0),(4,w/2)\}$, &$\{(2,0),(4,0),(5,w/2)\}$.
\end{tabular}}
\end{center}

The second part consists of $112s+63$ base blocks. All base blocks
can be obtained by developing following $16s+9$ initial base blocks
by $(+1,-)$ mod $(7,-)$, where $\infty+1=\infty$.

$s=0$.

\begin{center}
{\small
\begin{tabular}{lll}
$\{(0,0),(\infty,0),(2,1)\}$, &$\{(0,0),(3,0),(1,1)\}$,
&$\{(0,0),(3,1),(2,2)\}$,\\

$\{(0,0),(4,1),(\infty,2)\}$, &$\{(0,0),(3,2),(4,5)\}$,
&$\{(0,0),(5,2),(2,5)\}$,\\

$\{(0,0),(4,2),(\infty,6)\}$, &$\{(0,0),(1,2),(\infty,5)\}$,
&$\{(0,0),(6,2),(1,5)\}$.
\end{tabular}}
\end{center}

$s\geq1$.

\begin{center}
{ \small
\begin{tabular}{llll}
$\{(0,0),(1,s+1),(\infty,2s+2)\}$,
$\{(0,0),(1,5s+3),(\infty,11s+7)\}$,\\

$\{(0,0),(1,4s+2),(3,8s+4)\}$,
$\{(0,0),(4,2s+1),(\infty,9s+6)\}$,\\\vspace{0.1cm}

$\{(0,0),(4,4s+3),(\infty,10s+6)\}$,
$\{(0,0),(\infty,s),(3,10s+6)\}$,\\

$\{(0,0),(2,8s+6+i),(4,6s+5+2i)\}$, $i\in[0,2s]$,\\

$\{(0,0),(3,2s+4+2i),(\infty,3s+3+i)\}$, $i\in[0,s-1]$,\\

$\{(0,0),(3,4s+4+2i),(\infty,6s+5+i)\}$, $i\in[0,s-1]$,\\

$\{(0,0),(3,6s+5+2i),(\infty,5s+3+i)\}$, $i\in[0,s-1]$,\\

$\{(0,0),(3,8s+7+2i),(\infty,8s+6+i)\}$, $i\in[0,s-1]$,\\

$\{(0,0),(3,8s+6+2i),(\infty,4s+3+i)\}$, $i\in[0,s-1]$,\\

$\{(0,0),(1,10s+7-i),(3,10s+8+i)\}$, $i\in[0,4s+2]$,\\

$\{(0,0),(1,2s+2+i),(2,2i+2)\}$,
$i\in[0,4s+1]\setminus\{2s,3s+1\}$,\\

$\{(0,0),(3,6s+6+2i),(\infty,9s+7+i)\}$, $i\in[0,s-2]$, (null if
$s=1$).
\end{tabular}}
\end{center}

\begin{lemma}\label{w=12t+10}
There exists a $3$-SCGDP of type $w^8$ with $J^\ast(8\times 1\times
w,3,1)$ base blocks for any $w\equiv10\hspace{-0.3cm}\pmod{12}$.
\end{lemma}

\Proof Let $w=12s+10$. The required designs are constructed on
$I\times Z_w$ with group set $\{\{i\}\times Z_w:i\in I\}$, where
$I=Z_7\cup\{\infty\}$. The $112s+92$ base blocks are divided into
two parts. The first part consists of $8$ base blocks. In the second
part, we only list $16s+12$ initial base blocks. All other base
blocks are obtained by developing the initial base blocks by
$(+1,-)$ mod $(7,-)$, where $\infty+1=\infty$. \vspace{0.1cm}

Part I: \vspace{0.1cm}

{\small\hspace{2cm}$\{(2,0),(3,0),(4,0)\}$,
$\{(1,0),(3,0),(6,w/2)\}$, $\{(0,0),(3,w/2),(5,0)\}$,

\hspace{2cm}$\{(4,0),(5,0),(6,0)\}$, $\{(0,0),(2,w/2),(6,0)\}$,
$\{(1,0),(4,w/2),(6,0)\}$,

\hspace{2cm}$\{(0,0),(1,0),(2,0)\}$, $\{(1,0),(3,w/2),(5,w/2)\}$.}
\vspace{0.2cm}

Part II:\vspace{0.1cm}

\begin{center}
{ \small
\begin{tabular}{ll}
$\{(0,0),(3,6s+4),(\infty,7s+5)\}$,
$\{(0,0),(2,8s+7),(\infty,s)\}$,\\

$\{(0,0),(1,6s+5),(3,10s+8)\}$,
$\{(0,0),(1,s+1),(3,10s+9)\}$,\\\vspace{0.1cm}

$\{(0,0),(1,4s+3),(\infty,9s+7)\}$, $\{(0,0),(1,5s+4),(3,4s+3)\}$,\\

$\{(0,0),(3,8s+7+2i),(\infty,8s+6+i)\}$, $i\in[0,s]$,\\

$\{(0,0),(3,6s+6+2i),(\infty,9s+8+i)\}$, $i\in[0,s]$,\\

$\{(0,0),(2,8s+8+i),(4,6s+7+2i)\}$, $i\in[0,2s]\setminus\{s\}$,\\

$\{(0,0),(1,10s+8-i),(3,10s+10+i)\}$, $i\in[0,4s+2]$,\\

$\{(0,0),(1,2s+2+i),(2,1+2i)\}$,
$i\in[0,4s+2]\setminus\{2s+1,3s+2\}$,\\

$\{(0,0),(3,2s+4+2i),(\infty,3s+3+i)\}$, $i\in[0,s-1]$ (null if
$s=0$),\\

$\{(0,0),(3,4s+4+2i),(\infty,6s+5+i)\}$, $i\in[0,s-1]$ (null if
$s=0$),\\

$\{(0,0),(3,6s+7+2i),(\infty,5s+5+i)\}$, $i\in[0,s-1]$ (null if
$s=0$),\\

$\{(0,0),(3,8s+8+2i),(\infty,4s+3+i)\}$, $i\in[0,s-1]$ (null if
$s=0$).
\end{tabular}}
\end{center}

\begin{lemma}\label{w=6t+5,w^8}
There exists a $3$-SCGDP of type $w^{8}$ with $J^\ast(8\times
1\times w,3,1)$ base blocks for any
$w\equiv5\hspace{-0.3cm}\pmod{6}$.
\end{lemma}

\Proof Let $w=6s+5$. The required designs are constructed on
$I\times Z_w$ with group set $\{\{i\}\times Z_w:i\in I\}$, where
$I=Z_7\cup\{\infty\}$. The $56s+45$ base blocks are divided into two
parts. The first part consists of $3$ base blocks. In the second
part, we only list $8s+6$ initial base blocks. All other base blocks
are obtained by developing the initial base blocks by $(+1,-)$ mod
$(7,-)$, where $\infty+1=\infty$.\vspace{0.1cm}

\noindent $\bullet$ $s\in\{0,1,2\}$:\vspace{0.1cm}

Part I:
\begin{center}
{\small$\{(0,0),(1,0),(\infty,0)\}$, $\{(2,0),(3,0),(\infty,0)\}$,
$\{(4,0),(5,0),(\infty,0)\}$.}
\end{center}

Part II:\vspace{0.1cm}

$s=0:$

\begin{center}
{\small $\{(0,0),(2,0),(1,1)\}$, $\{(0,0),(3,0),(5,2)\}$,
$\{(0,0),(2,1),(\infty,2)\}$,

$\{(0,0),(4,1),(1,3)\}$, $\{(0,0),(3,1),(6,3)\}$,
$\{(0,0),(5,1),(\infty,4)\}$.}
\end{center}

$s=1:$

\begin{center}
{ \small
\begin{tabular}{llll}
$\{(0,0),(4,3),(1,8)\}$, &$\{(0,0),(1,4),(\infty,8)\}$,
&$\{(0,0),(2,0),(1,1)\}$, &$\{(0,0),(3,0),(5,5)\}$,\\
$\{(0,0),(2,1),(6,2)\}$, &$\{(0,0),(\infty,3),(3,7)\}$,
&$\{(0,0),(5,1),(2,9)\}$, &$\{(0,0),(\infty,1),(1,7)\}$,\\
$\{(0,0),(1,2),(4,7)\}$, &$\{(0,0),(2,2),(4,9)\}$,
&$\{(0,0),(4,2),(5,8)\}$, &$\{(0,0),(\infty,2),(5,7)\}$,\\
$\{(0,0),(1,3),(2,8)\}$, &$\{(0,0),(3,1),(\infty,10)\}$.
\end{tabular}}
\end{center}

$s=2:$

\begin{center}
{ \small
\begin{tabular}{llll}
$\{(0,0),(\infty,5),(3,9)\}$, &$\{(0,0),(3,4),(2,8)\}$,
&$\{(0,0),(2,0),(1,1)\}$, &$\{(0,0),(4,1),(\infty,16)\}$,\\
$\{(0,0),(2,1),(5,14)\}$, &$\{(0,0),(3,0),(4,5)\}$,
&$\{(0,0),(5,1),(1,12)\}$, &$\{(0,0),(\infty,1),(6,8)\}$,\\
$\{(0,0),(1,2),(4,7)\}$, &$\{(0,0),(3,1),(1,11)\}$,
&$\{(0,0),(2,2),(5,5)\}$, &$\{(0,0),(3,2),(\infty,14)\}$,\\
$\{(0,0),(4,2),(5,6)\}$, &$\{(0,0),(5,2),(2,13)\}$,
&$\{(0,0),(6,2),(\infty,8)\}$, &$\{(0,0),(\infty,2),(5,8)\}$,\\
$\{(0,0),(1,3),(2,10)\}$, &$\{(0,0),(4,3),(5,11)\}$,
&$\{(0,0),(\infty,3),(6,11)\}$, &$\{(0,0),(6,3),(\infty,7)\}$,\\
$\{(0,0),(5,3),(4,10)\}$, &$\{(0,0),(2,4),(4,9)\}$.
\end{tabular}}
\end{center}

\noindent $\bullet$ $s\equiv1\hspace{-0.3cm}\pmod{2}$ and
$s\geq3$:\vspace{0.1cm}

Part I:

\begin{center}
{\small$\{(4,0),(5,0),(\infty,(11s+9)/2)\}$,
$\{(0,0),(1,0),(\infty,(11s+9)/2)\}$,
$\{(2,0),(3,0),(\infty,(11s+9)/2)\}$.}
\end{center}

Part II:

\begin{center}
{ \small
\begin{tabular}{llll}
$\{(0,0),(3,4s+2),(\infty,4s+2)\}$,
$\{(0,0),(3,s+1),(\infty,5s+4)\}$,\\\vspace{0.1cm}

$\{(0,0),(3,3s+2),(\infty,5s+3)\}$,
$\{(0,0),(1,2s+2),(3,4s+4)\}$,\\

$\{(0,0),(3,4s+3),(\infty,(5s+3)/2)\}$,\\

$\{(0,0),(1,5s+3-i),(3,5s+4+i)\}$, $i\in[0,2s]$,\\

$\{(0,0),(2,4s+3+i),(4,3s+2+2i)\}$, $i\in[0,s]$,\\

$\{(0,0),(1,s+1+i),(2,2i)\}$, $i\in[0,2s+1]\setminus\{s+1\}$,\\

$\{(0,0),(3,s+2i),(\infty,(3s+1)/2+i)\}$, $i\in[0,(s-1)/2]$,\\

$\{(0,0),(3,2s+1+2i),(\infty,3s+2+i)\}$, $i\in[0,(s-1)/2]$,\\

$\{(0,0),(3,4s+6+2i),(\infty,2s+2+i)\}$, $i\in[0,(s-3)/2]$,\\

$\{(0,0),(3,4s+5+2i),(\infty,4s+4+i)\}$, $i\in[0,(s-3)/2]$,\\

$\{(0,0),(3,3s+4+2i),(\infty,(5s+5)/2+i)\}$, $i\in[0,(s-3)/2]$,\\

$\{(0,0),(3,3s+5+2i),(\infty,(9s+9)/2+i)\}$, $i\in[0,(s-5)/2]$ (null
if $s=3$).
\end{tabular}}
\end{center}

\noindent $\bullet$ $s\equiv0\hspace{-0.3cm}\pmod{2}$ and
$s\geq4$:\vspace{0.2cm}

Part I:

\begin{center}
{\small$\{(0,0),(1,0),(\infty,11s/2+5)\}$,
$\{(2,0),(3,0),(\infty,11s/2+5)\}$,
$\{(4,0),(5,0),(\infty,11s/2+5)\}$.}
\end{center}

Part II:

\begin{center}
{ \small
\begin{tabular}{llll}
$\{(0,0),(1,2s+1),(2,2s-2)\}$, $\{(0,0),(1,s+1),(2,1)\}$, \\

$\{(0,0),(1,5s+3),(3,5s+3)\}$, $\{(0,0),(1,2s),(3,4s)\}$,\\

$\{(0,0),(3,s+1),(\infty,3s+3)\}$,
$\{(0,0),(3,4s+2),(\infty,0)\}$,\\\vspace{0.1cm}

$\{(0,0),(1,2s+2),(3,4s+4)\}$,
$\{(0,0),(1,6s+4),(\infty,5s+2)\}$,\\

$\{(0,0),(3,5s+4),(\infty,9s/2+3)\}$,\\

$\{(0,0),(2,4s+3+i),(4,3s+2+2i)\}$, $i\in[0,s]$,\\

$\{(0,0),(3,s+2i),(\infty,3s/2+1+i)\}$, $i\in[0,s/2]$,\\

$\{(0,0),(1,s+3+2i),(2,2+4i)\}$, $i\in[0,s/2-2]$,\\

$\{(0,0),(1,s+2+2i),(2,4+4i)\}$, $i\in[0,s/2-2]$,\\

$\{(0,0),(3,2s+2+2i),(\infty,3s+4+i)\}$, $i\in[0,s/2]$,\\

$\{(0,0),(1,2s+3+i),(2,2s+4+2i)\}$, $i\in[0,s-1]$,\\

$\{(0,0),(1,5s+2-i),(3,5s+5+i)\}$, $i\in[0,2s-1]$,\\

$\{(0,0),(3,4s+6+2i),(\infty,2s+4+i)\}$, $i\in[0,s/2-2]$,\\

$\{(0,0),(3,4s+5+2i),(\infty,4s+4+i)\}$, $i\in[0,s/2-2]$,\\

$\{(0,0),(3,3s+5+2i),(\infty,5s/2+3+i)\}$, $i\in[0,s/2-1]$,\\

$\{(0,0),(3,3s+4+2i),(\infty,9s/2+4+i)\}$, $i\in[0,s/2-3]$ (null if
$s=4$).
\end{tabular}}
\end{center}

\begin{lemma}\label{w=6t+5,w^14}
There exists a $3$-SCGDP of type $w^{14}$ with $J^\ast(14\times
1\times w,3,1)$ base blocks for any
$w\equiv5\hspace{-0.3cm}\pmod{6}$.
\end{lemma}

\Proof When $w=5$, the required design is constructed on $I\times
Z_5$ with group set $\{\{i\}\times Z_5:i\in I\}$, where
$I=Z_{13}\cup\{\infty\}$. The $149$ base blocks are divided into two
parts. The first part consists of $6$ base blocks.

\begin{center}
{ \small
\begin{tabular}{lll}
$\{(0,0),(1,0),(\infty,0)\}$, &$\{(2,0),(3,0),(\infty,0)\}$,
&$\{(4,0),(5,0),(\infty,0)\}$,\\

$\{(6,0),(7,0),(\infty,0)\}$, &$\{(8,0),(9,0),(\infty,0)\}$,
&$\{(10,0),(11,0),(\infty,0)\}$.
\end{tabular}}
\end{center}

In the second part, we only list $11$ initial base blocks. All base
blocks are obtained by developing the initial base blocks by
$(+1,-)$ modulo $(13,-)$, where $\infty+1=\infty$.\vspace{0.2cm}

\begin{center}
{ \small
\begin{tabular}{llll}
$\{(0,0),(8,1),(2,3)\}$, &$\{(0,0),(2,0),(5,0)\}$,
&$\{(0,0),(4,1),(3,2)\}$, &$\{(0,0),(4,2),(\infty,4)\}$, \\

$\{(0,0),(4,0),(1,1)\}$, &$\{(0,0),(6,0),(2,1)\}$,
&$\{(0,0),(3,1),(1,2)\}$, &$\{(0,0),(\infty,1),(8,3)\}$,\\

$\{(0,0),(5,1),(7,3)\}$, &$\{(0,0),(6,1),(1,3)\}$,
&$\{(0,0),(7,1),(3,3)\}$.
\end{tabular}}
\end{center}

When $w\geq11$, let $X=I_{14}\times Z_w$ and
$\mathcal{G}=\{\{i\}\times Z_w:i\in I_{14}\}$. We divide the
construction of the $J^\ast(14\times 1\times w,3,1)=(91w-8)/3$ base
blocks into three parts.\vspace{0.2cm}

Part I: Let $(I_{14},\mathcal{B})$ be a PBD($14,\{3,4,5^*\}$) and
$I_5$ be the unique block of size $5$. For each
$B\in\mathcal{B}\setminus\{I_5\}$, construct a $3$-SCHGDD of type
$(|B|,1^w)$ on $B\times Z_w$ with group set $\{\{x\}\times Z_w:x\in
B\}$ and hole set $\{B\times \{i\}:i\in Z_w\}$. Denote the base
blocks by $\mathcal{A}_B$. Let
$\mathcal{A}_1=\sum_{B\in\mathcal{B}\setminus\{I_5\}}\mathcal{A}_B$.
It is readily checked that $|\mathcal{A}_1|=({14\choose 2}-{5\choose
2})(w-1)/3=27(w-1)$.\vspace{0.2cm}

Part II: Construct a $3$-GDP of type $1^{14}$ on
$I_{14}\times\{0\}$, which consists of the following $26$ blocks.
Let $\mathcal{A}_2$ denote the block set. For saving space, we omit
the second candidates.\vspace{0.2cm}

\hspace{2cm}{\small$\{0,8,10\}$, $\{0,9,12\}$, $\{2,7,11\}$,
$\{0,4,6\}$, $\{6,9,10\}$, $\{3,6,8\}$,

\hspace{2cm}$\{1,6,11\}$, $\{1,7,12\}$, $\{2,5,10\}$, $\{1,2,8\}$,
$\{2,6,12\}$, $\{3,10,11\}$,

\hspace{2cm}$\{3,4,13\}$, $\{3,5,12\}$, $\{4,7,10\}$, $\{3,7,9\}$,
$\{7,8,13\}$, $\{1,10,13\}$,

\hspace{2cm}$\{4,8,12\}$, $\{4,9,11\}$, $\{5,6,13\}$, $\{1,5,9\}$,
$\{5,8,11\}$, $\{0,11,13\}$,

\hspace{2cm}$\{2,9,13\}$, $\{0,5,7\}$.}\vspace{0.2cm}

Let
$M=\{\{0,1\},\{0,2\},\{0,3\},\{1,3\},\{1,4\},\{2,3\},\{2,4\},\{4,5\},\{6,7\},\{8,9\},\{10,12\}$,
$\{11,12\},\{12,13\}\}$. Note that all pairs in $M$ are not covered
by above blocks.

\vspace{0.2cm}Part III: Construct a $3$-SCGDP of type $w^5$ on
$I_{5}\times Z_w$ with group set $\{\{i\}\times Z_w:i\in I_{5}\}$,
which consists of $(10w-5)/3$ base blocks. Let $\mathcal{A}_3$
denote these base blocks.

We first list $2(w-8)/3$ initial base blocks. We can obtain
$10(w-8)/3$ base blocks by developing these blocks by $(+1,-)$
modulo $(5,-)$. Here, $2\leq i\leq(w-5)/3$.

\begin{center} {\small$\{(0,0),(1,i),(2,2(w-2)/3+2i)\}$,
$\{(0,0),(1,(w-2)/3+i),(3,2(w-2)/3-i+1)\}$.}
\end{center}

The remaining $25$ base blocks in this part are listed below.

\begin{center}
{ \small
\begin{tabular}{llll}
$\{(1,0),(2,1),(4,(w+1)/3)\}$, &$\{(0,0),(1,1),(3,(w-5)/3)\}$,\\

$\{(0,0),(4,1),(1,(w-2)/3)\}$, &$\{(0,0),(4,2),(2,(w+1)/3)\}$,\\

$\{(1,0),(4,(w-2)/3),(2,w-2)\}$,
&$\{(0,0),(3,2(w+1)/3),(1,w-1)\}$,\\

$\{(0,0),(4,2(w+1)/3),(1,w-2)\}$,
&$\{(0,0),(2,(2w+5)/3),(4,w-1)\}$,\\

$\{(0,0),(2,(w-5)/3),(3,(w-2)/3)\}$,
&$\{(0,0),(2,(w-2)/3),(1,(w+1)/3)\}$,\\

$\{(0,0),(2,(w-8)/3),(1,2(w-2)/3)\}$,
&$\{(0,0),(3,(w+1)/3),(1,(2w-1)/3)\}$,\\

$\{(0,0),(4,(w+1)/3),(2,(2w-1)/3)\}$,
&$\{(1,0),(3,(w-5)/3),(2,(w+1)/3)\}$,\\

$\{(1,0),(4,(w-5)/3),(3,(w+1)/3)\}$,
&$\{(1,0),(2,(w-2)/3),(4,(2w-1)/3)\}$,\\

$\{(1,0),(3,(w-2)/3),(2,(2w-1)/3)\}$,
&$\{(1,0),(3,(w+4)/3),(4,2(w+1)/3)\}$,\\

$\{(2,0),(4,(w-5)/3),(3,2(w-2)/3)\}$,
&$\{(2,0),(3,(w-2)/3),(4,(2w-1)/3)\}$,\\

$\{(0,0),(2,(w+4)/3),(3,(2w+5)/3)\}$,
&$\{(0,0),(4,(w+4)/3),(3,2(w+4)/3)\}$,\\

$\{(1,0),(4,2(w-2)/3),(3,(2w-1)/3)\}$,
&$\{(0,0),(3,2(w-2)/3),(4,(2w-1)/3)\}$,\\

$\{(0,0),(3,(2w-1)/3),(2,2(w+1)/3)\}$.
\end{tabular}}
\end{center}

Finally, let $\mathcal{A}=\sum_{i=1}^3\mathcal{A}_i$. It is readily
checked that $\mathcal{A}$ forms the base blocks of the required
design. Clearly $|\mathcal{A}|=(91w-8)/3$. \qed